\pdfoutput=1
\documentclass[]{aa}
\usepackage{graphicx}
\usepackage{txfonts}
\usepackage[utf8]{inputenc}
\usepackage{threeparttable}
\usepackage{amsmath}
\usepackage{hyperref}
\usepackage{float}
\usepackage[toc,page]{appendix} 
\usepackage[caption = false]{subfig}
\usepackage{newfloat}
\DeclareFloatingEnvironment[name={Supplementary Figure}]{suppfigure}

\begin{document}

   \title{The HST Key Project galaxies NGC 1326A, NGC 1425 and \\ NGC 4548: New variable stars and massive star population}

	\author{ Z.T. Spetsieri\inst{1,2} et al.}
   \author{Z.T. Spetsieri\inst{1,2}, A.Z. Bonanos \inst{1}, M. Yang\inst{1}, M. Kourniotis \inst{3}, D. Hatzidimitriou \inst{2,1}}

\authorrunning{Spetsieri et al.} 
\titlerunning{Massive star population in NGC 1326A, NGC 1425 and NGC 4548}
\institute{ IAASARS, National Observatory of Athens, 15236 Penteli, Greece\\
\email{zspetsieri@noa.gr}
 \and Department of Astrophysics, Astronomy \& Mechanics, Faculty of
Physics, University of Athens, 15783 Athens, Greece
\and Astronomick\'y \'ustav, Akademie v\v{e}d \v{C}esk\'e republiky, Fri\v{c}ova 298, 251\,65 Ond\v{r}ejov, Czech Republic}
%
 
\abstract {Studies on the massive star population in galaxies beyond the Local Group are the key to understand the link between their numbers and modes of star formation in different environments. We present the analysis of the massive star population of the galaxies NGC 1326A, NGC 1425 and NGC 4548 using archival Hubble Space Telescope Wide Field Planetary Camera 2 images in the F555W and F814W filters. Through high precision point spread function fitting photometry for all sources in the three fields we identified 7640 candidate blue supergiants, 2314 candidate yellow supergiants, and 4270 candidate red supergiants. We provide an estimation the ratio of blue to red supergiants for each field as a function of galactocentric radius. Using Modules for Experiments in Stellar Astrophysics (MESA) at solar metallicity, we defined the luminosity function and estimated the star formation history of each galaxy. We carried out a variability search in the \textit{V} and \textit{I} filters using three variability indexes: the median absolute deviation, the interquartile range, and the inverse von-Neumann ratio. This analysis yielded 243 new variable candidates with absolute magnitudes ranging from M$_{V}$ = $-$4 to $-$10 mag. We classified the variable stars based on their absolute magnitude and their position on the color-magnitude diagram using the MESA evolutionary tracks at solar metallicity. Our analysis yielded 8 candidate variable blue supergiants, 12 candidate variable yellow supergiants, 21 candidate variable red supergiants, and 4 candidate periodic variables.} 



    \maketitle
%

\section{Introduction}

\par Variability studies at large distances can be used as a powerful tool to study the pre-supernova state and evolved states of massive stars \citep[]{Kochanek2017}. The open questions of stellar evolution regarding mass loss, internal mixing and stellar winds \citep[]{Langer2012, Massey2003} have been frequently investigated in conjunction with photometric variability studies \citep{Bonanos2007, Ming, Kourniotis2016, Soraisam2018}. The evolution of very massive stars is quite uncertain both due to the limited observational samples and physical processes that trigger variability. For example, the origin of variability in the light curves of luminous blue variables (LBVs) in metal poor environments \citep{Kalari2018}, the physical mechanisms causing shear-induced instabilities in massive stars \citep[]{Maeder2000, Heger2000} or evolutionary paths leading to binarity in yellow and red supergiants \citep[]{Prieto2008, Moriya2018} are some of the open questions that are yet to be investigated. The short evolution timescale, combined with the rarity of massive stars are the biggest challenges in testing stellar evolutionary models against observations. A solution to overcome the problem of the limited observational sample of massive stars is to take advantage of the large amount of existing archival data. 
\par Archival data from the HST allow us to extend the studies of massive stars and trace star formation beyond the Local Group. Archives are widely recognized as a valuable resource for astronomy. The large sample of deep extragalactic HST observations, gives us the opportunity to study in detail the stellar population of blue supergiants (BSGs), yellow supergiants (YSGs) and red supergiants (RSGs) in galaxies beyond 10 Mpc. Studying massive stars at larger distances is the key to understand whether the evolution of massive stars is different in different environments. For example, the ratio of the BSGs to RSGs (B/R ratio) as a function of metallicity and galactocentric radius, is among the diagnostic tools applied to study metallicity, star formation, mass loss and convection processes \citep[]{brunish1986,LangerBsgs}. Our previous study on the massive star population of the Virgo cluster galaxy NGC 4535 \citep[Paper I]{Spetsieri2018} demonstrates the importance of archival HST data for extra-galactic studies on massive stars. NGC 4535 was studied by \citet{Macri99} for Cepheid variables as part of the HST Key project and 50 Cepheid variables were detected. Using data from the HST we conducted our own photometry for NGC 4535 aiming to the identification of more massive variable stars. Our study unveiled 120 new variable massive stars among which eight luminous candidate RSGs, four candidate YSGs and one candidate LBV.
\par In this study we aim to expand our search for massive stars, using the same method adopted in Paper I to three other galaxies studied in the HST Key project: NGC 1326A \citep{Prosser99}, NGC 1425 \citep{Mould2000} and the Virgo Cluster galaxy NGC 4548 \citep{Graham1999}. All three fields were previously studied for Cepheid variables with 15 Cepheids reported in NGC 1326A, 20 in NGC 1425, and 24 in NGC 4548. However, the existence of other objects that were not classified as Cepheids but showed variability in their light curve \citep{Graham1999} is yet to be investigated. These three galaxies were selected as metal rich galaxies with high star formation rates that can help in providing a better view of the variability displayed in the massive star population as massive stars serve as the primary source of carbon, nitrogen, and oxygen (CNO) enrichment of the interstellar medium (ISM) and are progenitors of supernovae \citep[]{Conroy2018, Maeder81}. 
\par The paper is structured as follows, the observations and data reduction are presented in Section~\ref{Observations}, the massive star population of NGC 1326A, NGC 1425, NGC 4548 and the blue to red supergiant ratio are described in Section~\ref{analysis1}. The selection of variable candidates and the classification of the new massive variable candidates are given in Section~\ref{selection}. The summary is given in Section~\ref{summary}.

\par 

\section{Observations and reduction}
\label{Observations}
\subsection{Observations}
We used archival observations of NGC 1326A, NGC 1425, and NGC 4548 taken with the HST WFPC2 as described in \citet{Freedman2001} as part of the HST Key Project. For the three galaxies, we used the observations available in the filters F555W (equivalent to Johnson V filter) and the F814W (equivalent to Kron-Cousins I). In particular, we used 13 epochs of observations in F555W and 8 epochs in F814W that are available 
for the galaxies NGC 1326A and NGC 4548, while for NGC 1425, we used the 14 available epochs in F555W and 8 in F814W . The F555W data consisted of 3x1200 $\sec$ exposures and 4x1300 $\sec$ exposures, while the F814W data consisted of 3x1300 $\sec$ exposures. Data were collected at four dithering positions, with one quarter of the observations (four pairs of F814W images and two pairs of F555W images) taken at each position. We retrieved the pre-reduced images through the Mikulsky Archive for Space Telescopes\footnote{\url{http://archive.stsci.edu/hst/search.php}} and performed point spread function (PSF) photometry using the latest WFPC2 module of \texttt{DOLPHOT} (updated in 2016), which is a modified version of \texttt{HSTphot} \citep{Dolphin2000}.

\subsection{Reduction}
\label{Reduction}
 We proceeded with the photometric reduction as described in the manual of the WFPC2 module of DOLPHOT and in Paper I. We applied the photometric quality criteria listed in the \texttt{DOLPHOT} manual to distinguish the isolated, point-like sources from the extended sources detected by \texttt{DOLPHOT}. The criteria are: S/N $\geq$ 10, $-$0.3$\leq$ sharpness $\leq$0.3, $\chi^{2}$ $\leq$ 2.5 and Object type=1.
 The resulting catalog for each field includes the coordinates of each source in RA, Dec (J2000), X, Y, and mean magnitude in V and I filters. The total number of stars after the quality cuts are: 5897 stars in NGC 1326A, 5232 stars in NGC 1425, and 5790 stars in NGC 4548. Tables~\ref{allsources1326}$-$ \ref{allsources4548} present the first 10 entries of each catalog listing the ID, RA (J2000), Dec (J2000), X, Y mean magnitudes and magnitude errors in V and I. The catalogs for the three galaxies are available in a machine-readable form at the Centre de Données Astronomiques de Strasbourg (CDS).
 
\par
\begin{figure*}[h!tb]
\hspace{-1.3 cm}
   \includegraphics[scale=0.49]{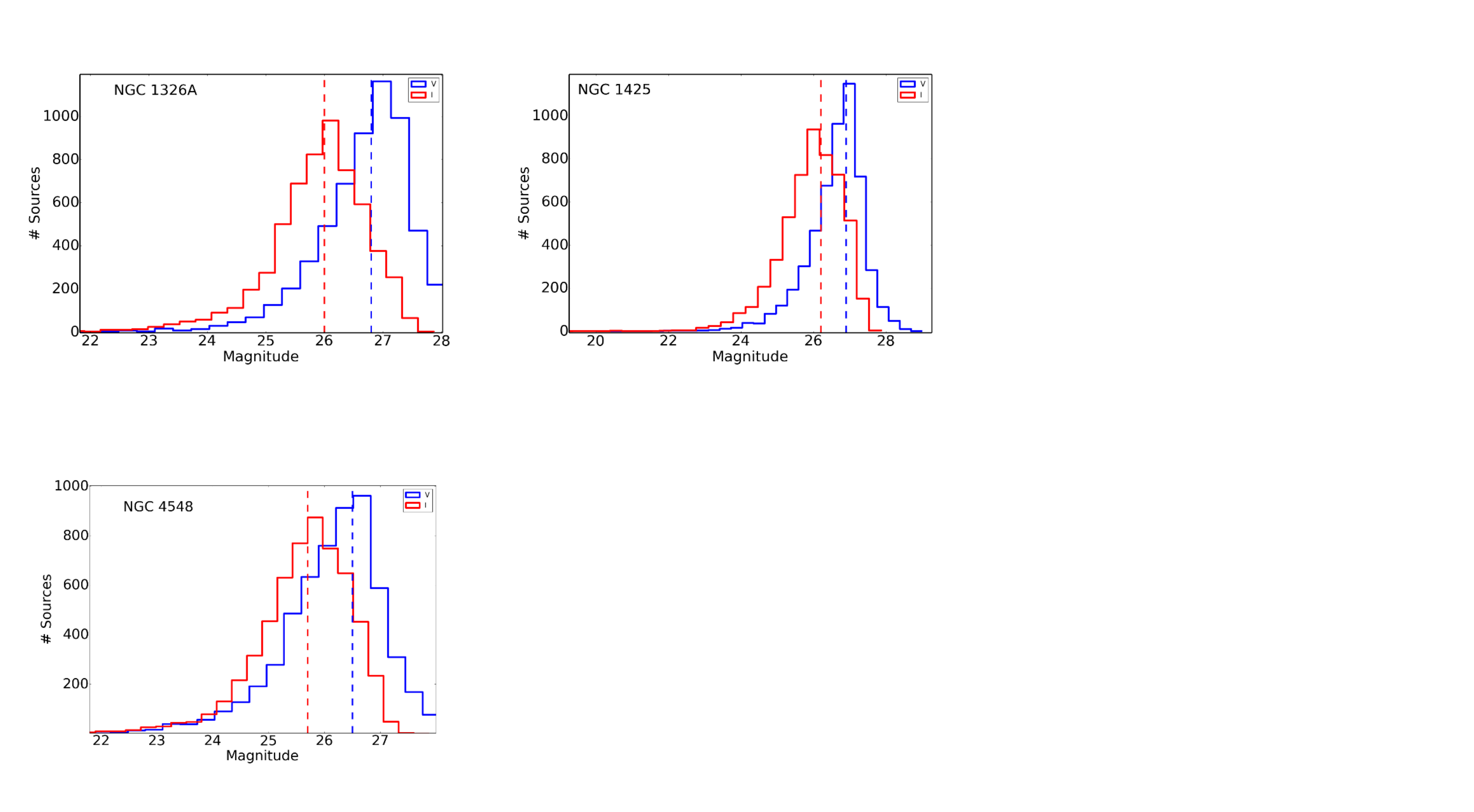}
   \caption{Histogram of the number of sources in filters V and I in NGC 1326A, NGC 1425, and NGC 4548. The blue and red dashed vertical line correspond to the magnitude in V and I, respectively where our sample reaches 50$\%$ completeness.}
\label{histogram}
\end{figure*}

\begin{figure*}[h!tb]
\hspace{-1.3 cm}
   \includegraphics[scale=0.55]{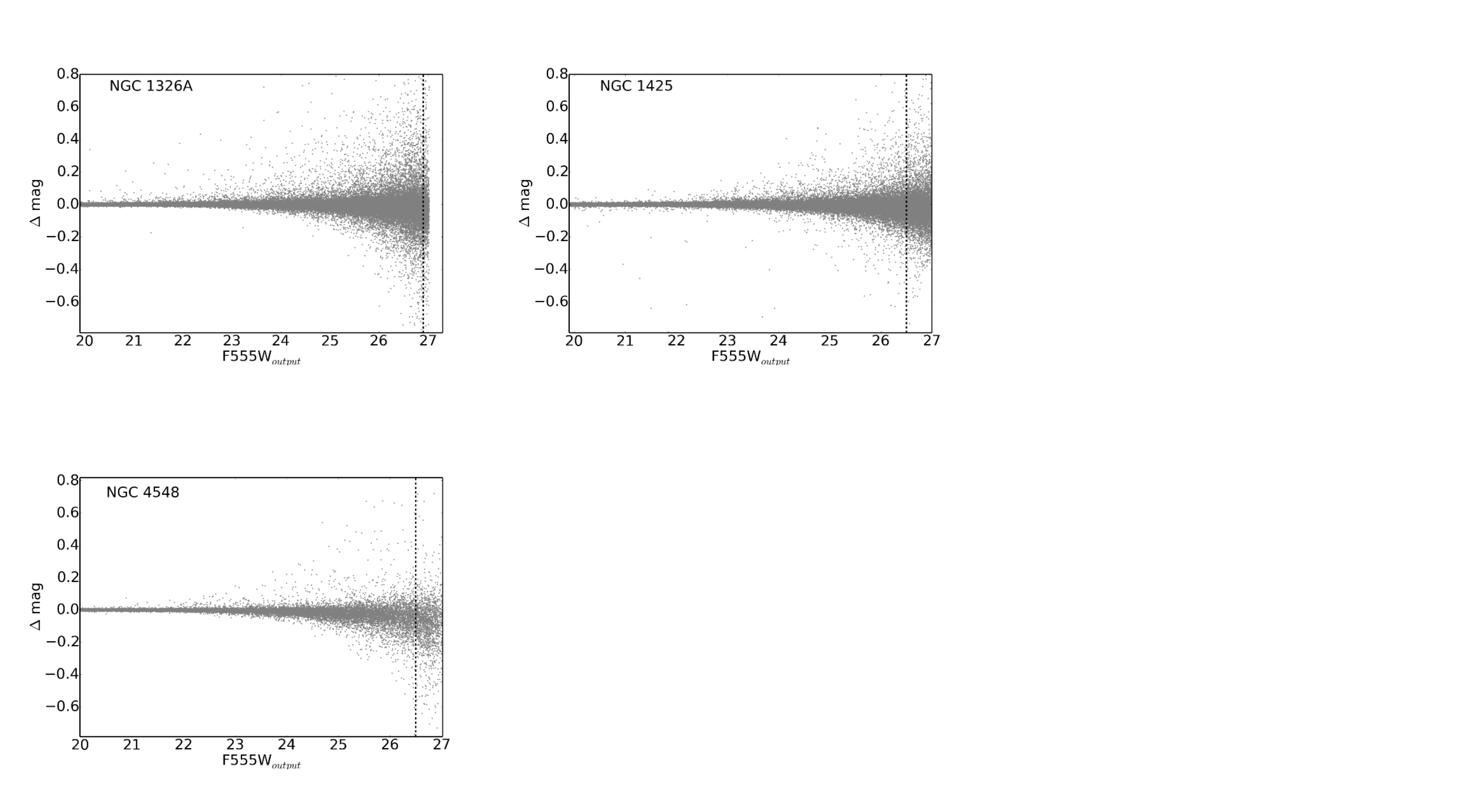}
   \caption{The difference between input and output magnitudes as a function of the output magnitude for NGC 1326A, NGC 1425, and NGC 4548. The dashed line indicates the 50$\%$ completeness magnitude.}
\label{artstars}
\end{figure*}

 We estimated the photometric errors and completeness by conducting artificial star tests as described in \citet{Lianou2013}. The magnitude and errors of the artificial stars inserted in each field  were measured at the same time as the field stars. We ran the tests injecting 5000 stars per chip ranging in magnitude from 19 to 27 mag in both filters for all three fields. The mean error was estimated using bins of 0.15 mag, within 0.05 to 0.06 mag in V and 0.04 to 0.05 mag in I. The number of stars used for the artificial star tests did not cause overcrowding as the program measured one star at the time. Fig.~\ref{histogram} shows a histogram with the magnitudes of the sources of each galaxy after the photometric quality cuts. Fig.~\ref{artstars} shows an example of the mean magnitude in the V filter as a function of the difference in magnitude between the artificial stars inserted and the ones recovered for each galaxy. The 50$\%$ completeness factor in V and I bands, measured from the artificial star tests occurs at $\sim$ 26.8 mag and 26.0 mag in NGC 1326A, $\sim$ 26.9 mag and $\sim$ 26.2 mag in NGC 1425, and $\sim$ 26.5 mag and $\sim$ 25.8 mag in NGC 4548 respectively.
Astrometry was performed by using the Hubble Source Catalog (HSC) version 3 \citep{Whitmore16} \footnote{\url{http://archive.stsci.edu/hst/hsc/}} and applying the astrometric correction suggested by \citet{Anderson2000} and described in detail in Paper I. We used stars from the HSC at various positions across each chip to calculate the transformations from the X, Y positions to the RA, Dec. Relative positions are good to $\sim$ 0.05 arcsec, while the accuracy of the absolute positions is limited by the HST coordinates, which are constrained by the accuracy of the HST guide star catalog to $\sim$ 0.1$\arcsec$.

\section{Massive star population}
\label{analysis1}
\subsection{Identification of the massive star candidates}
\label{analysis}

We examined the massive star population in each field using color and magnitude criteria to determine the blue, yellow, and red supergiant regions. 

Figures~\ref{totalplot1326} to \ref{totalplot4548} show the V$-$I versus M$_{V}$ CMD, for the stars that passed the quality tests mentioned in Sect.~\ref{Observations}, based on the distance modulus of each field (see Table~\ref{comparison}). The error-bars shown were derived from the artificial star tests. The total foreground interstellar reddening is E(V$-$I)= 0.15 $\pm$ 0.04 mag in NGC 1326A, E(V$-$I)= 0.16 $\pm$ 0.03 mag in NGC 1425, and E(V$-$I)= 0.18 $\pm$ 0.04 mag in NGC 4548 \citep{Freedman2001}. The faint limit in each field corresponds to the V magnitude up to which our sample reaches 50$\%$ completeness, as at this magnitude the photometric errors do not exceed 0.1 mag. 
We adopted the color and magnitude criteria by \citet{GrammerM101} to separate the regions of BSG, YSGs, and RSGs. The three populations for each field range from:

\begin{center}
\begin{tabular}{c c c c c c }
 &NGC 1326A\\
 \\
 BSGs: & (V$-$I) $\leq$ 0.25 mag & 21.5 $\leq$ V $\leq$ 25.5 mag &&\\ 
       & (V$-$I) $\leq$ 0.60 mag & 25.5 $<$ V $\leq$ 26.8 mag &&\\  
 YSGs: & 0.25 $<$ (V$-$I) $\leq$ 0.6 mag & 21.5 $\leq$ V $\leq$ 25.5 mag &&\\    
       & 0.6 $<$ (V$-$I) $\leq$ 1.3 mag & 21.5 $<$ V $\leq$ 26.8 mag &&\\    
 RSGs: & (V$-$I) $>$ 1.3 mag & 21.5 $\leq$ V $\leq$ 26.8 mag &&\\    

\end{tabular}
\end{center}

\begin{center}
\begin{tabular}{c c c c c c }
 &NGC 1425\\
 \\
 BSGs: & (V$-$I) $\leq$ 0.25 mag & 21.5 $\leq$ V $\leq$ 25.3 mag &&\\ 
       & (V$-$I) $\leq$ 0.60 mag & 25.3 $<$ V $\leq$ 26.9 mag &&\\  
 YSGs: & 0.25 $<$ (V$-$I) $\leq$ 0.6 mag & 21.5 $\leq$ V $\leq$ 25.3 mag &&\\    
       & 0.6 $<$ (V$-$I) $\leq$ 1.3 mag & 21.5 $<$ V $\leq$ 26.9 mag &&\\    
 RSGs: & (V$-$I) $>$ 1.3 mag & 21.5 $\leq$ V $\leq$ 26.9 mag &&\\    

\end{tabular}
\end{center}

\begin{center}
\begin{tabular}{ c c c c c c }
 &NGC 4548\\
 \\
 BSGs: & (V$-$I) $\leq$ 0.25 mag & 20.5 $\leq$ V $\leq$ 24.5 mag &&\\ 
       & (V$-$I) $\leq$ 0.60 mag & 24.5 $<$ V $\leq$ 26.5 mag &&\\  
 YSGs: & 0.25 $<$ (V$-$I) $\leq$ 0.6 mag & 20.5 $\leq$ V $\leq$ 24.5 mag &&\\    
       & 0.6 $<$ (V$-$I) $\leq$ 1.3 mag & 20.5 $<$ V $\leq$ 26.5 mag &&\\    
 RSGs: & (V$-$I) $>$ 1.3 mag & 20.5 $\leq$ V $\leq$ 26.5mag &&\\    

\end{tabular}
\end{center}

\par In total we identified 7640 candidate BSGs (2652 in NGC 1326A, 2833 NGC 1425, and 2155 in NGC 4548), 2314 candidate YSGs (868 in NGC 1326A, 939 in NGC 1425, and 507 in NGC 4548), and 4270 candidate RSGs (1868 in NGC 1326A, 1281 in NGC 1425, and 1121 in NGC 4548). 

\par We estimated the foreground contamination in the direction of the three fields through the Besan\c{c}on Galactic population synthesis model \citep{Besancon2003} and found that the majority of the foreground stars lie in the YSG and RSG regions. Foreground contamination has always been among the challenges in the identification of luminous massive stars such as YSGs and RSGs as these types of stars have similar magnitudes as Milky Way stars. Photometric errors near the completeness limits could also cause contamination, e.g. by YSGs that are part of the neighboring RSG population due to reddening, and extinction.

Table~\ref{properties} lists the properties (inclination, distance, distance modulus, metallicity, mophological type, redshift and absolute magnitude) of NGC 1326A, NGC 1425, and NGC 4548. Table~\ref{comparison} includes: the galaxy name, the numbers of candidate BSGs, YSGs, RSGs, along with those of their variable counterparts. As shown in Table~\ref{comparison} the galaxy with the highest foreground contamination is NGC 4548, followed by NGC 1425 and NGC 1326A.
 
\subsection{Blue to red supergiant ratio}

\par
\begin{figure*}[h!tb]
\centering
\includegraphics[trim=0cm 0cm 0.cm 0cm, width=0.85\textwidth]{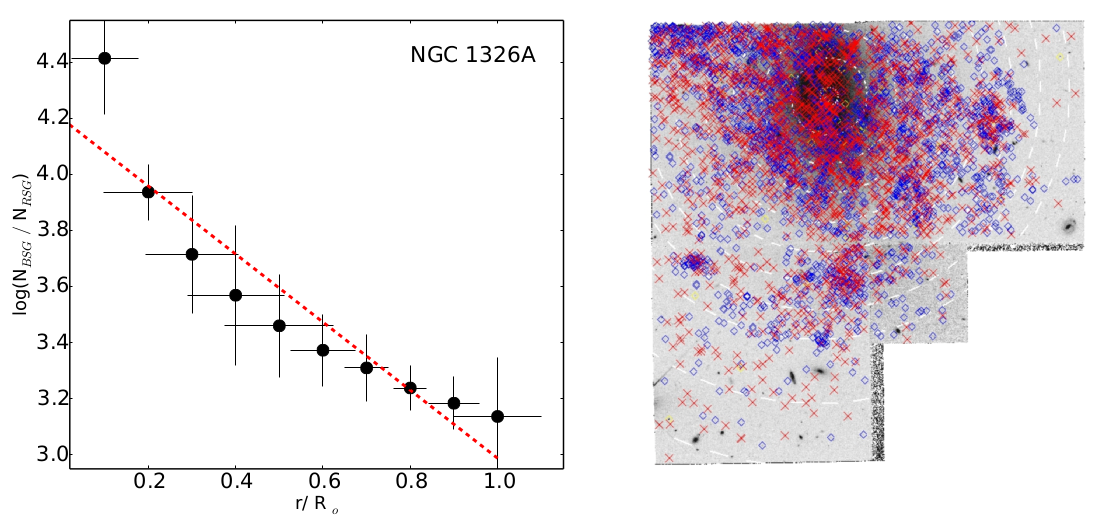}\\
\includegraphics[trim=0cm 0cm 0.cm 0cm, width=0.85\textwidth]{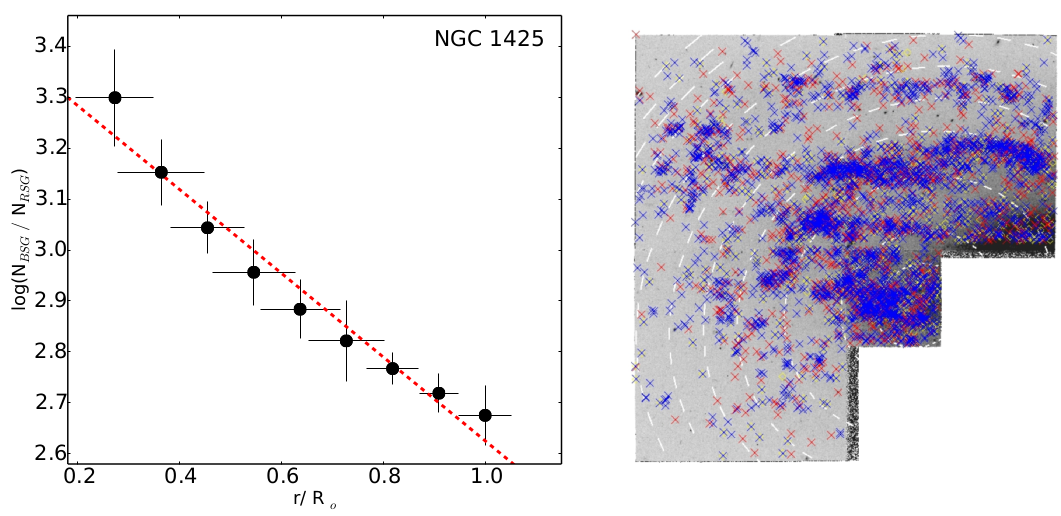}\\
\includegraphics[trim=0cm 0cm 0.cm 0cm, width=0.85\textwidth]{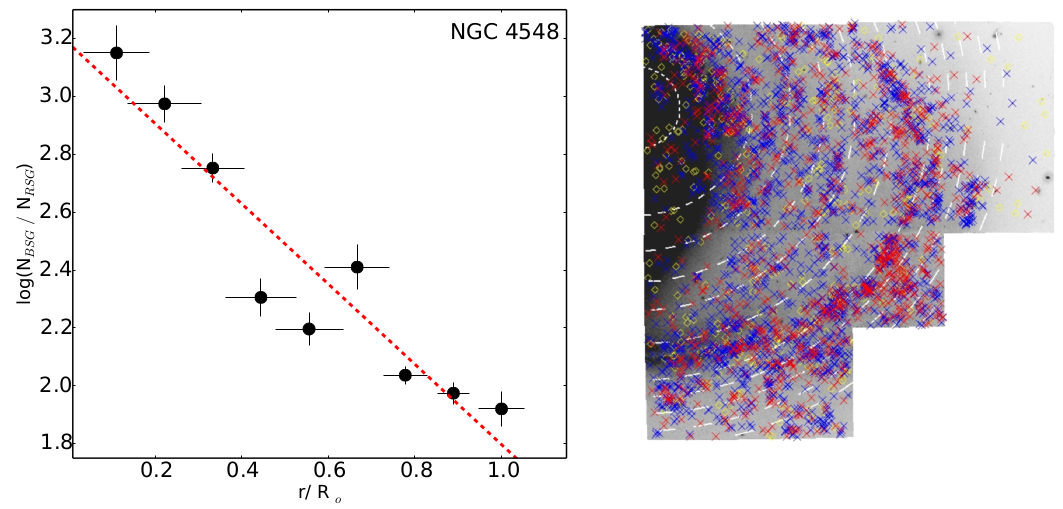} \\

   \caption{Left column: Blue to red supergiant ratio for NGC 1326A, NGC 1425 and, NGC 4548. The dashed red line is a linear fit indicating the monotonic radial decline of the blue to red supergiants. Right column: Spatial distribution of the blue and red supergiants. The radius of each of the white annuli displayed shows the distance in respect to the center of each galaxy.}
\label{b2rratiototal.pdf}
\end{figure*}

 We calculated the de-projected distances of the luminous stars with respect to the center of each field taking into account the inclination angle of each galaxy (see Table~\ref{comparison}). We split the massive stars in three groups: BSGs, YSGs, and RSGs, based on the color and magnitude criteria described in Section~\ref{analysis}. We set ten annuli of equal distance from the center of the galaxy, normalized the number of BSGs, YSGs, and RSGs to the same area taking into account the foreground contamination of each field, and propagated the related errors to calculate the B/R ratio. The spatial distribution of the BSGs and RSGs within the annuli in each field is shown in the right column of Fig.~\ref{b2rratiototal.pdf}. The left column of Fig.\ref{b2rratiototal.pdf} shows the trend of the B/R ratio versus galactocentric radius in NGC 1326A, NGC 1425, and NGC 4548. The B/R ratio in all galaxies declines radially indicating an evident drop of the number of BSGs at the outer regions of the field. The exponential decline of the B/R ratio in the first panel of Fig~\ref{b2rratiototal.pdf} is misleading as it is basically caused by the first data point that corresponds to the center of NGC 1326A which is included in the image of the galaxy. In NGC 4548 the B/R ratio shows a decreasing trend with a rise in the number of blue supergiants occurring at the sixth annulus (1.43$^\prime$). This rise may be explained by the coinciding of the annulus with the spiral arm of the field, where a larger number of BSGs are located due high gas and dust accumulation. A declining B/R ratio with increasing radius has been reported by previous studies \citep[]{HumphreysM33, Eggenberger2002, Spetsieri2018} and has been linked with the radial changes of the metallicity of the field. 
 We compared our results on the B/R ratio with our previous study and the study of \citet{GrammerM101} for NGC 1326A, NGC 1425 and NGC 4548, and assumed that the center of the galaxy is the region with the highest metallicity. The variation of the number of blue to red supergiants with metallicity is a test for stellar evolutionary models as it is highly linked with star formation events and star formation history \citep{Langer2012}. In areas with small coverage, variations in the B/R ratio can occur due to photometric errors and reddening in the outer regions of the field, however, we have eliminated this possibility, as we normalized our distances to the center of each field using galactic coordinates. Hence, the increase of metallicity with galactocentric radius is less likely to be due to systematic errors coming from reddening and photometry.

Our results are in agreement with our previous work on the massive star population in the Virgo Cluster Galaxy NGC 4535 in Paper I and in M101 \citep{GrammerM101}. The observations of M101 \citep{GrammerM101} covered the whole galaxy, while in this study the WFPC2 field of view only covered about a quarter of the galaxy. However, the radial decline of the B/R ratio with decreasing metallicity is common in all fields, confirming the trend already presented by previous studies \citep[]{Eggenberger2002, Maeder81}.

\subsection{Luminosity function and star formation rate}
\label{starformationratesection}
\par
\begin{figure*}[h!tb]
\centering
\includegraphics[trim=0cm 0cm 0.cm 0cm, width=1\textwidth]{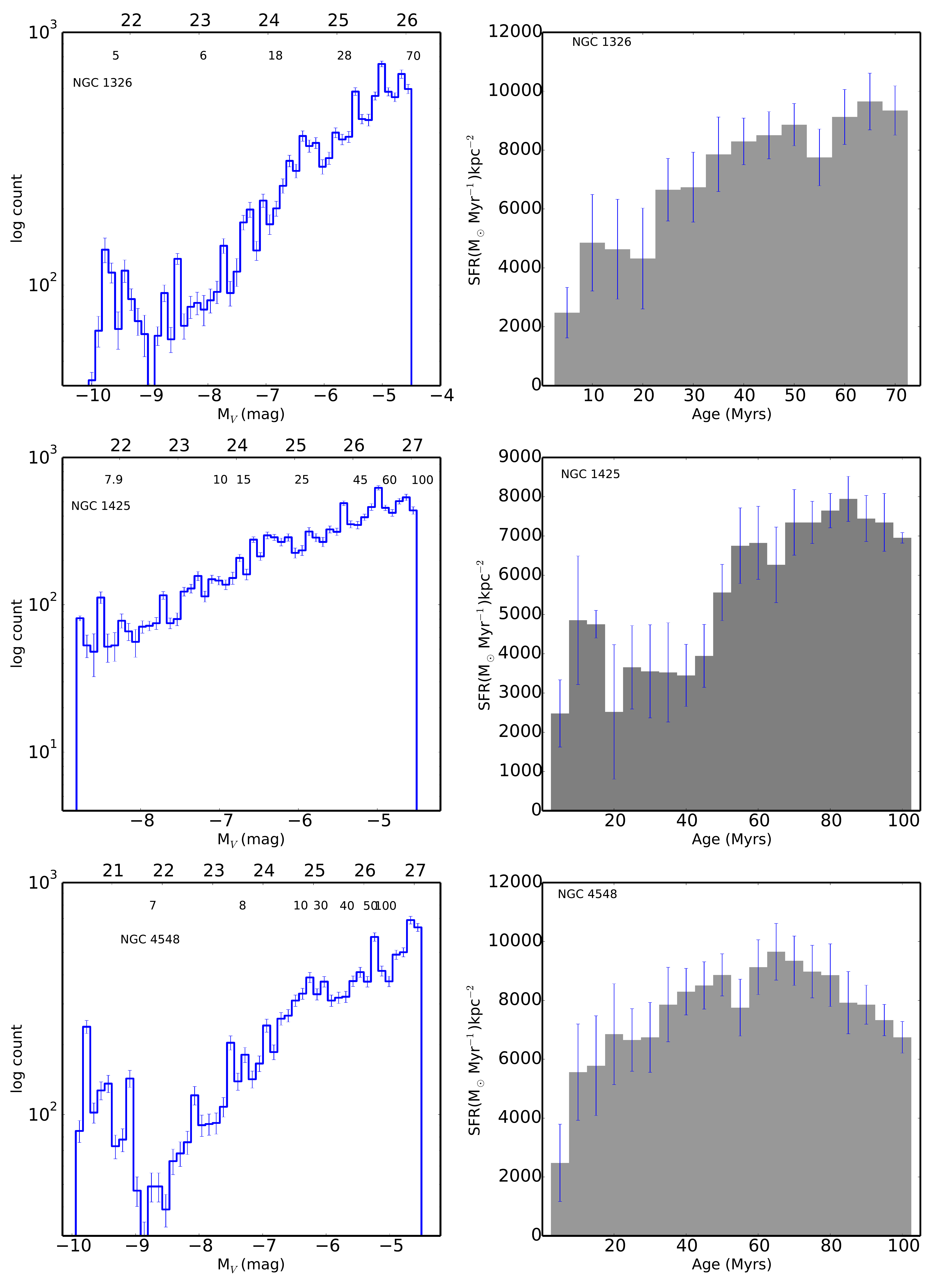}\\

   \caption{Left column: SFR in M$_{\odot}$ Myr$^{-1}$ kpc$^{-2}$ for NGC 1326A, NGC 1425, and NGC 4548 over the last 100 Myr, based on the blue HeB stars. The blue HeB luminosity function has been normalized to account for the IMF and the changing lifetime in this phase with mass. We used the Salpeter IMF slope and the MESA evolutionary tracks. Right column: Luminosity function of the Blue HeB stars in NGC 1326A, NGC 1425, and NGC 4548. The bin width is 0.18 mag. The uncertainties in magnitude are derived by the artificial star tests. Plotted above the histogram is the age in Myrs for the blue HeB stars as given by the MESA models.}
\label{sfr}
\end{figure*}

 We used the Modules for Experiments in Stellar Astrophysics (MESA) Isochrones and Stellar Tracks \citep[MIST;]{MESA2016} at the metallicity of each galaxy (see Table~\ref{comparison}), to estimate the evolutionary state of the massive stars. We used the method described in \citet{Palmer1997} to derive the V luminosity function for the blue helium-burning stars (blue HeB stars) in NGC 1326A, NGC 1425 and NGC 4548. For each field we set magnitude bins of equal magnitude (0.18 mag) from M$_{V}$= $-$3.8 up to $-$9 mag for NGC 1326A, NGC 1425, and NGC 4548 (i.e. the magnitude where the sample reaches 50 $\%$ completeness). The models indicate sources within the age of 5-100 Myrs. The luminosity function of the blue HeB stars in the V band in all three fields is plotted in Fig.~\ref{sfr} (right column). The errors were derived by the artificial star tests, while error propagation has been also used from one bin to another. The numbers plotted above the luminosity function are the ages of the isochrones in Myrs based on the MESA models. In order to estimate the star formation rate based on the counts given from the luminosity function, we made use of the equation given in \citet{Palmer1997}.

 \begin{equation}
C(M_{V}, V - I)= \int_{logm_{1}}^{logm_{2}} \int_{t_{1}(log m)} ^{t_{2}(logm)} \phi (logm,t)\times R(t) dt dlogm
\end{equation}

\noindent adjusted for the stars, where $\phi$ is the initial mass function (IMF) normalized to unity, m is mass, R is the star formation rate in units of M$_{\odot}$ yr$^{-1}$, and t is time. We used the Salpeter slope for the IMF and the normalized IMF is

\begin{equation}
\phi(logm)d logm=0.394 (\frac{m}{M_\odot})^{-1.35} dlogm
\end{equation}

\noindent For each galaxy we derived the area of the filed of view of our observations to find the SFR/kpc$^{2}$. The SFR/kpc$^{2}$ is shown in Fig.~\ref{sfr}.
\par The star formation history of NGC 1326A shows a gradual increase in look-back time at M$_{\odot}$ kpc$^{-2}$ indicating high star formation rates over the past 70 Myrs. In NGC 1425 the star formation history reveals a recent peak $\sim$ 10 Myrs followed by a drop, which could be a sign of a star formation event. The high number of YSGs in NGC 1425 presented in Table~\ref{comparison} could explain a star formation event over which massive stars formed and evolved simultaneously. The star formation rate appears constant between 25-40 Myrs and increases between 40-70 Myrs. From 70-100 Myrs there is no significant variation in the star formation history. 
The field with the highest star formation rate is the Virgo Cluster Galaxy NGC 4548. The star formation rate in NGC 4548 indicates a constant increase with small fluctuations between 5 and 65 Myrs with the star formation rate reaching its peak at 9000 M$_{\odot}$ kpc $^{-2}$ at 70 Myrs. Between 75 and 100 Myrs the star formation rate shows a slight decrease. NGC 4548 has the largest estimated value of metallicity compared to the other two fields. This could be an indication of how star formation proceeds as a function of metallicity, since the role of massive stars regarding star formation remains unclear. The recent star formation history estimated for the three fields does not show signs of significant star burst events that could result in the formation of 100 M$_{\odot}$ stars. As massive stars are characterized by their brief lifetimes we do not rule out the existence of stars more massive than 30 M$_{\odot}$,  as they may have evolved and exploded by the time of the observations. Additionally, very young massive stars are born in highly contaminated star forming regions. As a result, resolving such stars photometrically is challenging as they may appear as diffuse or blended sources. 

\section{Variability}
\label{selection}
After identifying the massive stars in the three fields our next step was to investigate whether those sources displayed any type of variability. We adopted the methodology described in Paper I and considered only the stars with more than five measurements in either the V or I filters. We used the three variability indexes described in Paper I: the median absolute deviation (MAD), the interquartile range (IQR), and the inverse von Neumann ratio (1/$\eta$) \citep{Sokolovsky16}. 

\par The variability indexes were measured in each one of the WFPC2 chips in all three fields. To derive the variability indexes we sorted the sources included in each WFPC2 chip by increasing magnitude and set bins with at least 5$\%$ of the total number of sources per bin to maintain statistical significance. For each bin, we calculated the median value of the index and the standard deviation ($\sigma$), we set a 4$\sigma$ threshold and considered as variable candidates all sources above that threshold in either F555W or F814W.
Fig.~\ref{indexes} shows an example of the index versus magnitude diagram for the MAD, in the WF3 chip of WFPC2 in the V filter for NGC 1326A, NGC 1425, and NGC 4548. Candidate variable sources are shown as red squares while the known Cepheids are shown as blue triangles. All published Cepheid variables in all three fields were recovered by our algorithm. We created index vs. magnitude plots for all chips in all three galaxies and determined the final number of variables.
\par The total number of new variable candidates is 243, which are distributed as follows: 48 variable candidates in NGC 1326A, 102 in NGC 1425 and 93 in NGC 4548. We present light curves for all the new variable candidates identified in Figs.~\ref{lcs1326}$-$\ref{lcs45483}. The figures show that the variable candidates vary with the same trend in both filters and display variations of similar amplitude. Tables~\ref{vars1326} to \ref{vars45482} contain the full list with IDs, coordinates, mean magnitude, magnitude errors, variability indexes, HSC v3 MatchIDs and notes of the variables in all three fields. The tables also include the amplitude of variability and classification based on their V$-$I color and mass as estimated by the models. In Tables~\ref{vars1326} $-$ \ref{vars45482} we use digit 1 to flag the sources that appeared variable in either of the three variability indexes and digit 0 for the sources that were not selected as variables by an index. In the case that a star is flagged as variable in only one index and filter, we assume that it could be due to the wavelength-dependence of variability, or due to the sensitivity of the index to the type of the variability displayed.
\par Among the variable sources primarily detected by the algorithm were the published Cepheids of the three fields. In particular, there were 15 reported Cepheids in NGC 1326A, 20 in NGC 1425 and 24 in NGC 4548.
These sources, are not included in Tables~\ref{vars1326} $-$ \ref{vars45482} as they are known sources, however, we used the mean magnitudes by \citet[]{Prosser99, Mould2000, Graham1999} to test the quality of our photometry. We found a negligible discrepancy between the latter studies and our measurements. The differences in magnitude are namely 0.015 $\pm$ 0.004 mag for NGC 1326A, 0.033 $\pm$ 0.009 mag for NGC 1425, and 0.018 $\pm$ 0.006 mag for NGC 4548. 

\par All variables in all three fields were identified by MAD in at least one filter, while 211 out of 243 variables were detected in at least one filter in IQR. The inverse von-Neumann ratio index is mostly sensitive to correlation-based than scatter-based observations and 37$\%$ of the variable candidates were selected by this index. 

\par
\begin{figure*}[h!tb]
\centering
\begin{tabular}{l c}
\includegraphics[trim=0cm 0cm 0.cm 0cm, width=1\textwidth]{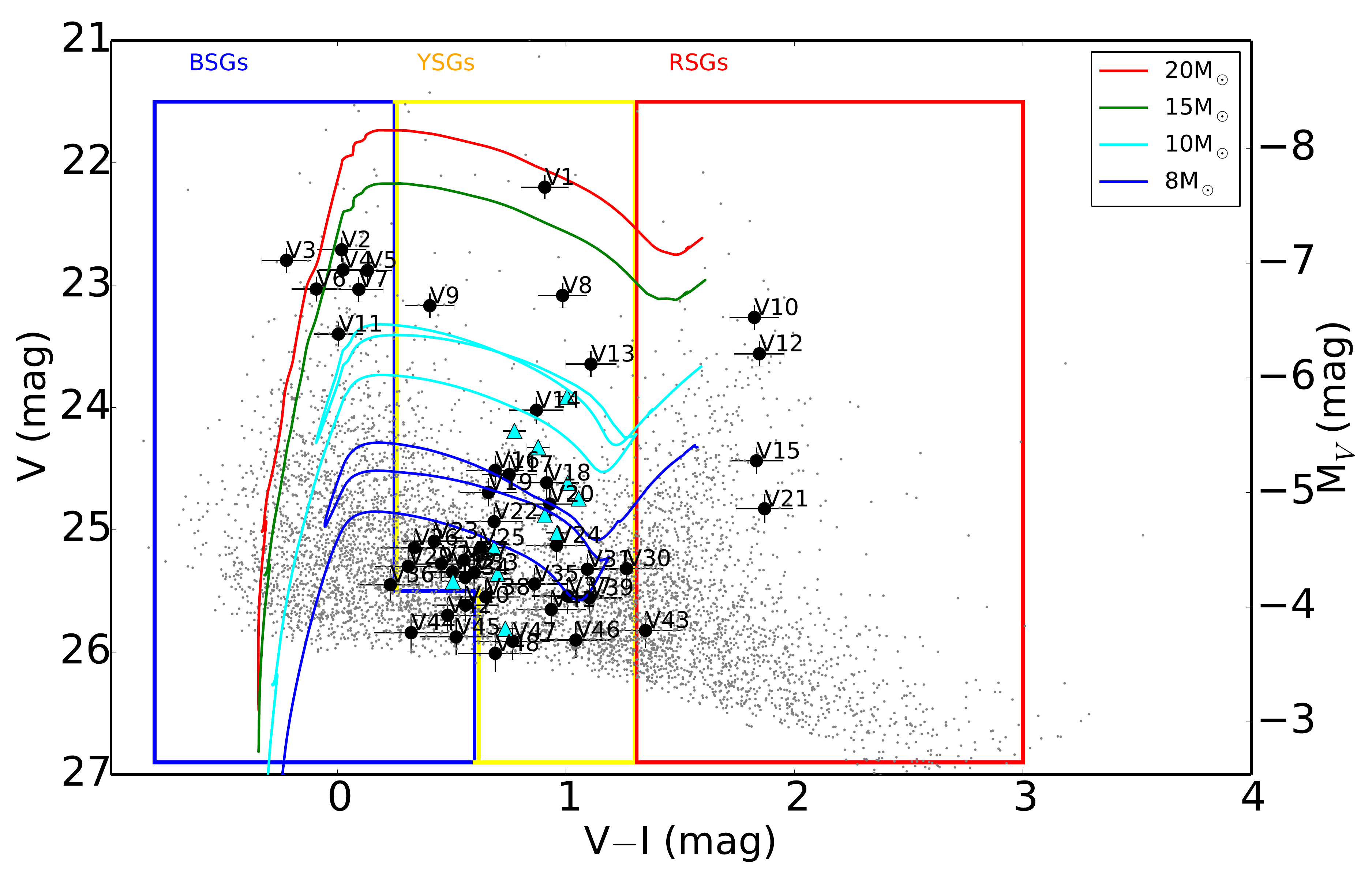}\\
\hspace{2.0 cm}
\includegraphics[trim=0cm 0cm 0.cm 0cm, width=0.55\textwidth]{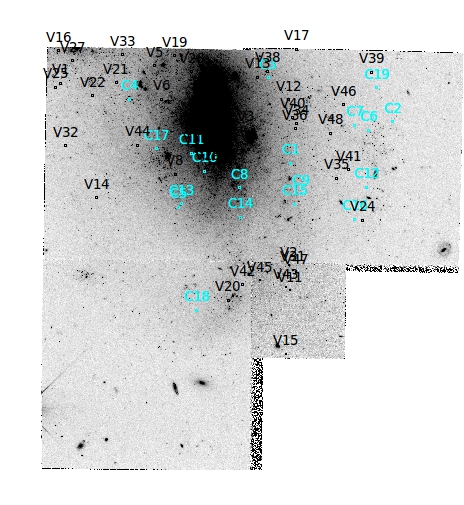}\\
\end{tabular}
 \caption{Top panel: V$-$I versus V color-magnitude diagram of NGC 1326A. Both absolute and mean magnitudes of the stars are plotted on the y-axis, based on the distance modulus 31.04 mag. The constant stars are shown in gray, the known Cepheids as blue triangles and the new candidate variables as black dots. Bottom panel: Spatial distribution of the candidate variable sources in NGC 1326A.}
\label{totalplot1326}
\end{figure*}

\begin{figure*}[h!tb]
\centering
\begin{tabular}{l c}
\includegraphics[trim=0cm 0cm 0.cm 0cm, width=1\textwidth]{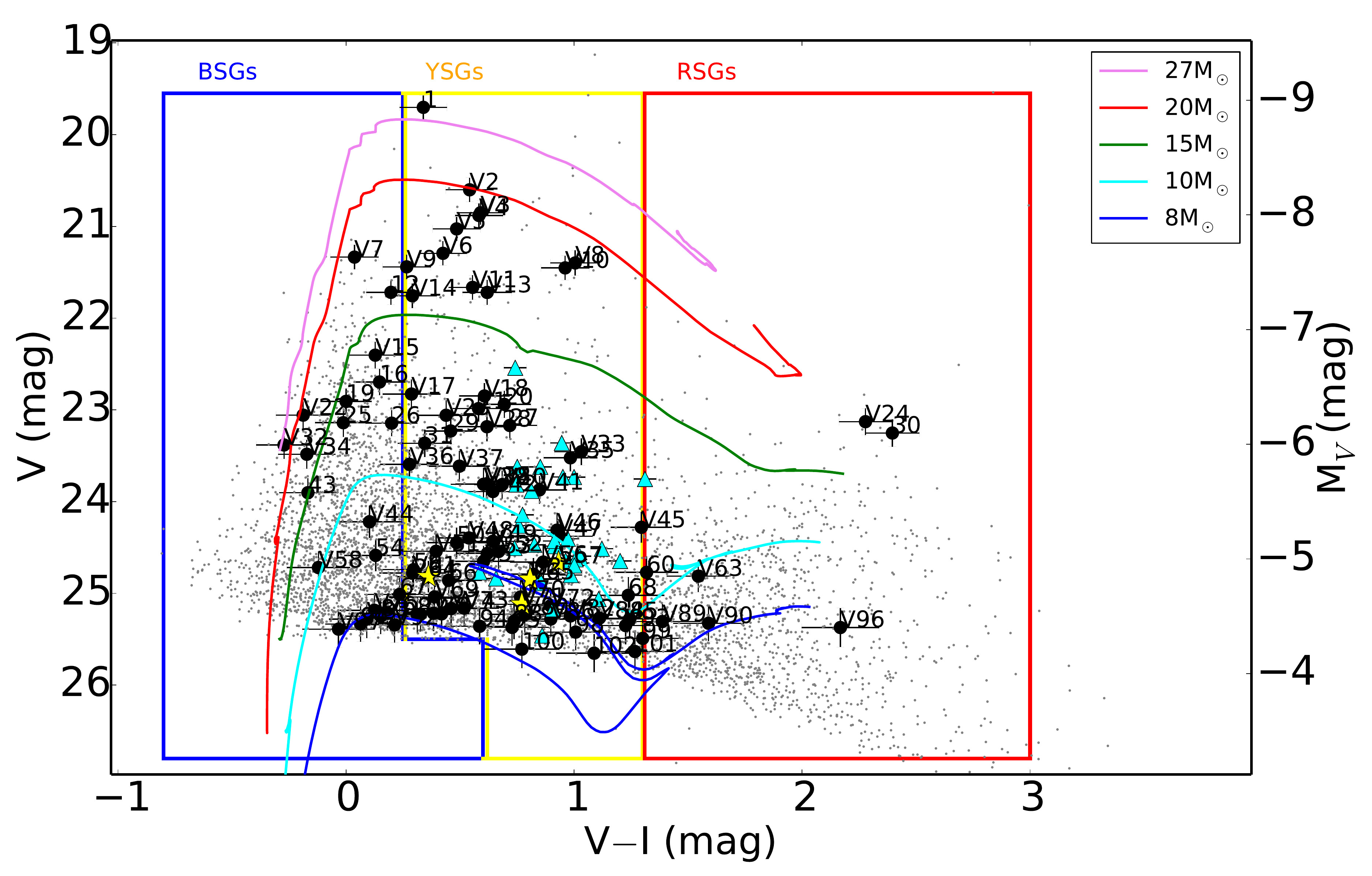}\\
\hspace{1.0 cm}
\includegraphics[trim=0cm 0cm 0.cm 0cm, width=0.7\textwidth]{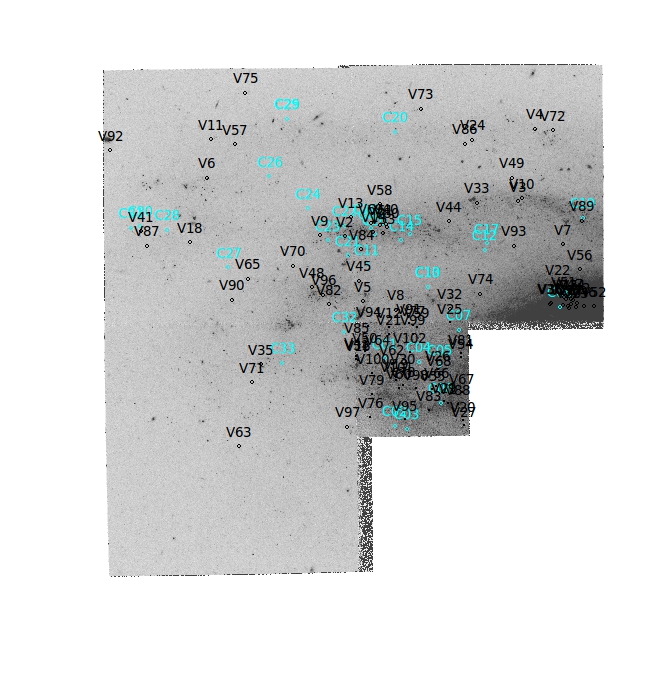}\\

\end{tabular}
 \caption{Same as Fig~\ref{totalplot1326}, for NGC 1425.}
\label{totalplot1425}
\end{figure*}

\begin{figure*}[h!tb]
\centering
\begin{tabular}{l c}
\includegraphics[trim=0cm 0cm 0.cm 0cm, width=1\textwidth]{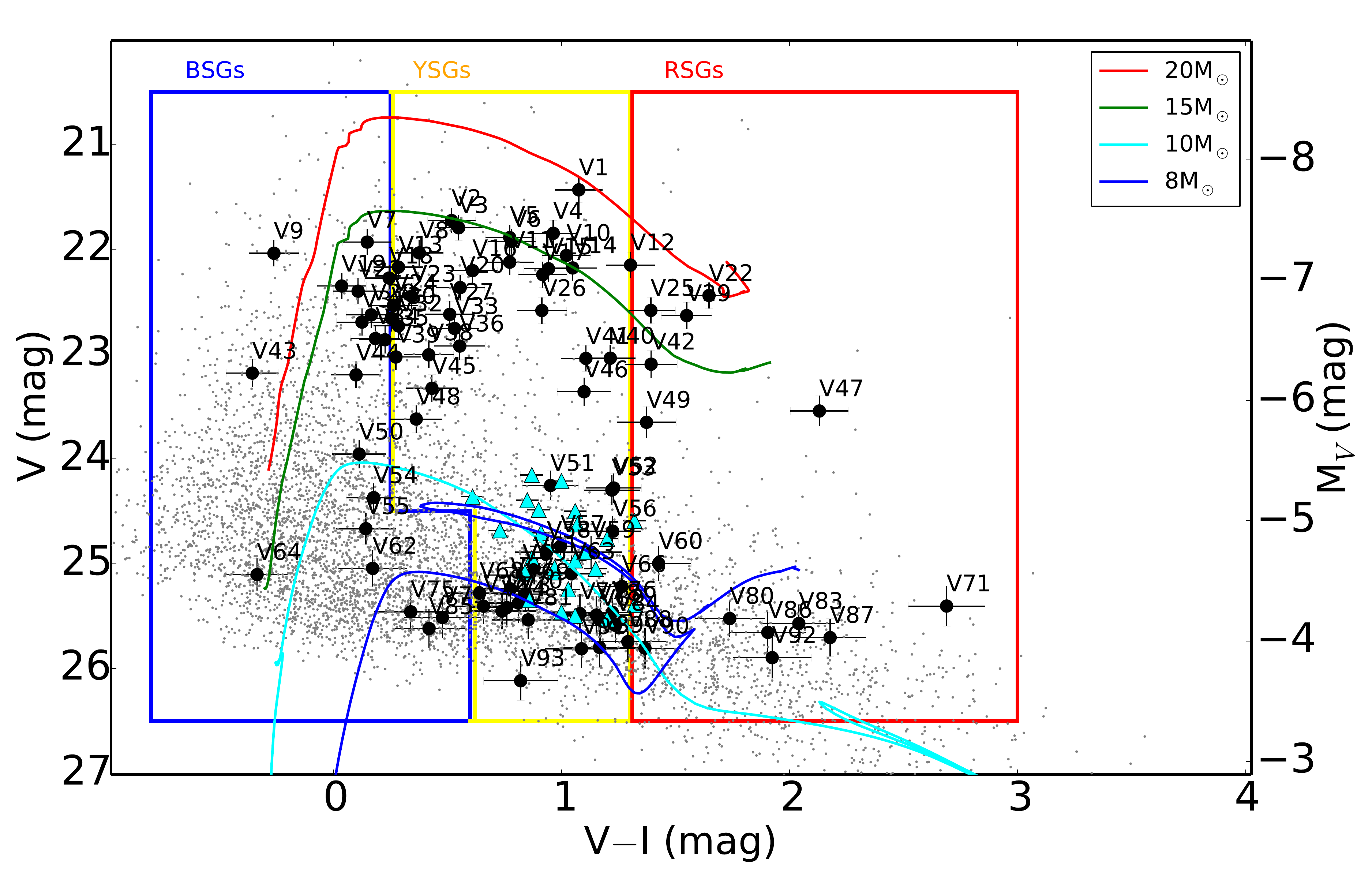}\\
\hspace{2.0 cm}
\includegraphics[trim=0cm 0cm 0.cm 0cm, width=0.6\textwidth]{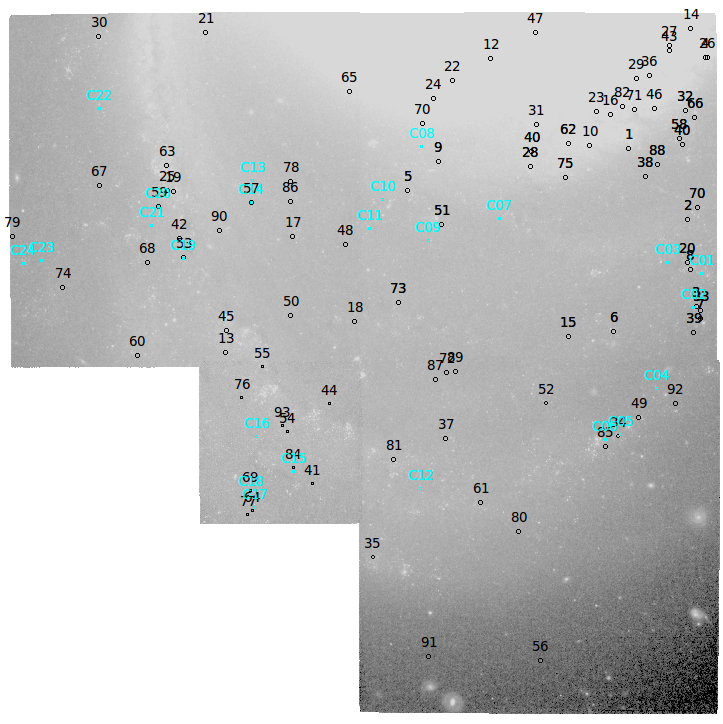} \\

\end{tabular}
 \caption{Same as Fig~\ref{totalplot1326}, for NGC 4548.}
\label{totalplot4548}
\end{figure*}

 We used MESA evolutionary tracks to estimate the masses of the new variables in each field. The models were generated assuming rotation at 40$\%$ of the critical speed and values of metallicity for each field based on \citet{Freedman2001}. We adopted solar metallicity for NGC 1326A and super-solar values for metallicity, for NGC 1425 and NGC 4548. Figs.~\ref{totalplot1326}, \ref{totalplot1425} and \ref{totalplot4548} show both the CMDs and finder charts of the new candidate variables and evolutionary tracks over-plotted along with the spatial distribution of the candidate variable stars for each galaxy, respectively. The masses corresponding to the tracks span from $\sim$ 8$-$20 M$_\odot$ and the magnitude ranges from $-$3.8 mag $-$9 mag. The error bars correspond to the uncertainty due to reddening E(V$-$I), photometric errors and the distance modulus. We took in to account the E(V$-$I) reddening in each field, derived the reddening vector based on the E(V$-$I) values mentioned in Section~\ref{analysis}, and adopted a ratio of total to selective absorption R$_{V}$= A$_{V}$/(A$_{V}$ $-$ A$_{I}$) =3.1 \citep[]{Cardelli89,Freedman2001}. The evolutionary tracks shown in each plot correspond to 8, 10, 15 and 20 M$_{\odot}$ for the metallicities implied by \cite{Freedman2001}. The degeneracy of the evolutionary tracks of different M$_{\odot}$ does not allow us to make a solid estimation of the mass of the candidate massive stars. For example in Figure~\ref{totalplot4548} the 8 M$_{\odot}$ track overlaps with the 10 M$_{\odot}$ track, making it challenging to distinguish stars within that mass range. However, the mass range proposed for the candidate massive stars seems fully compliant with the values of metallicity and extinction of each field.

\par Based on the colors (V$-$I) and mass range of the published variables shown in Fig.~\ref{totalplot1326}, \ref{totalplot1425} and \ref{totalplot4548} we proceed to classify the newly discovered candidate massive variable stars. In the three fields, the new candidate variables lie between $\sim$ 8$-$20 M$_{\odot}$, while in NGC 1326A 17 out of the 48 newly identified candidate variables display masses below 8 M$_{\odot}$. The majority of the variable stars in the fields show V$-$I colors between $-$0.5 to 2.5 mag.
The CMD in Fig.~\ref{totalplot1326} displays fewer candidate massive variable stars, compared to the CMDs in Figs.~\ref{totalplot1425} and \ref{totalplot4548}. This is explained by the different parts of each galaxy covered during the observations of the HST Key Project. In NGC 1425 and NGC 4548 the observations cover a larger part of the spiral arms compared to NGC 1326A.

\par The subsections below provide information about the new variable candidate BSGs, YSGs and RSGs identified in the three galaxies.

 \subsection{Candidate variable red supergiants}
Our analysis yielded in total 21 candidate variable RSGs, three in NGC 1326A, six in NGC 1425 and ten in NGC 4548. The magnitude range of the candidates is $-$7.0 $\leq$ M$_{V}$ $\leq$ $-$4.0 mag and the V$-$I index is within 1.6$-$2.1 mag. According to the MESA models, the masses and temperatures of the stars are between 8$-$16 M$_{\odot}$ and 3000 $<$ T$_{eff}$ $<$ 4000 K. However, stars between 7 M$_{\odot}$ and 8 M$_{\odot}$ may also be super-AGB stars \citep[]{Becker1980,Groenewegen2018}. Super-AGB stars are equally luminous to RSGs and with a similar color, making it difficult to distinguish sources of those two types given the uncertainty in distance. The basic feature used to identify RSGs is their bolometric magnitude, (M$_{bol}$ $\leq$ $-$7.1 mag) \citep {Wood1983}. However, stars with M$_V$ < $-$7 mag displaying masses above 16 M$_{\odot}$ are less likely to be super-AGB stars.
\par Although most RSGs do not usually show variability or periodic variability, the candidate RSGs in the three fields vary in both filters and in some cases display amplitudes between 0.2-0.6 mag. The studies of \citet{Ming} and \citet{Ming2018} have studied periodic variability of the RSGs in the LMC and SMC, howevwer, in our case it is difficult to define a period for the light curve of the candidate RSGs due to the small timescale and the low number of observations. Based on the studies of \citet{Levesque2006}, \citet{Szczygiel2010}, and Yang et al. (2018), claiming periodic variability in RSGs is more robust with a large number of observations and a large timescale. In our work, the amplitude of the light variation is between 0.2$-$0.6 $\pm$ 0.03 mag over $\sim$ 100 days. Variability in red supergiants has many underlying causes such as radial pulsations \citep[]{Wood1983, Stothers1996, Heger1997} photospheric changes, huge convection cells \citep[]{Schwarzschild75, Antia1984} mass transfer and stellar winds.

\subsection{Candidate variable yellow supergiants/hypergiants} 

Among the massive stars of our sample are 12 variable candidate YSGs. The stars show M$_V$ brighter than $-$ 5 mag, while the models imply masses and temperatures in the range 15 $<$ M $<$ 20 M$_\odot$ and 5900 $<$ T$_{eff}$ $<$ 7500 K. They also display a variation in their light curve in both filters within a small timescale. The galaxy with the largest number of candidate variable YSGs is NGC 1425, which also contains the largest number of candidate variable stars. The YSG region is often affected by foreground contamination. In our study foreground contamination occurs between 0.8$<$ V$-$I $<$1.3 mag.
YSGs are among the most visually luminous stars, with absolute magnitude M$_{V}$ reaching $-$9 mag, evolving from the main sequence towards the RSG state, or following the opposite path \citep{Humphreys2016}. Another short evolutionary state of post-RSGs with masses between 20$-$40 M$_\odot$, is the yellow hypergiant phase (YHGs) \citep{deJager98}. YHGs display temperatures ranging between 4000$-$7000 K and $\log{L/L_{o}}$ > 5.4 and are characterized by atmospheric instability and extended turbulent structures. During the YHG phase, massive stars undergo high mass losses shed as expanding pseudo-photospheric shells. Such sources are among the SN type IIn progenitors \citep{Smith2014}, which makes their identification and observational follow-up valuable. The brightest source in the YSG region is V1 in NGC 1425 that displays M$_{V}$ $\sim$ $-$9 mag, $\sim$ 30 M$_{\odot}$ and T$_{eff}$ $\sim$ 7000K. Based on its characteristics, this source is classified as a candidate variable YSG/YHG. Variability in YSGs, has been previously reported in the studies of \citet[]{Humphreys1979,Percy2014,Humphreys2016,Kourniotis2016,Spetsieri2018}, in the region with M$_\odot$ $\geq$ 20 M$_\odot$ and $\log{L/L_{o}} $ > 5.4. and is typical of amplitude 0.3-0.6 magnitudes similar to the amplitude of V1. As spectroscopic classification is essential for accurate discrimination between YHGs and YSGs, we cannot further infer on the state of our yellow stars. 

\subsection{Candidate variable blue supergiants}

\par Among the candidate variable massive stars identified in the three fields are 8 blue supergiants. We expect no foreground contamination in that region, according to the Besan\c{c}on models. The mass range implied by the models for the candidate variable BSGs is 15 to 20 M$_\odot$, the magnitudes range between $-$8 $<$ M$_{V}$ $<$ $-$4.9 mag, and the temperatures, based on the models, range from  5900$<$ T$_{eff}$ $<$17400 K. The field with the largest number of BSGs (64) is the Virgo Cluster galaxy NGC 4548. As shown in section~\ref{starformationratesection} this galaxy is considered to be star forming. The amplitude of variability of the BSGs reaches 1.2 mag for the low-mass candidate BSGs (e.g. V97 in NGC 1425) and 0.1 mag for brighter and more massive BSGs (e.g. V7 in NGC 1425).

\par The existence of a large number of BSGs in a galaxy is indicative of recent star formation as BSGs are young stellar objects that are found in the galactic spiral arms in regions rich in gas and dust. Based on stellar evolution theory \citep{Ekstrom2012} a star in the BSG region could be either coming directly from the main sequence to the supergiant region or returning back to higher temperatures before central helium exhaustion, after the RSG phase. In order to distinguish whether a source in the BSG region is evolving red-ward after the main sequence or is undergoing a blue-loop, one needs to know about the internal mixing in the radiative layers, the strength of stellar winds (if evident) and the metallicity. Variability of BSGs and in particular pulsations of such sources, has been used as a diagnostic tool for the evolutionary stages before and after helium core ignition \citep{Ostrowski2015}. \citet{Kraus2015} studied the blue supergiant 55 Cygnus over a five year baseline to link its pulsational activity with mass-loss episodes and the formation of clumps in the stellar wind.

\par To summarize, we identified extragalactic candidate variable BSGs, and report their variability over a short base line of $\sim$ 100 days. As BSGs are pre-collapse SN progenitors, large amplitude variations in their light curves could be indicative of an unstable stage before stellar death.

\subsection{New candidate periodic variables}
We conducted a period search to define the periodic variable sources in all three fields (see Table~\ref{vars45482}). We identified four Cepheid variables in NGC 1425 within a time-baseline of 100 days. We estimated the period of the variable sources based on their light curve, using the online tool of the NASA Exoplanet Archive \footnote{\url{http://exoplanetarchive.ipac.caltech.edu}}. The periods estimated for the periodic variables range from 25 to 60 days and are marked in Table~\ref{vars14251}. The periodic variables are located in the YSG region between 8 to 9 M$_{\odot}$ near the known Cepheids of each field, i.e. the instability strip \citep[]{Sandage1971, Macri99}. Their magnitude and color ranges are: $-$4 $<$ M$_{V}$ $<$ $-$5 mag and 0.5$<$ V$-$I $<$ 1.2 mag. The light curves of all sources (Fig.~\ref{lcs45483}) show pulsations similar to the Cepheid variables. They are shown as yellow stars in the CMD in Figure~\ref{totalplot1425}. We propose follow-up photometric observations to better identify the nature of these sources. 

\section{Summary}
\label{summary}

\par We performed PSF photometry on archival HST WFPC2 images and created a catalog of luminous stars for the HST Key project galaxies NGC 1326A, NGC 1425 and NGC 4548. In all three fields we studied the massive star population and created three separate sub-catalogs of massive stars with M$_{V}$ $<$ $-$4.0 mag. In particular the number of massive stars identified in each field were: 5388 in NGC 1326A, 5053 in NGC 1425, and 3783 in NGC 4548. Using color criteria on the CMD, we separated the massive star candidates into three subsets: BSGs, YSGs, and RSGs. We calculated the foreground contamination in the direction of each galaxy using the Besan\c{c}on Galactic population synthesis model.
The region with the lowest foreground contamination in the three galaxies was the BSG region with  the percentage of foreground stars in that region being: $\sim$1.3$\%$ in NGC 1326A, $\sim$0.8$\%$ in NGC 1425, and $\sim$3.7$\%$ in NGC 4548. The region most affected by foreground contamination was the YSG region with $\sim$25$\%$ in NGC 1326A, $\sim$29$\%$ in NGC 1425, and $\sim$30$\%$ in NGC 4548 being foreground stars. 
\par We calculated the de-projected distances of the massive stars and estimated the blue to red supergiant ratio at various radial distances for the three galaxies. The B/R ratio decreases monotonically with increasing distance and decreasing metallicity. We examined the recent star formation history of the field over the last 100 Myrs within the WFPC2 field of view. We derived the luminosity function based on the blue HeB stars and estimated the SFR using the Salpeter slope for the IMF. NGC 4548 includes the largest number of stars and has the highest star formation rate compared to NGC 1326A and NGC 1425.
The star formation history of NGC 1326A shows an increase indicating high star formation rates over the past 70 Myrs, while in NGC 1425 the star formation history has a recent peak at $\sim$ 10 Myrs, an indication of a star formation event. 
\par We conducted a variability search among 5789 sources in NGC 1326A, 5232 sources in NGC 1425, and 5790 sources in NGC 4548 using three different methods: the MAD, IQR, and the inverse von Neumann ratio. The number of stars used for the variability search include all sources that fulfilled the quality criteria mentioned in Section~\ref{Reduction}. The number of stars in Table~\ref{comparison} imply to the massive stars in each field. The variability search yielded 243 new variable sources in addition to the 87 known Cepheid variables in the three galaxies. We used the MESA models and evolutionary tracks to model our results and classify the new variable candidates. Among the luminous massive variable sources are 138 variable candidate BSGs, 86 variable candidate YSGs, and 19 variable candidate RSGs. In addition to the candidate variable massive stars we identified four periodic variable candidates in NGC 1425, which we suggest for follow-up observations.

\par This work provides a census of the massive star population and variable massive stars in three HST Key Project galaxies using archival data from the Hubble Legacy Archive (HLA). Future work on the massive star population in distant fields using the combination of deep HST data and high-precision PSF photometry, will shed light on the link of massive stars with the galactic star formation history. Increasing the number of photometric and spectroscopic observations of massive stars could help identify SN Type II progenitors and constrain the evolutionary models of massive stars unveiling important information about their properties and evolution. 
\begin{acknowledgements}
We would like to thank the anonymous referee for the insightful comments that helped improve this paper.
We acknowledge financial support by the European Space Agency (ESA) under the ‘Hubble Catalog of Variables’ program, contract no. 4000112940. This research has made use of the VizieR catalog access tools, CDS, Strasbourg, France and NASA's Astrophysics Data System (ADS). The Virtual Observatory tools (VO) TOPCAT \citep{Taylor2005} and Aladin were used for image downloading and table manipulation. The data used in this study were downloaded from the Mikulski Archive for Space Telescopes (MAST). All figures in this work were produced using Matplotlib a Python library for publication graphics.
\end{acknowledgements}

\bibliographystyle{aa}
\bibliography{aanda.bib}

\begin{table*}
\caption{Catalog of 5897 stars in NGC 1326A.}
\begin{tabular}{|r|l|l|r|r|r|r|r|r|}
\hline
  \multicolumn{1}{|c|}{ID} &
  \multicolumn{1}{c|}{RA (J2000)} &
  \multicolumn{1}{c|}{Dec (J2000)} &
  \multicolumn{1}{c|}{X (pixels)} &
  \multicolumn{1}{c|}{Y(pixels)} &
  \multicolumn{1}{c|}{<V> (mag)} &
 \multicolumn{1}{c|}{$\sigma_{<V>}$ (mag)} &
  \multicolumn{1}{c|}{<I> (mag)} &
  \multicolumn{1}{c|}{$\sigma_{<I>}$ (mag)} \\
\hline
  1 & 3:25:03.205 & -36:21:57.02 & 88.87 & 490.39 & 24.539 & 0.010 & 22.788 & 0.008\\
  2 & 3:25:03.591 & -36:22:01.12 & 68.74 & 355.53 & 23.924 & 0.008 & 23.905 & 0.013\\
  3 & 3:25:03.154 & -36:21:54.00 & 56.39 & 549.66 & 27.395 & 0.077 & 23.164 & 0.011\\
  4 & 3:25:03.005 & -36:22:06.73 & 266.77 & 361.29 & 24.659 & 0.008 & 24.654 & 0.002\\
  5 & 3:25:02.120 & -36:21:59.13 & 338.64 & 640.10& 25.348 & 0.016 & 23.865 & 0.015\\
  6 & 3:25:01.619 & -36:22:22.16 & 766.54 & 338.78 & 25.764 & 0.017 & 23.930 & 0.012\\
  7 & 3:25:03.805 & -36:22:12.63 & 188.22 & 125.41 & 25.168 & 0.012 & 25.458 & 0.039\\
  8 & 3:25:01.933 & -36:22:04.16 & 447.80& 587.63 & 25.414 & 0.014 & 25.467 & 0.040\\
  9 & 3:25:03.533 & -36:22:17.63 & 314.34 & 87.91 & 25.913 & 0.019 & 24.266 & 0.016\\
  10 & 3:25:03.570 & -36:22:00.98 & 71.06 & 361.40 & 25.504 & 0.017 & 25.695 & 0.053\\

\hline

\end{tabular}
\begin{tablenotes}
\item \textbf{Note.} -- Units of right ascension are hours, minutes, and seconds, and units of declination are degrees, arcminutes, and arcseconds. This table is available in its entirety in an electronic version. The ten first entries are shown as guidance for the content of the tables.Units of X and Y are pixels coordinates. 

\end{tablenotes}
\label{allsources1326}

\end{table*} 

\begin{table*}
\caption{Catalog of 5232 stars in NGC 1425.}
\begin{tabular}{|r|l|l|r|r|r|r|r|r|}
\hline
 \multicolumn{1}{|c|}{ID} &
  \multicolumn{1}{c|}{RA (J2000)} &
  \multicolumn{1}{c|}{Dec (J2000)} &
  \multicolumn{1}{c|}{X (pixels)} &
  \multicolumn{1}{c|}{Y (pixels)} &
  \multicolumn{1}{c|}{<V> (mag)} &
 \multicolumn{1}{c|}{$\sigma_{<V>}$ (mag)} &
  \multicolumn{1}{c|}{<I> (mag)} &
  \multicolumn{1}{c|}{$\sigma_{<I>}$ (mag)} \\
\hline
  1 & 3:42:15.534 & -29:54:51.60 & 607.53 & 672.60 & 22.677  & 0.004 & 22.341 & 0.005\\
  2 & 3:42:17.574 & -29:55:05.22 & 303.50& 92.98 & 23.386  & 0.004 & 22.939 & 0.007\\
  3 & 3:42:16.400 & -29:54:48.58 & 367.20& 584.80& 23.513  & 0.005 & 22.702 & 0.007\\
  4 & 3:42:15.199 & -29:54:52.36 & 695.40& 714.03 & 23.715 & 0.005 & 22.933 & 0.007\\
  5 & 3:42:16.777 & -29:55:06.89 & 510.48 & 193.74 & 23.664 & 0.005 & 23.495 & 0.009\\
  6 & 3:42:17.543 & -29:54:56.66 & 202.43 & 251.88 & 24.152 & 0.007 & 23.758 & 0.013\\
  7 & 3:42:16.751 & -29:55:00.70 & 438.47 & 309.33 & 24.300& 0.007 & 23.418 & 0.010\\
  8 & 3:42:16.720 & -29:54:57.16 & 400.82 & 377.97 & 25.471 & 0.019 & 23.111 & 0.007\\
  9 & 3:42:15.594 & -29:54:51.75 & 595.27 & 660.14 & 24.100 & 0.010 & 24.123 & 0.018\\
  10 & 3:42:15.771 & -29:54:56.63 & 615.72 & 543.30& 25.451 & 0.020 & 23.219 & 0.008\\
\hline
\end{tabular}
\begin{tablenotes}
\item \textbf{Note.} -- Units of right ascension are hours, minutes, and seconds, and units of declination are degrees, arcminutes, and arcseconds. This table is available in its entirety in an electronic version. The ten first entries are shown as guidance for the content of the tables.

\end{tablenotes}
\label{allsources1425}

\end{table*}

\begin{table*}
\caption{Catalog of 5790 stars in NGC 4548.}
\begin{tabular}{|r|l|l|r|r|r|r|r|r|}
\hline
  \multicolumn{1}{|c|}{ID} &
  \multicolumn{1}{c|}{RA (J2000)} &
  \multicolumn{1}{c|}{Dec (J2000)} &
  \multicolumn{1}{c|}{X (pixels)} &
  \multicolumn{1}{c|}{Y (pixels)} &
  \multicolumn{1}{c|}{<V> (mag)} &
 \multicolumn{1}{c|}{$\sigma_{<V>}$ (mag)} &
  \multicolumn{1}{c|}{<I> (mag)} &
  \multicolumn{1}{c|}{$\sigma_{<I>}$ (mag)} \\
\hline
  1 & 3:25:03.205 & -36:21:57.02 & 88.87 & 490.39 & 24.539 & 0.009 & 22.788 & 0.008\\
  2 & 3:25:03.591 & -36:22:01.12 & 68.74 & 355.53 & 23.924 & 0.007 & 23.905 & 0.013\\
  3 & 3:25:03.154 & -36:21:54.00 & 56.39 & 549.66 & 27.395 & 0.074 & 23.164 & 0.011\\
  4 & 3:25:03.005 & -36:22:06.73 & 266.77 & 361.29 & 24.659 & 0.008 & 24.654 & 0.020\\
  5 & 3:25:02.120 & -36:21:59.13 & 338.64 & 640.10 &25.348  & 0.017 & 23.865 & 0.014\\
  6 & 3:25:01.619 & -36:22:22.16 & 766.54 & 338.78 & 25.764 & 0.020 & 23.930 & 0.014\\
  7 & 3:25:03.805 & -36:22:12.63 & 188.22 & 125.41 & 25.168 & 0.011 & 25.458 & 0.036\\
  8 & 3:25:01.933 & -36:22:04.16 & 447.80& 587.63 & 25.414 & 0.020 & 25.467 & 0.015\\
  9 & 3:25:03.533 & -36:22:17.63 & 314.34 & 87.91 & 25.913  & 0.014 & 24.266 & 0.041\\
  10 & 3:25:03.570& -36:22:00.98 & 71.06 & 361.40& 25.504  & 0.015 & 25.695 & 0.053\\
\hline
\end{tabular}

\label{allsources4548}
\begin{tablenotes}
\item \textbf{Note.} -- Units of right ascension are hours, minutes, and seconds, and units of declination are degrees, arcminutes, and arcseconds. This table is available in its entirety in an electronic version. The ten first entries are shown as guidance for the content of the tables. 

\end{tablenotes}
\end{table*}

\begin{table*}
\caption{Properties of NGC 1326A, NGC1425 and NGC 4548.}
 \begin{tabular}{l c c c c c c c c c c c } 
 \hline
 {Galaxy}&{i} &{D$_{0}$} &{$\mu_{0}$} &{Z} & {Morphological type}& {Redshift} &{Absolute Magnitude }\\
 & {(deg)} &{(Mpc)} &{(mag)} &{(dex)} & &{z} &{(mag)} \\
 
\hline\hline
 NGC 1326A & 50 &16.1 $\pm${0.04} &31.04 $\pm${0.002} &8.50 & Sb & 0.006108 & $-$18.0 $\pm{0.13}$\\
 NGC 1425 &77 &21.8 $\pm${0.08} &31.70 $\pm${0.200} &9.00& Sb & 0.005037& $-$21.9 $\pm{0.10}$\\
 NGC 4548& 35& 16.2 $\pm${0.50} &31.05$\pm${0.080} &9.34& Sb& 0.001621 & $-$22.3 $\pm{0.61}$\\
 
\hline
\end{tabular}

\begin{tablenotes}

\item \textbf{Note.} -- Values of inclination (i), distance (D), distance modulus ($\mu$ $_{0}$) and metallicity (Z), redshift,\\ morphological type and absolute magnitude are from the study of \citet{Freedman2001} and \url{https://cdsportal.u-strasbg.fr}. 

\end{tablenotes}
\label{properties}
\end{table*}

\begin{table*}
\caption{Massive star population in NGC 1326A, NGC1425 and NGC 4548.}
 \begin{tabular}{l c c c c c c c c c c c } 
 \hline
 {Galaxy}&{$\#$Stars} &{$\#$BSGs} & {$\#$variable BSGs} &{$\#$YSGs} & {$\#$variable YSGs} & {$\#$RSGs} & {$\#$variable RSGs} \\

\hline\hline
 NGC 1326A& 5388  (8.0$\%$) &2652 (1.0$\%$)&21 &868 (25.0$\%$)&22 &1868 (10.6$\%$)&5 \\
 NGC 1425& 5053 (10.0$\%$) & 2833 (0.8$\%$)&53 & 939 (29.0$\%$)&41 & 1281 (17.0$\%$)&8 \\
 NGC 4548& 3783 (11.0$\%$)& 2155 (3.7$\%$)&64 & 507  (30.0$\%$)&23  & 1121 (3.21$\%$)&6 \\
 Total&  16811 & 7640 & 138 &2314 &85& 4270 &19 \\
\hline
\end{tabular}

\begin{tablenotes}

\item \textbf{Note.} -- The percentages next to columns 1$-$7, indicate the percentage of foreground contamination in the corresponding population. 

\end{tablenotes}
\label{comparison}
\end{table*}

\begin{sidewaystable*} [ht!]

\caption{Properties of the variable candidates in NGC 1326A ordered by average magnitudes. For convenience we use the letter "V" to name each variable in this field.}
\scalebox{0.8}{
\begin{tabular}{l c c c c c c c c c c c c c c c c c c}
\hline
{ID} & RA& Dec& {X}& {Y} &{V}& {$\rm \sigma$ $_{V}$} &{I}&{$\rm \sigma$ $_{I}$}&  {MAD$_{V}$} & {MAD$_{I}$} & {IQR$_{V}$} & {IQR $_{I}$} & {1/$\eta$ $_{V,I}$} & {$\Delta$V} & {$\Delta$I}&{HSC MatchID}&{Notes}  \\ 

& (J2000)&(J2000)& (pixels) & (pixels) & (mag) & (mag) & (mag)& (mag) & & &  &  & & (mag) & (mag) & \\
\hline

   V1 & 3:25:12.065 & -36:22:23.02 & 662.8 & 743.68 & 23.378 & 0.004 & 22.47 & 0.005 & 1 & 1 & 1 & 1 & 1 & 1.0 & 0.3 & 100711291 & candidate YSG\\
  V2 & 3:25:03.610 & -36:22:01.14 & 68.74 & 355.53 & 23.924 & 0.008 & 23.905 & 0.013 & 1 & 1 & 1 & 1 & 1 & 0.2 & 0.2 & 21471881 \\
  V3 & 3:25:07.386 & -36:21:40.45 & 506.12 & 60.04 & 24.016 & 0.009 & 24.239 & 0.018 & 1 & 0 & 1 & 1 & 0 & 0.2 & 0.2 & 51171750 & candidate BSG \\
  V4 & 3:25:07.578 & -36:21:55.72 & 426.93 & 181.9 & 24.1 & 0.007 & 24.075 & 0.015 & 1 & 1 & 1 & 1 & 1 & 0.8 & 0.2 &  & candidate BSG \\
  V5 & 3:25:10.705 & -36:21:52.99 & 731.98 & 404.25 & 24.104 & 0.006 & 23.972 & 0.013 & 1 & 1 & 1 & 1 & 0 & 0.5 & 0.1 &  & candidate BSG \\
  V6 & 3:25:09.799 & -36:21:58.66 & 608.12 & 357.28 & 24.266 & 0.008 & 24.358 & 0.019 & 1 & 1 & 1 & 1 & 0 & 0.1 & 0.3 &  & candidate BSG \\
  V7 & 3:25:06.806 & -36:21:53.24 & 367.55 & 99.03 & 24.269 & 0.008 & 24.175 & 0.017 & 1 & 1 & 0 & 1 & 0 & 0.4 & 0.2 &  & candidate BSG \\
  V8 & 3:25:07.950 & -36:22:13.76 & 365.24 & 360.44 & 24.322 & 0.007 & 23.336 & 0.008 & 1 & 1 & 1 & 1 & 0 & 0.3 & 0.3 & 20795427 \\
  V9 & 3:25:07.029 & -36:21:46.15 & 436.12 & 63.09 & 24.412 & 0.008 & 24.007 & 0.014 & 1 & 1 & 1 & 1 & 0 & 0.4 & 0.2 \\
  V10 & 3:25:08.102 & -36:21:58.96 & 454.4 & 248.09 & 24.514 & 0.008 & 22.689 & 0.006 & 1 & 1 & 0 & 1 & 1 & 0.8 & 0.1 & 21211 & candidate RSG \\
  V11 & 3:25:03.026 & -36:22:06.69 & 266.77 & 361.29 & 24.659 & 0.008 & 24.654 & 0.02 & 1 & 1 & 1 & 1 & 0 & 0.7 & 0.3\\
  V12 & 3:25:07.410 & -36:21:23.83 & 172.48 & 609.38 & 24.831 & 0.01 & 22.984 & 0.006 & 1 & 1 & 1 & 1 & 1 & 1.0 & 0.2 & 37669517 & candidate RSG\\
  V13 & 3:25:08.504 & -36:21:27.26 & 57.23 & 689.89 & 24.921 & 0.011 & 23.811 & 0.012 & 1 & 1 & 1 & 1 & 0 & 0.3 & 0.3 & 59943606 \\
  V14 & 3:25:08.771 & -36:22:39.17 & 252.25 & 606.32 & 25.323 & 0.018 & 24.452 & 0.023 & 0 & 0 & 1 & 0 & 1 & 1.2 & 0.5 & 57563267 \\
  V15 & 3:25:01.644 & -36:22:22.11 & 766.54 & 338.78 & 25.764 & 0.017 & 23.930& 0.012 & 1 & 1 & 1 & 1 & 0 & 1.0 & 0.3 & 18558392 \\
  V16 & 3:25:12.860 & -36:22:16.25 & 782.54 & 754.69 & 25.847 & 0.027 & 25.156 & 0.033 & 0 & 1 & 0 & 1 & 0 & 0.8 & 0.6 \\
  V17 & 3:25:08.418 & -36:21:10.11 & 200.05 & 792.88 & 25.885 & 0.02 & 25.133 & 0.031 & 1 & 1 & 0 & 1 & 0 & 0.2 & 0.5 \\
  V18 & 3:25:07.017 & -36:21:54.42 & 382.63 & 126.3 & 25.956 & 0.028 & 25.04 & 0.033 & 1 & 0 & 1 & 0 & 1 & 1.0 & 0.5 \\
  V19 & 3:25:10.567 & -36:21:45.34 & 768.89 & 333.98 & 26.039 & 0.024 & 25.378 & 0.042 & 1 & 0 & 1 & 0 & 0 & 1.0 & 1.0 \\
  V20 & 03:25:03.95 & -36:22:25.92 & 124.52 & 181.62 & 26.141 & 0.03 & 25.211 & 0.039 & 0 & 0 & 1 & 0 & 0 & 1.0 & 0.6 \\
  V21 & 3:25:10.396 & -36:22:07.09 & 667.28 & 544.7 & 26.182 & 0.024 & 24.312 & 0.015 & 1 & 1 & 1 & 1 & 0 & 0.6 & 0.4 & 69513998 & candidate RSG \\
  V22 & 3:25:11.189 & -36:22:17.08 & 621.13 & 627.76 & 26.295 & 0.028 & 25.609 & 0.047 & 1 & 1 & 1 & 1 & 1 & 0.3 & 0.5 \\
  V23 & 3:25:08.800 & -36:21:48.59 & 586.72 & 223.21 & 26.465 & 0.043 & 26.041 & 0.104 & 1 & 1 & 1 & 1 & 1 & 0.8 & 0.8 \\
  V24 & 3:25:03.257 & -36:21:30.63 & 448.95 & 182.82 & 26.501 & 0.036 & 25.541 & 0.06 & 1 & 1 & 1 & 1 & 1 & 0.6 & 0.5\\
  V25 & 3:25:12.079 & -36:22:25.40 & 651.41 & 771.25 & 26.521 & 0.032 & 25.892 & 0.056 & 1 & 1 & 1 & 1 & 1 & 0.1 & 0.8 & 85805459 \\
  V26 & 3:25:08.078 & -36:21:51.65 & 499.75 & 190.01 & 26.522 & 0.048 & 26.185 & 0.099 & 1 & 1 & 1 & 1 & 0 & 0.7 & 0.9 & 86993890 \\
  V27 & 3:25:12.363 & -36:22:14.59 & 747.67 & 697.66 & 26.63 & 0.037 & 26.075 & 0.071 & 1 & 0 & 0 & 1 & 0 & 0.2 & 0.6\\
  V28 & 3:25:09.870 & -36:21:44.19 & 717.8 & 271.42 & 26.664 & 0.045 & 26.209 & 0.097 & 1 & 0 & 1 & 0 & 1 & 0.4 & 0.8\\
  V29 & 3:25:07.342 & -36:22:01.93 & 365.33 & 218.62 & 26.684 & 0.045 & 26.374 & 0.114 & 1 & 1 & 1 & 1 & 1 & 0.3 & 1.2 & 10543145 \\
  V30 & 3:25:08.432 & -36:21:49.18 & 548.75 & 198.86 & 26.705 & 0.061 & 25.439 & 0.063 & 0 & 1 & 0 & 1 & 0 & 0.8 & 0.6 \\
  V31 & 3:25:03.430 & -36:22:01.00 & 103.3 & 389.2 & 26.71 & 0.036 & 25.616 & 0.048 & 1 & 0 & 1 & 0 & 0 & 1.2 & 1.2\\
  V32 & 3:25:10.554 & -36:22:35.97 & 458.04 & 702.58 & 26.733 & 0.038 & 26.229 & 0.079 & 1 & 0 & 1 & 1 & 0 & 1.2 & 1.2 & 2433197\\
  V33 & 3:25:11.556 & -36:21:59.34 & 771.17 & 517.85 & 26.739 & 0.042 & 26.14 & 0.075 & 1 & 1 & 1 & 0 & 1 & 1.2 & 1.0\\
  V34 & 3:25:06.693 & -36:21:26.83 & 205.93 & 524.64 & 26.781 & 0.038 & 26.221 & 0.068 & 1 & 1 & 0 & 0 & 0 & 1.2 & 1.2\\
  V35 & 3:25:04.707 & -36:21:28.13 & 352.99 & 327.51 & 26.838 & 0.041 & 25.975 & 0.058 & 1 & 0 & 1 & 0 & 0 & 1.0 & 1.2 & 58705971\\
  V36 & 3:25:06.615 & -36:21:28.50 & 196.58 & 515.31 & 26.845 & 0.04 & 26.613 & 0.102 & 0 & 1 & 0 & 1 & 0 & 0.8 & 0.8\\
  V37 & 3:25:07.951 & -36:22:01.13 & 426.33 & 252.91 & 26.943 & 0.057 & 25.935 & 0.08 & 1 & 0 & 1 & 0 & 0 & 0.8 & 0.4 & 11471347\\
  V38 & 3:25:08.459 & -36:21:23.01 & 96.63 & 712.02 & 26.951 & 0.051 & 26.301 & 0.085 & 1 & 0 & 1 & 0 & 0 & 0.8 & 0.3 & 56314445\\
  V39 & 3:25:06.465 & -36:20:54.33 & 472.02 & 716.92 & 26.959 & 0.043 & 25.856 & 0.047 & 0 & 1 & 0 & 1 & 1 & 0.8 & 0.6\\
  V40 & 3:25:06.931 & -36:21:26.58 & 197.36 & 546.34 & 27.023 & 0.045 & 26.462 & 0.09 & 0 & 1 & 1 & 1 & 0 & 1.2 & 0.7\\
  V41 & 3:25:04.678 & -36:21:23.05 & 400.36 & 358.78 & 27.061 & 0.052 & 26.125 & 0.079 & 1 & 0 & 1 & 0 & 1 & 0.8 & 0.6 & 90123409\\
  V42 & 3:25:03.167 & -36:22:06.92 & 64.05 & 134.93 & 27.111 & 0.056 & 26.628 & 0.106 & 0 & 1 & 0 & 1 & 0 & 1.2 & 0.8\\
  V43 & 3:25:03.167 & -36:22:06.92 & 241.1 & 333.8 & 27.244 & 0.053 & 25.895 & 0.053 & 1 & 1 & 0 & 1 & 0 & 1.5 & 1.4\\
  V44 & 3:25:08.145 & -36:22:11.45 & 447.81 & 459.36 & 27.263 & 0.063 & 26.94 & 0.14 & 1 & 0 & 1 & 0 & 0 & 1.2 & 1.2\\
  V45 & 3:25:03.829 & -36:22:21.71 & 199.76 & 109.74 & 27.3 & 0.059 & 26.78 & 0.12 & 0 & 1 & 0 & 1 & 1 & 0.6 & 0.6 & 82802949 & \\
  V46 & 3:25:07.091 & -36:21:16.73 & 376.76 & 344.07 & 27.327 & 0.058 & 26.283 & 0.071 & 0 & 1 & 1 & 1 & 1 & 0.8 & 0.6 & 53129508\\
  V47 & 3:25:03.320 & -36:22:00.80 & 122.09 & 411.43 & 27.337 & 0.063 & 26.57 & 0.109 & 1 & 0 & 1 & 1 & 1 & 0.8 & 0.6\\
  V48 & 3:25:05.855 & -36:21:19.77 & 324.31 & 488.78 & 27.443 & 0.062 & 26.752 & 0.119 & 1 & 0 & 1 & 0 & 0 & 1.2 & 1 & 93149850\\

\hline
\end{tabular}}
\begin{tablenotes}
\item \textbf{Note.} --Units of right ascension are hours, minutes, and seconds, and units of declination are degrees, arcminutes, and arcseconds. Numbers 1 or 0 in columns MAD, IQR and inverse von Neumann ratio show whether the star was characterized as a variable in the respective index. Units of X and Y are pixels coordinates. Columns $\Delta_ {V}$ and $\Delta_{I}$ show the amplitude of variability.
\end{tablenotes}
\label{vars1326}
\end{sidewaystable*}

\begin{sidewaystable*} [ht!]

\caption{Properties of the variable candidates in NGC 1425 ordered by average magnitudes. For convenience we use the letter "V" to name each variable in this field.}
\scalebox{0.8}{
\begin{tabular}{l c c c c c c c c c c c c c c c c c c}
\hline
{ID} &RA &Dec &{X}& {Y} & {V}& {$\rm \sigma$ $_{V}$} &{I}&{$\rm \sigma$ $_{I}$}&  {MAD$_{V}$} & {MAD$_{I}$} & {IQR$_{V}$} & {IQR $_{I}$} & {1/$\eta$ $_{V,I}$} & {$\Delta$V} & {$\Delta$I}&{HSC MatchID}&{Notes}  \\ 

& (J2000) & (J2000)& (pixels)&(pixels) & (mag) & (mag) & (mag)& (mag) & & &  &  & & (mag) & (mag) & \\
\hline

  V1 & 3:42:15.539 & -29:54:51.62 & 607.53 & 672.6 & 22.662 & 0.004 & 22.323 & 0.005 & 1 & 0 & 1 & 0 & 1 & 0.12 & 0.2 &  & \\
  V2 & 3:42:20.029 & -29:54:47.57 & 288.98 & 72.49 & 23.381 & 0.005 & 22.839 & 0.008 & 1 & 1 & 1 & 1 & 1 & 0.2 & 0.2 & 90652043 & candidate YSG \\
  V3 & 3:42:18.351 & -29:53:59.39 & 547.65 & 376.16 & 23.581 & 0.005 & 22.991 & 0.007 & 1 & 1 & 1 & 1 & 1 & 0.2 & 0.1 & 9637663 & candidate YSG\\
  V4 & 3:42:19.469 & -29:53:42.77 & 596.66 & 402.37 & 23.605 & 0.005 & 23.022 & 0.007 & 1 & 1 & 1 & 1 & 1 & 0.1 & 0.2 & 41675618 & candidate YSG\\
  V5 & 3:42:18.570 & -29:54:54.44 & 78.41 & 72.46 & 23.722 & 0.005 & 23.237 & 0.008 & 1 & 1 & 1 & 1 & 1 & 0.2 & 0.2 & 7739340 & candidate YSG\\
  V6 & 3:42:22.972 & -29:55:10.64 & 469.5 & 485.1 & 23.935 & 0.005 & 23.51 & 0.009 & 1 & 0 & 1 & 1 & 1 & 0.1 & 0.3 & 58838018 & candidate YSG\\
  V7 & 3:42:16.946 & -29:53:56.02 & 679.1 & 256.56 & 23.968 & 0.006 & 23.931 & 0.015 & 1 & 1 & 1 & 1 & 0 & 0.3 & 0.3 & 19102580 & candidate BSG\\
  V8 & 3:42:17.970 & -29:54:48.04 & 175.09 & 59.46 & 24.019 & 0.009 & 23.014 & 0.01 & 1 & 0 & 1 & 0 & 1 & 0.3 & 0.3 & 68385761 & candidate YSG\\
  V9 & 3:42:20.380 & -29:54:53.31 & 292.63 & 146.83 & 24.053 & 0.006 & 23.787 & 0.011 & 1 & 1 & 0 & 1 & 0 & 0.2 & 0.2 & 65467227 & candidate YSG\\
  V10 & 3:42:18.356 & -29:53:58.01 & 548.53 & 376.8 & 24.062 & 0.006 & 23.101 & 0.008 & 1 & 0 & 1 & 1 & 0 & 0.1 & 0.6 & 20177712 & candidate YSG\\
  V11 & 3:42:23.636 & -29:55:02.87 & 585.08 & 471.66 & 24.232 & 0.006 & 23.677 & 0.01 & 1 & 1 & 0 & 0 & 1 & 0.1 & 0.3 & 20606278 & candidate YSG\\
  V12 & 3:42:17.749 & -29:54:52.61 & 79.59 & 282.61 & 24.275 & 0.008 & 24.078 & 0.016 & 1 & 1 & 1 & 1 & 1 & 0.6 & 0.5 & 95116497 & candidate BSG\\
  V13 & 3:42:20.302 & -29:54:42.75 & 345.69 & 54.01 & 24.276 & 0.008 & 23.657 & 0.012 & 1 & 1 & 1 & 1 & 0 & 0.3 & 0.2 & 90537371 & candidate YSG\\
  V14 & 3:42:19.723 & -29:54:39.85 & 107.7 & 294.22 & 24.305 & 0.007 & 24.014 & 0.014 & 1 & 1 & 1 & 1 & 0 & 0.2 & 0.7 & 36741455 & candidate YSG\\
  V15 & 3:42:15.877 & -29:54:04.96 & 692.26 & 78.84 & 24.823 & 0.016 & 24.695 & 0.032 & 1 & 1 & 1 & 1 & 1 & 0.4 & 0.4 & 73522719\\ 
  V16 & 3:42:17.557 & -29:55:06.01 & 316.94 & 71.03 & 25.057 & 0.014 & 24.91 & 0.031 & 1 & 1 & 1 & 1 & 1 & 0.4 & 0.7 & 37383109\\ 
  V17 & 3:42:15.719 & -29:54:04.63 & 707.22 & 63.73 & 25.162 & 0.027 & 24.875 & 0.04 & 1 & 1 & 1 & 1 & 0 & 0.4 & 0.8 & 500234\\ 
  V18 & 3:42:21.999 & -29:55:25.97 & 276.1 & 597.5 & 25.179 & 0.011 & 24.572 & 0.018 & 1 & 1 & 1 & 1 & 0 & 0.3 & 0.2 & 107344064 \\ 
  V19 & 3:42:16.721 & -29:54:59.73 & 431.0 & 326.95 & 25.226 & 0.012 & 25.226 & 0.035 & 1 & 1 & 1 & 1 & 1 & 0.3 & 0.6 & 45871608 \\ 
  V20 & 3:42:15.025 & -29:54:51.11 & 734.87 & 768.76 & 25.252 & 0.017 & 24.558 & 0.029 & 1 & 1 & 1 & 1 & 0 & 0.3 & 0.4 & 45578344 \\ 
  V21 & 3:42:17.621 & -29:54:53.79 & 125.44 & 282.59 & 25.289 & 0.013 & 24.708 & 0.024 & 0 & 1 & 0 & 1 & 1 & 0.3 & 0.3 & 52701106 \\
  V22 & 3:42:16.262 & -29:54:04.50 & 666.15 & 123.03 & 25.346 & 0.019 & 25.534 & 0.057 & 1 & 0 & 1 & 0 & 1 & 0.3 & 0.6 & 16420841\\ 
  V23 & 3:42:15.596 & -29:54:03.99 & 722.16 & 54.1 & 25.347 & 0.029 & 24.908 & 0.045 & 1 & 0 & 0 & 0 & 0 & 0.3 & 0.4 & 40931113 \\
  V24 & 3:42:20.106 & -29:53:59.76 & 407.59 & 570.31 & 25.403 & 0.016 & 23.125 & 0.007 & 1 & 1 & 1 & 1 & 1 & 0.4 & 0.8 & 83509791 & candidate RSG\\
  V25 & 3:42:17.018 & -29:54:37.00 & 51.99 & 686.92 & 25.412 & 0.024 & 25.424 & 0.065 & 1 & 0 & 1 & 0 & 1 & 0.6 & 0.5 & 85512615 \\
  V26 & 3:42:16.301 & -29:54:48.17 & 362.69 & 600.18 & 25.416 & 0.014 & 25.216 & 0.037 & 1 & 1 & 1 & 1 & 1 & 0.3 & 0.6 & 92064031 \\
  V27 & 3:42:14.909 & -29:54:51.69 & 768.39 & 782.04 & 25.436 & 0.02 & 24.718 & 0.028 & 0 & 1 & 0 & 1 & 0 & 0.4 & 0.3 & 62147934 \\
  V28 & 3:42:15.821 & -29:54:06.59 & 683.16 & 63.23 & 25.447 & 0.028 & 24.829 & 0.051 & 1 & 0 & 0 & 0 & 1 & 0.3 & 0.5 \\
  V29 & 3:42:15.629 & -29:54:52.45 & 589.29 & 650.79 & 25.484 & 0.016 & 25.025 & 0.037 & 0 & 1 & 0 & 0 & 0 & 0.4 & 0.6 & 100312212\\ 
  V30 & 3:42:16.726 & -29:54:57.24 & 400.82 & 377.97 & 25.502 & 0.019 & 23.106 & 0.007 & 1 & 1 & 1 & 1 & 0 & 0.6 & 0.7 &  & candidate RSG\\ 
  V31 & 3:42:17.595 & -29:55:06.13 & 310.97 & 68.1 & 25.591 & 0.019 & 25.246 & 0.041 & 1 & 1 & 1 & 1 & 1 & 0.5 & 0.9 & 51633672 \\
  V32 & 3:42:17.272 & -29:54:34.70 & 341.29 & 53.92 & 25.606 & 0.022 & 25.878 & 0.074 & 1 & 0 & 0 & 0 & 0 & 0.5 & 0.5 & 9181183 \\
  V33 & 3:42:18.872 & -29:54:09.60 & 424.6 & 371.7 & 25.665 & 0.018 & 24.633 & 0.023 & 1 & 1 & 1 & 1 & 1 & 0.5 & 0.6 & 74867700 \\
  V34 & 3:42:15.772 & -29:54:03.34 & 713.67 & 77.0 & 25.687 & 0.026 & 25.859 & 0.096 & 1 & 1 & 1 & 1 & 0 & 0.6 & 0.4 & \\
  V35 & 3:42:18.782 & -29:55:30.24 & 328.4 & 170.58 & 25.718 & 0.018 & 24.734 & 0.022 & 1 & 1 & 1 & 1 & 1 & 0.6 & 0.4 & 86921818 \\
  V36 & 3:42:16.020 & -29:54:09.67 & 642.41 & 66.7 & 25.775 & 0.028 & 25.497 & 0.063 & 1 & 0 & 1 & 0 & 0 & 0.6 & 0.5\\
  V37 & 3:42:15.732 & -29:54:05.62 & 698.08 & 59.29 & 25.791 & 0.034 & 25.294 & 0.072 & 1 & 0 & 0 & 0 & 0 & 0.9 & 0.5 & 63484288 \\
  V38 & 3:42:16.012 & -29:54:09.42 & 645.04 & 67.31 & 25.945 & 0.036 & 25.329 & 0.052 & 1 & 1 & 1 & 1 & 1 & 0.8 & 0.9\\
  V39 & 3:42:15.690 & -29:54:05.88 & 699.16 & 53.21 & 25.949 & 0.043 & 25.346 & 0.072 & 0 & 1 & 0 & 0 & 0 & 0.3 & 0.3 & 63484288 \\
  V40 & 3:42:19.711 & -29:54:35.47 & 149.28 & 309.16 & 25.951 & 0.023 & 25.266 & 0.036 & 1 & 1 & 1 & 1 & 1 & 0.3 & 0.8 \\
  V41 & 3:42:22.870 & -29:55:36.10 & 296.42 & 690.0 & 25.998 & 0.021 & 25.15 & 0.029 & 1 & 1 & 1 & 1 & 0 & 0.6 & 0.5 & 99010888 \\
  V42 & 3:42:15.667 & -29:54:02.82 & 726.22 & 68.72 & 26.012 & 0.035 & 25.368 & 0.064 & 1 & 1 & 1 & 1 & 0 & 0.6 & 0.6 & 52737352 & \\
  V43 & 3:42:17.636 & -29:55:05.68 & 294.38 & 73.19 & 26.022 & 0.026 & 26.189 & 0.084 & 1 & 1 & 1 & 1 & 0 & 0.6 & 0.6 & 22885126 & \\
  V44 & 3:42:18.911 & -29:54:19.81 & 338.36 & 316.11 & 26.276 & 0.041 & 26.173 & 0.129 & 1 & 0 & 0 & 0 & 0 & 0.5 & 0.6\\
  V45 & 3:42:19.012 & -29:54:51.80 & 64.66 & 135.92 & 26.324 & 0.034 & 25.029 & 0.037 & 1 & 0 & 1 & 0 & 0 & 0.6 & 0.6 & 57959110 &\\ 
  V46 & 3:42:15.505 & -29:54:01.83 & 746.99 & 56.93 & 26.349 & 0.052 & 25.424 & 0.072 & 1 & 0 & 0 & 0 & 0 & 0.6 & 0.9 & 70111043 &\\ 
  V47 & 3:42:15.818 & -29:54:03.74 & 706.8 & 79.67 & 26.402 & 0.052 & 25.472 & 0.067 & 1 & 0 & 1 & 0 & 0 & 0.6 & 0.8 & 20780705 & \\
  V48 & 3:42:19.524 & -29:55:04.43 & 137.06 & 176.73 & 26.417 & 0.033 & 25.878 & 0.062 & 1 & 1 & 1 & 1 & 0 & 0.8 & 1.2 & 91641235 & \\
  V49 & 3:42:18.878 & -29:53:56.81 & 539.26 & 434.64 & 26.446 & 0.034 & 25.802 & 0.057 & 1 & 1 & 1 & 1 & 0 & 1.2 & 0.6\\
  V50 & 3:42:17.618 & -29:55:02.63 & 238.44 & 125.9 & 26.458 & 0.031 & 25.973 & 0.066 & 1 & 0 & 1 & 0 & 0 & 0.2 & 0.5 & 73383575 & \\

\hline

\end{tabular}}
\begin{tablenotes}
\item \textbf{Note.}-- Units of right ascension are hours, minutes, and seconds, and units of declination are degrees, arcminutes, and arcseconds. Numbers 1 or 0 in columns MAD, IQR and inverse von Neumann ratio show whether the star was characterized as a variable in the respective index. Units of X and Y are pixels coordinates. Columns $\Delta_ {V}$ and $\Delta_{I}$ show the amplitude of variability.
\end{tablenotes}
\label{vars14251}
\end{sidewaystable*}

\begin{sidewaystable*} [ht!]

\caption{Table~\ref{vars14251} continued.}
\scalebox{0.8}{
\begin{tabular}{l c c c c c c c c c c c c c c c c c c}
\hline
{ID} & RA& Dec& {X}& {Y} &  {V}& {$\rm \sigma$ $_{V}$} &{I}&{$\rm \sigma$ $_{I}$}&  {MAD$_{V}$} & {MAD$_{I}$} & {IQR$_{V}$} & {IQR $_{I}$} & {1/$\eta$ $_{V,I}$} & {$\Delta$V} & {$\Delta$I}&{HSC MatchID}&{Notes}  \\ 

& (J2000)& (J2000)&(pixels)& (pixels)& (mag) & (mag) & (mag)& (mag) & & &  &  & & (mag) & (mag) & \\
\hline

    V51 & 3:42:15.955 & -29:54:05.15 & 684.61 & 86.25 & 26.533 & 0.049 & 26.138 & 0.114 & 1 & 1 & 1 & 0 & 0 & 0.5 & 0.5 & 3333322 & \\
  V52 & 3:42:15.373 & -29:53:59.48 & 776.83 & 56.28 & 26.535 & 0.055 & 25.866 & 0.109 & 1 & 1 & 1 & 1 & 0 & 0.8 & 0.7\\
  V53 & 3:42:19.577 & -29:54:37.80 & 144.27 & 285.7 & 26.551 & 0.038 & 25.931 & 0.065 & 1 & 0 & 1 & 0 & 1 & 0.4 & 0.5\\
  V54 & 3:42:16.216 & -29:54:40.69 & 303.37 & 768.7 & 26.569 & 0.038 & 26.439 & 0.107 & 1 & 0 & 1 & 0 & 1 & 0.4 & 0.6\\
  V55 & 3:42:16.005 & -29:54:53.00 & 515.36 & 570.3 & 26.621 & 0.035 & 26.018 & 0.066 & 1 & 0 & 1 & 0 & 1 & 0.8 & 0.9\\
  V56 & 3:42:16.250 & -29:53:56.30 & 741.44 & 168.6 & 26.627 & 0.046 & 25.762 & 0.07 & 1 & 1 & 1 & 0 & 0 & 0.8 & 0.5 & 79220173 & \\
  V57 & 3:42:23.221 & -29:54:57.96 & 569.9 & 400.4 & 26.635 & 0.035 & 25.703 & 0.046 & 1 & 1 & 1 & 0 & 0 & 1.0 & 0.9 &  & candidate periodic variable (P=50 days)\\
  V58 & 3:42:20.149 & -29:54:33.33 & 125.6 & 373.7 & 26.675 & 0.039 & 26.796 & 0.129 & 1 & 1 & 1 & 1 & 0 & 1.2 & 0.5 & 84479115 & \\
  V59 & 3:42:17.378 & -29:54:45.94 & 56.3 & 462.0 & 26.693 & 0.047 & 26.397 & 0.123 & 1 & 0 & 0 & 0 & 0 & 0.5 & 0.5\\
  V60 & 3:42:16.481 & -29:55:00.84 & 503.19 & 351.06 & 26.717 & 0.039 & 25.399 & 0.041 & 1 & 0 & 0 & 0 & 1 & 1.2 & 0.6 & 75007472 & \\
  V61 & 3:42:19.765 & -29:54:36.86 & 122.72 & 302.24 & 26.722 & 0.046 & 26.43 & 0.11 & 1 & 1 & 1 & 1 & 0 & 0.6 & 0.5 & 85716829 & \\
  V62 & 3:42:17.026 & -29:54:58.47 & 363.29 & 287.41 & 26.733 & 0.04 & 25.911 & 0.061 & 1 & 1 & 0 & 0 & 0 & 1.2 & 0.8 & 73155751 & \\
  V63 & 3:42:17.550 & -29:55:49.95 & 397.92 & 414.84 & 26.75 & 0.039 & 25.205 & 0.03 & 1 & 0 & 1 & 0 & 0 & 0.8 & 0.7 &  & candidate RSG\\
  V64 & 3:42:17.401 & -29:55:00.00 & 285.91 & 202.02 & 26.754 & 0.038 & 26.393 & 0.093 & 1 & 1 & 1 & 1 & 1 & 0.8 & 1.1 &  & candidate periodic variable (P=29 days) \\
  V65 & 3:42:20.538 & -29:55:18.40 & 154.16 & 361.87 & 26.777 & 0.04 & 25.97 & 0.061 & 1 & 0 & 1 & 0 & 0 & 1.5 & 1.2 &  & candidate periodic variable (P=29 days)\\
  V66 & 3:42:16.005 & -29:54:51.58 & 471.15 & 558.42 & 26.786 & 0.046 & 26.336 & 0.094 & 0 & 0 & 1 & 0 & 0 & 1.2 & 1.2 & 1612324 & \\
  V67 & 3:42:15.551 & -29:54:46.35 & 533.22 & 746.67 & 26.909 & 0.054 & 26.674 & 0.146 & 1 & 0 & 1 & 1 & 0 & 0.8 & 0.6\\
  V68 & 3:42:16.194 & -29:54:48.81 & 400.48 & 599.78 & 26.917 & 0.058 & 25.679 & 0.062 & 0 & 0 & 1 & 1 & 0 & 0.9 & 0.8\\
  V69 & 3:42:19.922 & -29:54:39.02 & 92.26 & 310.95 & 26.931 & 0.054 & 26.543 & 0.116 & 1 & 0 & 0 & 0 & 1 & 0.9 & 0.6 & 94862026 & \\
  V70 & 3:42:20.176 & -29:55:05.26 & 192.59 & 237.09 & 26.932 & 0.046 & 26.167 & 0.072 & 1 & 0 & 1 & 0 & 0 & 1.2 & 0.8 &  \\
  V71 & 3:42:18.563 & -29:55:35.54 & 369.78 & 328.29 & 26.995 & 0.048 & 26.224 & 0.073 & 1 & 1 & 1 & 1 & 0 & 1.2 & 0.8 & 45614411 & candidate periodic variable (P=60 days)\\
  V72 & 3:42:19.199 & -29:53:38.50 & 642.29 & 604.24 & 27.008 & 0.057 & 26.111 & 0.088 & 0 & 0 & 1 & 0 & 0 & 0.8 & 1.2 & 97364472 &  & \\
  V73 & 3:42:21.380 & -29:54:06.84 & 230.86 & 685.52 & 27.029 & 0.053 & 26.511 & 0.11 & 1 & 1 & 1 & 1 & 0 & 0.9 & 0.8 & 103267462 & \\
  V74 & 3:42:17.137 & -29:54:24.72 & 466.385 & 46.96 & 27.032 & 0.058 & 26.571 & 0.133 & 1 & 1 & 0 & 0 & 0 & 0.9 & 1.2 & 7043734 & \\
  V75 & 3:42:24.036 & -29:54:46.47 & 760.46 & 392.97 & 27.048 & 0.052 & 26.923 & 0.131 & 1 & 0 & 1 & 0 & 1 & 1.2 & 1.2 & 17813499 & \\
  V76 & 3:42:16.328 & -29:55:12.83 & 672.69 & 89.74 & 27.073 & 0.052 & 26.891 & 0.154 & 0 & 1 & 1 & 1 & 0 & 0.9 & 0.8 & 3845446 & \\
  V77 & 3:42:17.463 & -29:54:46.40 & 58.81 & 458.61 & 27.074 & 0.055 & 26.69 & 0.164 & 1 & 1 & 1 & 1 & 0 & 1.0 & 1 &  & candidate periodic variable (P=42 days)\\
  V78 & 3:42:16.474 & -29:54:59.58 & 498.47 & 378.83 & 27.079 & 0.05 & 26.66 & 0.124 & 1 & 1 & 1 & 1 & 0 & 0.9 & 0.7\\
  V79 & 3:42:16.735 & -29:55:08.41 & 540.07 & 172.78 & 27.082 & 0.052 & 26.755 & 0.148 & 1 & 1 & 1 & 0 & 0 & 0.9 & 0.8\\
  V80 & 3:42:19.627 & -29:54:35.76 & 153.35 & 298.56 & 27.084 & 0.058 & 26.3 & 0.09 & 1 & 1 & 1 & 1 & 1 & 1.2 & 1.2\\
  V81 & 3:42:16.291 & -29:54:39.87 & 287.02 & 781.71 & 27.085 & 0.058 & 26.773 & 0.14 & 0 & 1 & 0 & 1 & 0 & 0.9 & 0.8\\
  V82 & 3:42:18.981 & -29:55:03.33 & 84.54 & 122.62 & 27.097 & 0.057 & 26.113 & 0.077 & 1 & 1 & 1 & 0 & 0 & 0.8 & 0.9 & 83539125 & \\
  V83 & 3:42:15.673 & -29:54:57.54 & 650.81 & 543.61 & 27.099 & 0.056 & 26.947 & 0.166 & 1 & 1 & 1 & 1 & 0 & 0.5 & 0.3 & 80153594 & \\
  V84 & 3:42:19.563 & -29:54:45.77 & 75.79 & 232.78 & 27.116 & 0.06 & 26.005 & 0.072 & 1 & 0 & 1 & 0 & 0 & 0.9 & 0.2 & 16527370 & \\
  V85 & 3:42:17.915 & -29:55:02.81 & 199.54 & 78.3 & 27.121 & 0.056 & 25.877 & 0.061 & 1 & 1 & 1 & 1 & 0 & 0.6 & 0.3 & 57938375 & \\
  V86 & 3:42:20.141 & -29:54:02.37 & 366.41 & 567.74 & 27.124 & 0.054 & 26.226 & 0.084 & 1 & 0 & 0 & 0 & 0 & 1.2 & 1.2 & 29608543 & \\
  V87 & 3:42:22.508 & -29:55:37.02 & 254.6 & 663.74 & 27.124 & 0.066 & 27.033 & 0.181 & 1 & 0 & 0 & 0 & 0 & 1.0 & 1.0\\
  V88 & 3:42:15.388 & -29:54:49.37 & 607.68 & 727.97 & 27.143 & 0.057 & 26.406 & 0.113 & 1 & 0 & 0 & 0 & 0 & 0.8 & 0.3\\
  V89 & 3:42:17.130 & -29:53:47.30 & 742.54 & 322.8 & 27.146 & 0.067 & 25.759 & 0.061 & 0 & 1 & 0 & 1 & 0 & 0.8 & 0.9 & 34302839 & candidate RSG\\
  V90 & 3:42:20.369 & -29:55:25.90 & 92.12 & 410.64 & 27.161 & 0.056 & 25.571 & 0.041 & 0 & 1 & 0 & 1 & 0 & 1.2 & 0.6 & 27428939 & candidate RSG\\
  V91 & 3:42:17.558 & -29:54:47.12 & 67.77 & 424.13 & 27.167 & 0.055 & 27.103 & 0.18 & 0 & 1 & 1 & 1 & 0 & 1.0 & 0.3\\
  V92 & 3:42:24.786 & -29:55:29.20 & 552.78 & 781.16 & 27.175 & 0.054 & 26.962 & 0.129 & 0 & 1 & 0 & 1 & 1 & 1.2 & 1.2\\
  V93 & 3:42:17.589 & -29:54:08.30 & 533.02 & 243.01 & 27.181 & 0.063 & 25.955 & 0.066 & 0 & 1 & 0 & 1 & 0 & 0.9 & 0.6\\
  V94 & 3:42:18.051 & -29:54:57.42 & 69.18 & 146.2 & 27.187 & 0.057 & 26.601 & 0.14 & 1 & 0 & 1 & 0 & 0 & 0.2 & 0.0\\
  V95 & 3:42:15.810 & -29:55:05.01 & 714.27 & 383.1 & 27.197 & 0.061 & 26.469 & 0.106 & 1 & 0 & 1 & 0 & 0 & 0.5 & 0.0\\
  V96 & 3:42:19.226 & -29:55:02.94 & 112.49 & 137.05 & 27.198 & 0.07 & 25.03 & 0.029 & 1 & 1 & 1 & 1 & 1 & 0.4 & 0.7 & 101548221 & candidate RSG\\
  V97 & 3:42:16.448 & -29:55:20.24 & 72.06 & 359.9 & 27.212 & 0.059 & 27.245 & 0.186 & 1 & 0 & 1 & 0 & 1 & 1.2 & 1.2 & 87402582 & \\
  V98 & 3:42:16.256 & -29:54:56.89 & 493.4 & 444.97 & 27.238 & 0.059 & 26.231 & 0.081 & 1 & 0 & 1 & 0 & 0 & 1.1 & 1.1 & 39509540 & \\
  V99 & 3:42:17.302 & -29:54:48.13 & 106.73 & 451.16 & 27.293 & 0.067 & 25.992 & 0.072 & 1 & 0 & 0 & 0 & 1 & 1.0 & 0.5 & 11666137 & \\
  V100 & 3:42:17.121 & -29:55:04.67 & 392.81 & 181.35 & 27.387 & 0.064 & 26.616 & 0.107 & 1 & 0 & 1 & 0 & 0 & 1.2 & 0.6 & 24815704 & \\
  V101 & 3:42:16.660 & -29:55:00.14 & 461.21 & 336.5 & 27.407 & 0.067 & 26.14 & 0.087 & 1 & 1 & 1 & 1 & 0 & 0.6 & 1.1 & 48554067 & \\
  V102 & 3:42:17.010 & -29:54:52.01 & 259.73 & 435.11 & 27.422 & 0.066 & 26.334 & 0.09 & 1 & 0 & 1 & 0 & 0 & 1.2 & 0.0\\

\hline

\end{tabular}}
\begin{tablenotes}
\item \textbf{Note.}-- Units of right ascension are hours, minutes, and seconds, and units of declination are degrees, arcminutes, and arcseconds. Numbers 1 or 0 in columns MAD, IQR and inverse von Neumann ratio show whether the star was characterized as a variable in the respective index. Units of X and Y are pixels coordinates. Columns $\Delta_ {V}$ and $\Delta_{I}$ show the amplitude of variability.
\end{tablenotes}
\label{vars14252}
\end{sidewaystable*}

\begin{sidewaystable*} [ht!]

\caption{Properties of the variable candidates in NGC 4548 ordered by average magnitudes. For convenience we use the letter "V" to name each variable in this field.}
\scalebox{0.8}{
\begin{tabular}{l c c c c c c c c c c c c c c c c c c}
\hline
{ID} &RA& Dec& {X}& {Y} &  {V}& {$\rm \sigma$ $_{V}$} &{I}&{$\rm \sigma$ $_{I}$}&  {MAD$_{V}$} & {MAD$_{I}$} & {IQR$_{V}$} & {IQR $_{I}$} & {1/$\eta$ $_{V,I}$} & {$\Delta$V} & {$\Delta$I}&{HSC MatchID}&{Notes}  \\

& (J2000)& (J2000) &(pixels) &(pixels) & (mag) & (mag) & (mag)& (mag) & & &  &  & & (mag) & (mag) & \\
\hline

   V1 & 12:35:24.398 & +14:29:02.32 & 615.07 & 508.51 & 23.13 & 0.005 & 22.554 & 0.007 & 1 & 1 & 1 & 1 & 1 & 0.8 & 0.3 &  & candidate YSG\\
  V2 & 12:35:23.771 & +14:28:44.86 & 740.39 & 350.46 & 23.385 & 0.005 & 23.367 & 0.011 & 1 & 1 & 1 & 1 & 1 & 0.3 & 0.2 &  & candidate YSG\\
  V3 & 12:35:23.915 & +14:28:26.29 & 761.15 & 162.91 & 23.445 & 0.006 & 23.396 & 0.012 & 1 & 1 & 1 & 1 & 1 & 0.2 & 0.4 & 37659734 & candidate YSG\\
  V4 & 12:35:23.013 & +14:29:17.79 & 783.82 & 701.88 & 23.491 & 0.007 & 23.027 & 0.013 & 1 & 1 & 1 & 1 & 1 & 0.3 & 0.2 & 101298266 & candidate YSG\\
  V5 & 12:35:27.705 & +14:29:03.65 & 137.36 & 412.85 & 23.527 & 0.006 & 23.254 & 0.013 & 1 & 1 & 1 & 1 & 1 & 0.3 & 0.3 &  & candidate YSG\\
  V6 & 12:35:25.186 & +14:28:24.83 & 580.28 & 108.85 & 23.558 & 0.005 & 23.277 & 0.01 & 1 & 0 & 1 & 0 & 1 & 0.3 & 0.1 & 38090981 & candidate YSG\\
  V7 & 12:35:23.904 & +14:28:23.57 & 768.82 & 135.72 & 23.564 & 0.007 & 23.916 & 0.016 & 0 & 1 & 0 & 0 & 0 & 0.15 & 0.15\\
  V8 & 12:35:23.891 & +14:28:34.28 & 746.26 & 243.11 & 23.654 & 0.006 & 23.778 & 0.015 & 1 & 1 & 1 & 1 & 1 & 0.1 & 0.1 & 37481267 & candidate YSG\\
  V9 & 12:35:27.176 & +14:29:08.27 & 203.58 & 474.39 & 23.657 & 0.009 & 24.418 & 0.032 & 1 & 1 & 1 & 1 & 1 & 0.3 & 0.3 &  & candidate BSG\\
  V10 & 12:35:24.934 & +14:29:04.23 & 529.68 & 510.46 & 23.675 & 0.006 & 23.153 & 0.011 & 1 & 1 & 1 & 1 & 1 & 0.16 & 0.3 & 9143332 & \\
  V11 & 12:35:26.099 & +14:29:27.33 & 403.75 & 498.11 & 23.731 & 0.007 & 23.458 & 0.015 & 1 & 1 & 1 & 1 & 0 & 0.12 & 0.2 &  & candidate YSG\\
  V12 & 12:35:30.830 & +14:28:38.22 & 315.33 & 697.37 & 23.755 & 0.007 & 22.952 & 0.011 & 1 & 1 & 1 & 1 & 1 & 0.1 & 0.12 & 80158243 & candidate YSG\\
  V13 & 12:35:23.131 & +14:29:24.51 & 70.65 & 336.35 & 23.772 & 0.009 & 23.987 & 0.022 & 1 & 1 & 0 & 1 & 0 & 0.2 & 0.1 & 17221257\\
  V14 & 12:35:25.854 & +14:28:25.77 & 748.84 & 764.24 & 23.779 & 0.007 & 23.23 & 0.013 & 1 & 1 & 1 & 1 & 0 & 0.2 & 0.15 & 4800577 & candidate YSG\\
  V15 & 12:35:24.552 & +14:29:10.34 & 483.21 & 97.27 & 23.786 & 0.008 & 23.343 & 0.015 & 1 & 1 & 1 & 1 & 0 & 0.2 & 0.1 &  & candidate YSG\\
  V16 & 12:35:29.507 & +14:28:59.25 & 573.55 & 578.13 & 23.801 & 0.009 & 23.691 & 0.024 & 1 & 1 & 1 & 1 & 0 & 0.2 & 0.3 & 67101111\\
  V17 & 12:35:28.881 & +14:28:38.72 & 317.71 & 190.5 & 23.835 & 0.007 & 23.417 & 0.016 & 1 & 0 & 1 & 0 & 1 & 0.3 & 0.3 & \\
  V18 & 12:35:31.087 & +14:29:14.22 & 133.16 & 57.38 & 23.863 & 0.007 & 24.119 & 0.021 & 1 & 1 & 1 & 0 & 1 & 0.1 & 0.2 & 54253155\\
  V19 & 12:35:23.911 & +14:28:35.87 & 417.96 & 446.84 & 23.929 & 0.008 & 24.393 & 0.027 & 0 & 1 & 1 & 1 & 0 & 0.6 & 0.4 & 34392135\\
  V20 & 12:35:30.115 & +14:29:46.01 & 739.91 & 257.91 & 23.942 & 0.007 & 23.886 & 0.016 & 1 & 0 & 1 & 0 & 1 & 0.3 & 0.2 & 37053207\\
  V21 & 12:35:26.724 & +14:29:24.68 & 760.98 & 373.09 & 23.972 & 0.008 & 24.364 & 0.028 & 1 & 1 & 1 & 1 & 0 & 0.3 & 0.3 & 26608022\\
  V22 & 12:35:24.753 & +14:29:11.54 & 232.4 & 650.49 & 24.008 & 0.009 & 22.861 & 0.011 & 1 & 0 & 1 & 0 & 0 & 0.2 & 0.3 &  & candidate RSG\\
  V23 & 12:35:27.062 & +14:29:21.28 & 542.24 & 582.31 & 24.014 & 0.008 & 24.171 & 0.028 & 1 & 1 & 0 & 0 & 1 & 0.7 & 0.7 & 81341135\\
  V24 & 12:35:31.169 & +14:29:14.72 & 189.01 & 611.95 & 24.091 & 0.009 & 24.328 & 0.036 & 1 & 1 & 0 & 0 & 0 & 0.7 & 0.7 & 1156563 & \\
  V25 & 12:35:22.983 & +14:29:17.73 & 420.4 & 459.66 & 24.132 & 0.008 & 23.24 & 0.011 & 1 & 1 & 0 & 0 & 1 & 0.18 & 0.1 & 80781554 & candidate RSG\\
  V26 & 12:35:23.487 & +14:29:21.98 & 780.41 & 701.32 & 24.133 & 0.009 & 23.719 & 0.018 & 1 & 1 & 0 & 0 & 1 & 0.8 & 0.2 & 101298266\\
  V27 & 12:35:25.874 & +14:29:02.98 & 707.19 & 738.07 & 24.165 & 0.009 & 24.155 & 0.027 & 1 & 1 & 1 & 1 & 0 & 0.1 & 0.2 & 24979858 & \\
  V28 & 12:35:24.065 & +14:29:16.59 & 400.61 & 464.72 & 24.169 & 0.01 & 24.503 & 0.034 & 1 & 1 & 0 & 0 & 0 & 0.1 & 0.2 & 16384262\\
  V29 & 12:35:31.669 & +14:29:49.91 & 630.82 & 655.48 & 24.175 & 0.01 & 23.126 & 0.013 & 1 & 1 & 0 & 0 & 0 & 0.7 & 0.1 &  & candidate RSG\\
  V30 & 12:35:25.647 & +14:29:11.57 & 753.48 & 604.28 & 24.197 & 0.011 & 24.442 & 0.027 & 1 & 1 & 1 & 0 & 1 & 0.6 & 0.4 & 93054781\\
  V31 & 12:35:23.461 & +14:29:07.74 & 410.96 & 562.46 & 24.229 & 0.012 & 24.604 & 0.04 & 1 & 0 & 1 & 0 & 1 & 0.5 & 0.4 & 62222454 \\
  V32 & 12:35:23.869 & +14:28:25.26 & 733.43 & 591.73 & 24.261 & 0.009 & 24.476 & 0.03 & 1 & 1 & 1 & 1 & 1 & 0.7 & 0.5\\
  V33 & 12:35:25.452 & +14:28:02.84 & 770.28 & 154.13 & 24.282 & 0.008 & 24.251 & 0.021 & 1 & 1 & 1 & 1 & 1 & 0.18 & 0.2 & 61828036\\
  V34 & 12:35:29.360 & +14:27:48.80 & 211.81 & 572.49 & 24.366 & 0.009 & 24.682 & 0.025 & 1 & 1 & 1 & 1 & 1 & 0.5 & 0.3\\
  V35 & 12:35:23.873 & +14:29:16.63 & 472.7 & 55.62 & 24.376 & 0.014 & 24.65 & 0.037 & 1 & 0 & 1 & 1 & 0 & 0.4 & 0.6 &  & \\
  V36 & 12:35:27.937 & +14:28:10.10 & 658.1 & 661.55 & 24.428 & 0.012 & 24.374 & 0.035 & 1 & 1 & 1 & 1 & 1 & 0.3 & 0.5\\
  V37 & 12:35:24.244 & +14:28:55.63 & 217.02 & 212.81 & 24.444 & 0.009 & 23.216 & 0.01 & 1 & 1 & 1 & 1 & 0 & 0.5 & 0.6\\
  V38 & 12:35:24.039 & +14:28:20.85 & 649.63 & 442.5 & 24.501 & 0.009 & 24.583 & 0.027 & 1 & 0 & 1 & 0 & 1 & 0.3 & 0.7 & 82850503\\
  V39 & 12:35:25.798 & +14:29:06.23 & 755.59 & 104.12 & 24.519 & 0.01 & 24.745 & 0.031 & 1 & 1 & 1 & 0 & 1 & 0.4 & 0.6\\
  V40 & 12:35:23.614 & +14:29:00.70 & 729.19 & 512.11 & 24.529 & 0.011 & 23.815 & 0.016 & 1 & 1 & 1 & 1 & 0 & 0.3 & 0.3 & 1066919\\
  V41 & 12:35:29.994 & +14:28:06.78 & 265.61 & 635.21 & 24.531 & 0.009 & 23.924 & 0.015 & 1 & 1 & 1 & 1 & 0 & 0.8 & 0.2\\
  V42 & 12:35:31.146 & +14:29:04.10 & 316.68 & 434.65 & 24.579 & 0.015 & 23.686 & 0.03 & 1 & 1 & 1 & 1 & 0 & 0.3 & 0.3 &  & candidate RSG\\
  V43 & 12:35:23.513 & +14:29:20.84 & 702.07 & 725.89 & 24.653 & 0.014 & 25.509 & 0.085 & 1 & 1 & 1 & 1 & 0 & 0.6 & 0.6 & 24979858\\
  V44 & 12:35:29.497 & +14:28:22.82 & 184.2 & 255.9 & 24.668 & 0.01 & 25.069 & 0.033 & 1 & 0 & 1 & 0 & 0 & 0.25 & 0.2 & 34077973\\
  V45 & 12:35:30.754 & +14:28:42.64 & 116.86 & 334.59 & 24.779 & 0.013 & 24.847 & 0.059 & 1 & 1 & 1 & 1 & 1 & 0.4 & 0.4 & 45607617 & \\
  V46 & 12:35:23.904 & +14:29:09.55 & 665.77 & 595.38 & 24.808 & 0.017 & 24.209 & 0.028 & 1 & 1 & 1 & 1 & 0 & 0.3 & 0.4 & 10916330\\
  V47 & 12:35:25.380 & +14:29:30.78 & 384.46 & 774.15 & 24.968 & 0.027 & 23.337 & 0.018 & 1 & 1 & 1 & 1 & 0 & 0.4 & 0.7 & 101291878 & candidate RSG\\
  V48 & 12:35:28.770 & +14:28:55.25 & 300.13 & 76.32 & 25.035 & 0.014 & 25.172 & 0.054 & 1 & 0 & 1 & 0 & 0 & 1.0 & 1.0\\
  V49 & 12:35:25.094 & +14:28:05.68 & 168.47 & 629.23 & 25.062 & 0.03 & 24.189 & 0.02 & 1 & 1 & 0 & 1 & 0 & 0.4 & 0.4 &  & candidate RSG\\
  V50 & 12:35:29.779 & +14:28:42.86 & 147.87 & 195.42 & 25.326 & 0.018 & 25.713 & 0.076 & 1 & 0 & 1 & 0 & 1 & 0.2 & 0.2\\
\hline
\end{tabular}}
\begin{tablenotes}
\item \textbf{Note.} --Units of right ascension are hours, minutes, and seconds, and units of declination are degrees, arcminutes, and arcseconds. Numbers 1 or 0 in columns MAD, IQR and inverse von Neumann ratio show whether the star was characterized as a variable in the respective index. Units of X and Y are pixels coordinates. Columns $\Delta_ {V}$ and $\Delta_{I}$ show the amplitude of variability.
\end{tablenotes}
\label{vars45481}
\end{sidewaystable*}

\begin{sidewaystable*} [ht!]

\caption{Table~\ref{vars45481} continued.}
\scalebox{0.8}{
\begin{tabular}{l c c c c c c c c c c c c c c c c c c}
\hline
{ID} &RA& Dec& {X}& {Y} &  {V}& {$\rm \sigma$ $_{V}$} &{I}&{$\rm \sigma$ $_{I}$}&  {MAD$_{V}$} & {MAD$_{I}$} & {IQR$_{V}$} & {IQR $_{I}$} & {1/$\eta$ $_{V,I}$} & {$\Delta$V} & {$\Delta$I}&{HSC MatchID}&{Notes}  \\ 
& (J2000) & (J2000) & (pixels)& (pixels)&(mag) & (mag) & (mag)& (mag) & & &  &  & & (mag) & (mag) & \\
\hline
    V51 & 12:35:27.325 & +14:28:54.99 & 210.52 & 339.69 & 25.587 & 0.023 & 25.135 & 0.051 & 1 & 1 & 1 & 1 & 0 & 0.2 & 0.8 & 75035962 & \\
  V52 & 12:35:26.377 & +14:28:13.03 & 137.44 & 429.0 & 25.608 & 0.022 & 24.879 & 0.032 & 1 & 1 & 1 & 1 & 0 & 0.8 & 0.6 & 91122684\\
  V53 & 12:35:31.141 & +14:28:59.88 & 275.38 & 425.32 & 25.625 & 0.023 & 24.904 & 0.038 & 1 & 1 & 1 & 1 & 1 & 1.2 & 0.8 & 31722844\\
  V54 & 12:35:30.198 & +14:28:18.98 & 385.91 & 386.15 & 25.688 & 0.019 & 26.012 & 0.073 & 1 & 1 & 1 & 1 & 1 & 0.6 & 0.6 & 34197730\\
  V55 & 12:35:30.355 & +14:28:33.71 & 503.13 & 78.94 & 25.946 & 0.031 & 26.304 & 0.107 & 1 & 1 & 1 & 1 & 0 & 0.25 & 0.25 & 36455381 & \\
  V56 & 12:35:27.270 & +14:27:19.47 & 693.9 & 419.3 & 25.966 & 0.025 & 25.242 & 0.037 & 1 & 1 & 1 & 1 & 1 & 0.5 & 0.6\\
  V57 & 12:35:29.997 & +14:29:08.38 & 452.94 & 361.4 & 26.091 & 0.033 & 25.595 & 0.068 & 1 & 1 & 1 & 1 & 0 & 1.2 & 1.2 & 55406013 & \\
  V58 & 12:35:23.638 & +14:29:02.13 & 719.99 & 530.64 & 26.141 & 0.038 & 25.706 & 0.08 & 1 & 1 & 1 & 1 & 1 & 1.2 & 0.6 & 42502132 & \\
  V59 & 12:35:31.340 & +14:29:11.74 & 386.03 & 478.0 & 26.141 & 0.036 & 25.511 & 0.068 & 1 & 1 & 1 & 1 & 1 & 0.6 & 0.7\\
  V60 & 12:35:32.108 & +14:28:41.69 & 66.55 & 526.41 & 26.235 & 0.043 & 25.309 & 0.055 & 1 & 0 & 1 & 1 & 0 & 1.2 & 1.2 & 32643206 & candidate RSG\\
  V61 & 12:35:27.631 & +14:27:55.20 & 354.35 & 289.31 & 26.271 & 0.039 & 25.891 & 0.076 & 1 & 1 & 1 & 0 & 1 & 0.8 & 0.8 & 34523449\\
  V62 & 12:35:25.254 & +14:29:06.10 & 480.85 & 515.56 & 26.276 & 0.05 & 26.604 & 0.206 & 0 & 1 & 0 & 0 & 0 & 0.3 & 0.5\\
  V63 & 12:35:31.095 & +14:29:20.04 & 474.94 & 460.05 & 26.32 & 0.051 & 25.777 & 0.097 & 1 & 0 & 0 & 0 & 0 & 0.7 & 0.5\\
  V64 & 12:35:30.950 & +14:28:03.96 & 552.02 & 759.99 & 26.328 & 0.034 & 27.163 & 0.192 & 1 & 1 & 1 & 1 & 1 & 0.8 & 0.2 &  & \\
  V65 & 12:35:28.238 & +14:29:27.10 & 630.57 & 64.66 & 26.332 & 0.051 & 26.002 & 0.124 & 1 & 0 & 0 & 0 & 0 & 0.6 & 0.5 & 72878727\\
  V66 & 12:35:23.349 & +14:29:05.78 & 753.83 & 575.9 & 26.427 & 0.047 & 25.662 & 0.087 & 1 & 1 & 0 & 0 & 0 & 0.2 & 0.8 & 95680813\\
  V67 & 12:35:32.131 & +14:29:18.85 & 479.41 & 538.61 & 26.446 & 0.048 & 26.17 & 0.105 & 1 & 0 & 0 & 0 & 0 & 0.7 & 0.8 & 1345627\\
  V68 & 12:35:31.675 & +14:29:00.57 & 340.17 & 594.99 & 26.482 & 0.047 & 26.341 & 0.13 & 1 & 1 & 1 & 1 & 0 & 0.8 & 0.8 & 26832370\\
  V69 & 12:35:30.909 & +14:28:08.19 & 558.55 & 666.14 & 26.49 & 0.038 & 26.148 & 0.079 & 1 & 1 & 1 & 1 & 0 & 0.5 & 0.9\\
  V70 & 12:35:27.149 & +14:29:17.72 & 184.36 & 574.13 & 26.564 & 0.067 & 26.253 & 0.163 & 1 & 1 & 1 & 1 & 0 & 1.0 & 1.6\\
  V71 & 12:35:23.592 & +14:28:46.83 & 625.31 & 587.7 & 26.589 & 0.068 & 24.4 & 0.035 & 1 & 1 & 1 & 1 & 0 & 0.6 & 0.2 & 107997074 & candidate RSG\\
  V72 & 12:35:24.195 & +14:29:10.13 & 74.0 & 214.24 & 26.59 & 0.047 & 26.432 & 0.125 & 1 & 1 & 0 & 1 & 0 & 0.8 & 0.2\\
  V73 & 12:35:27.714 & +14:28:23.79 & 125.42 & 174.45 & 26.602 & 0.054 & 26.342 & 0.121 & 1 & 1 & 0 & 0 & 0 & 0.7 & 0.7\\
  V74 & 12:35:28.198 & +14:28:40.69 & 214.01 & 686.13 & 26.631 & 0.051 & 26.392 & 0.139 & 1 & 1 & 1 & 1 & 0 & 0.8 & 0.2\\
  V75 & 12:35:32.971 & +14:28:59.25 & 474.57 & 446.04 & 26.637 & 0.061 & 26.798 & 0.207 & 1 & 1 & 1 & 1 & 0 & 0.2 & 0.2 & 1080337\\
  V76 & 12:35:25.396 & +14:28:59.14 & 597.15 & 225.25 & 26.641 & 0.049 & 25.918 & 0.073 & 1 & 1 & 1 & 0 & 1 & 0.8 & 0.4\\
  V77 & 12:35:30.742 & +14:28:28.12 & 572.92 & 778.77 & 26.652 & 0.063 & 26.071 & 0.091 & 1 & 1 & 0 & 0 & 1 & 0.8 & 0.8\\
  V78 & 12:35:31.022 & +14:28:03.40 & 436.1 & 193.41 & 26.664 & 0.055 & 26.011 & 0.105 & 1 & 0 & 0 & 0 & 0 & 1.0 & 0.9 & 98942748\\
  V79 & 12:35:29.361 & +14:29:10.84 & 323.91 & 794.9 & 26.684 & 0.069 & 26.706 & 0.2 & 1 & 0 & 0 & 0 & 0 & 0.4 & 0.8 \\
  V80 & 12:35:33.540 & +14:29:12.13 & 416.72 & 370.9 & 26.693 & 0.054 & 25.455 & 0.052 & 1 & 1 & 1 & 1 & 1 & 0.6 & 0.0\\
  V81 & 12:35:27.177 & +14:27:47.31 & 254.41 & 95.08 & 26.704 & 0.06 & 26.35 & 0.134 & 1 & 1 & 1 & 1 & 1 & 0.2 & 0.8\\
  V82 & 12:35:28.751 & +14:28:08.09 & 613.92 & 607.01 & 26.706 & 0.073 & 26.036 & 0.142 & 1 & 1 & 1 & 1 & 1 & 1.0 & 0.9 & 50783506\\
  V83 & 12:35:24.356 & +14:29:11.46 & 762.25 & 376.03 & 26.734 & 0.059 & 25.193 & 0.049 & 1 & 1 & 1 & 0 & 0 & 0.75 & 0.75 & 103939316 & candidate RSG\\
  V84 & 12:35:30.224 & +14:28:11.03 & 356.71 & 558.55 & 26.75 & 0.045 & 26.013 & 0.074 & 1 & 1 & 0 & 0 & 0 & 0.6 & 0.7 &  & \\
  V85 & 12:35:25.665 & +14:28:01.16 & 232.23 & 556.86 & 26.776 & 0.06 & 26.857 & 0.181 & 1 & 0 & 0 & 0 & 0 & 0.8 & 0.7 & 96500353\\
  V86 & 12:35:29.430 & +14:29:06.83 & 394.62 & 195.08 & 26.808 & 0.067 & 25.403 & 0.058 & 1 & 1 & 0 & 0 & 0 & 0.8 & 0.9 &  & candidate RSG\\
  V87 & 12:35:27.902 & +14:28:22.92 & 89.12 & 188.84 & 26.851 & 0.057 & 25.173 & 0.043 & 1 & 0 & 1 & 0 & 0 & 0.8 & 0.2 &  & candidate RSG \\
  V88 & 12:35:24.036 & +14:28:57.75 & 672.09 & 474.96 & 26.885 & 0.072 & 26.094 & 0.099 & 1 & 0 & 1 & 0 & 0 & 0.4 & 0.6 & 59753525\\
  V89 & 12:35:27.591 & +14:28:23.79 & 79.33 & 229.97 & 26.933 & 0.067 & 26.266 & 0.109 & 1 & 0 & 0 & 0 & 0 & 0.8 & 0.6\\
  V90 & 12:35:30.546 & +14:29:04.07 & 334.28 & 348.89 & 26.939 & 0.069 & 26.073 & 0.104 & 1 & 1 & 0 & 0 & 0 & 0.9 & 0.8 \\
  V91 & 12:35:28.862 & +14:27:25.30 & 688.36 & 179.47 & 26.944 & 0.061 & 26.356 & 0.092 & 1 & 0 & 1 & 1 & 0 & 0.2 & 0.6 \\
  V92 & 12:35:24.528 & +14:28:07.01 & 136.77 & 707.95 & 27.018 & 0.072 & 25.594 & 0.062 & 1 & 0 & 0 & 0 & 1 & 0.5 & 0.5 &  & candidate RSG\\
  V93 & 12:35:30.246 & +14:28:20.44 & 411.91 & 347.92 & 27.21 & 0.063 & 26.889 & 0.148 & 1 & 0 & 0 & 0 & 0 & 0.5 & 0.5 & 83982657\\

\hline

\end{tabular}}
\begin{tablenotes}
\item \textbf{Note.}-- Units of right ascension are hours, minutes, and seconds, and units of declination are degrees, arcminutes, and arcseconds. Numbers 1 or 0 in columns MAD, IQR and inverse von Neumann ratio show whether the star was characterized as a variable in the respective index. Units of X and Y are pixels coordinates. Columns $\Delta_ {V}$ and $\Delta_{I}$ show the amplitude of variability.
\end{tablenotes}
\label{vars45482}
\end{sidewaystable*}

\clearpage


\begin{figure*}[h!tb]
\centering
\includegraphics[trim=0cm 0cm 0.cm 0cm, width=0.60\textwidth]{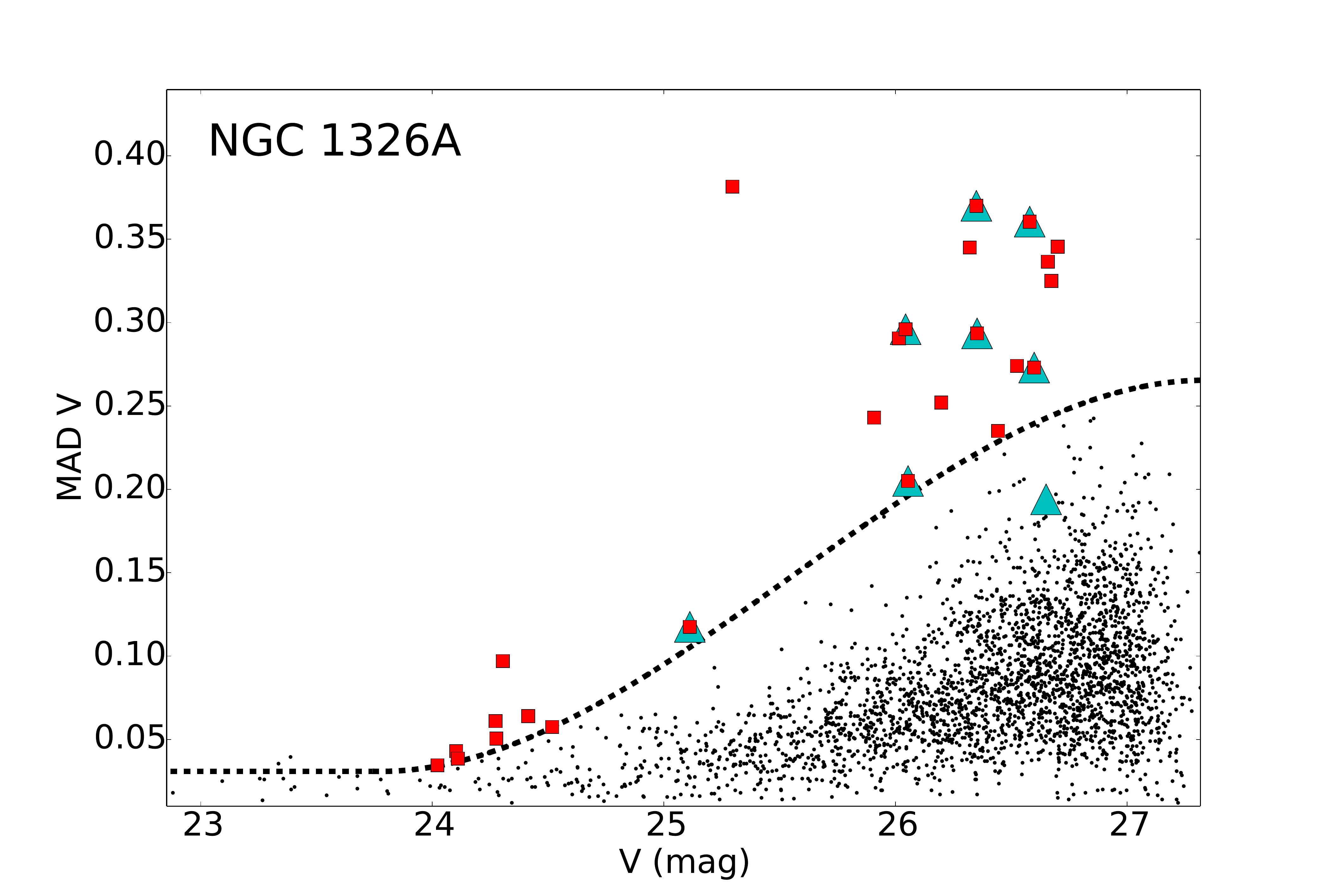}\\
\includegraphics[trim=0cm 0cm 0.cm 0cm, width=0.60\textwidth]{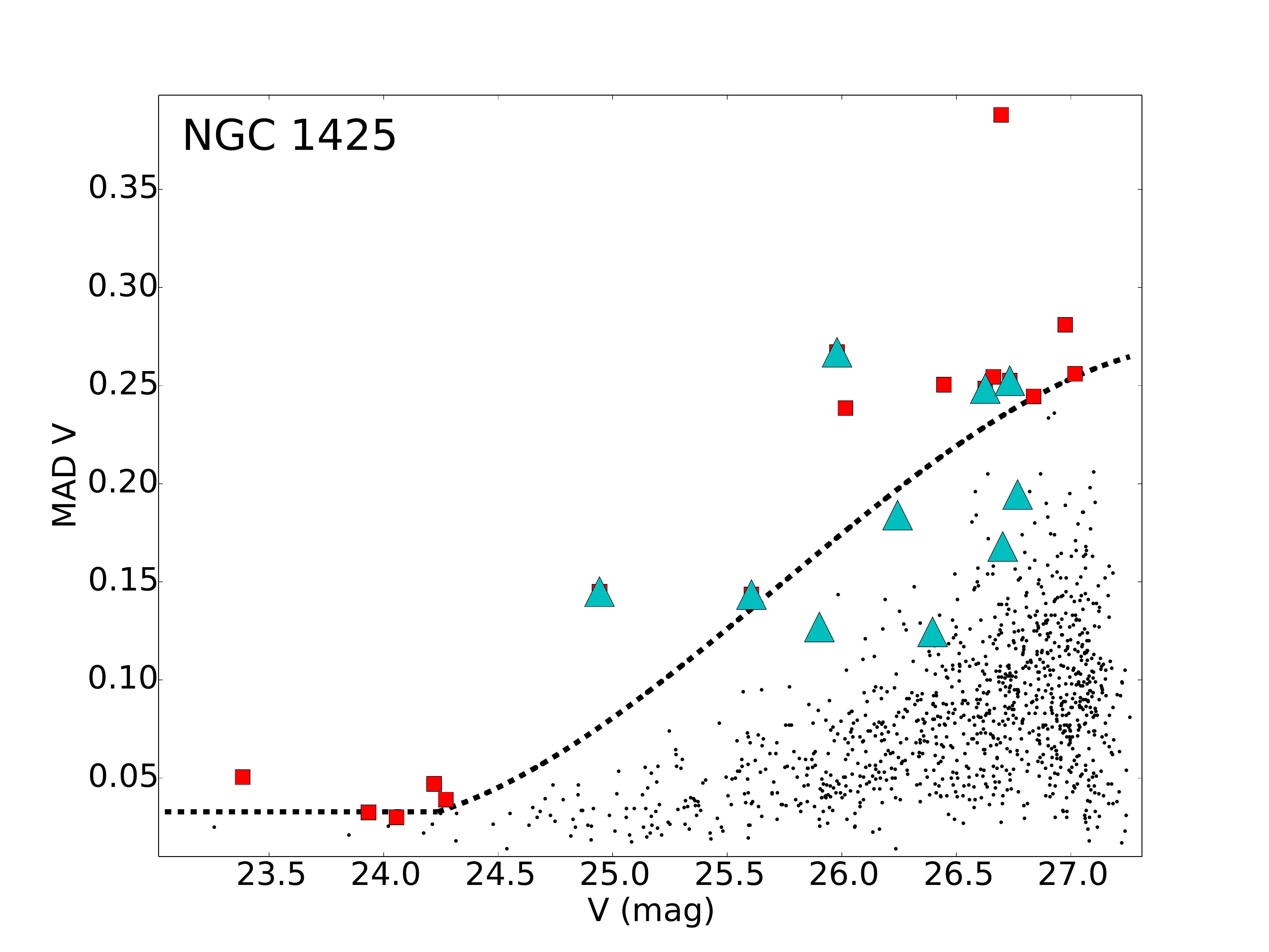}\\
\includegraphics[trim=0cm 0cm 0.cm 0cm, width=0.60\textwidth]{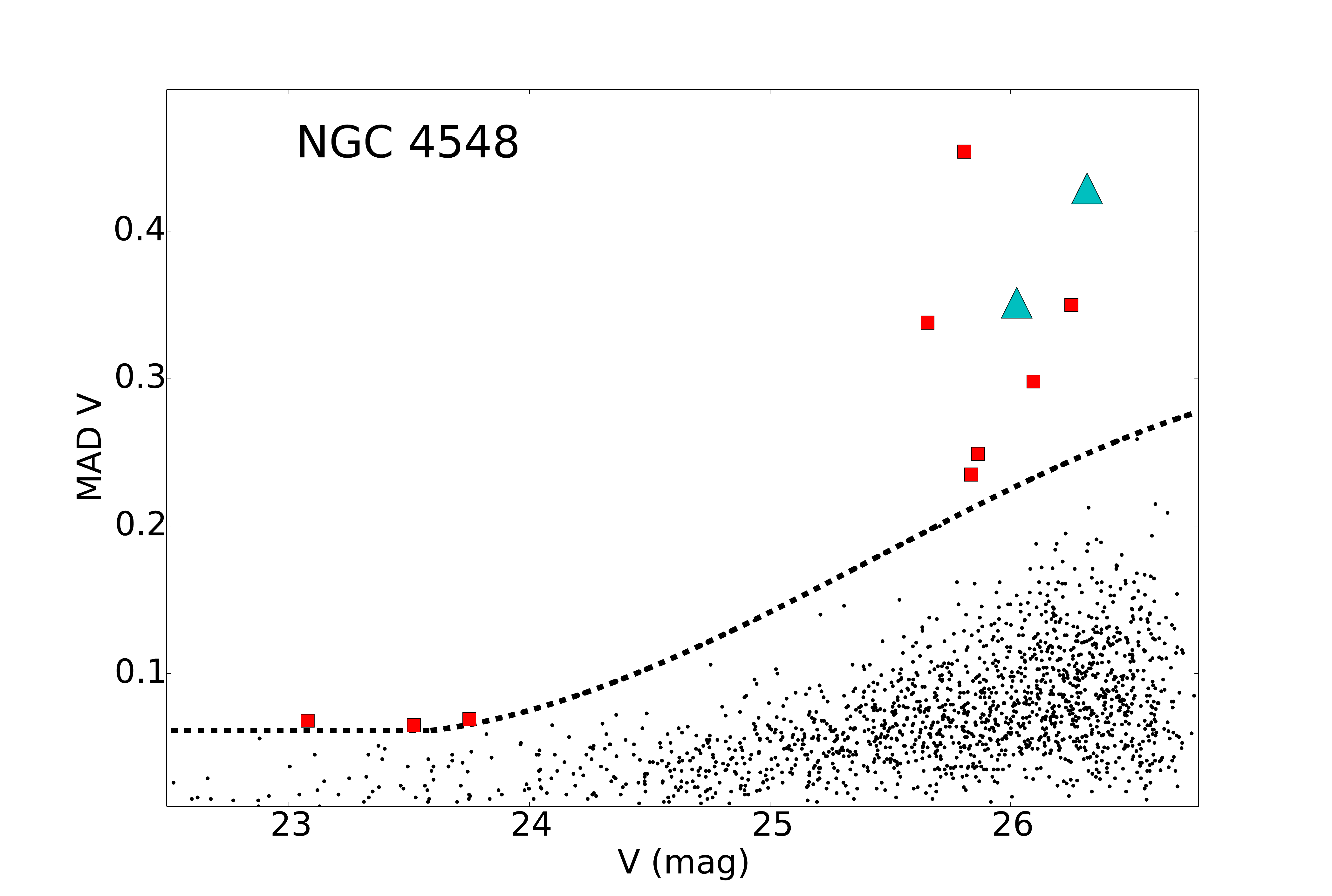} \\
   \caption{Mean absolute deviation versus mean magnitude for the WF3 chip in NGC 1326A, NGC 1425 and NGC 4548. The dashed line shows the stars over the 4$\sigma$ threshold. The variable candidates are shown in red squares and the known variables are shown in blue triangles.}
\label{indexes}
\end{figure*}

\begin{figure*}[h!tb]
\centering
\begin{tabular}{l c c }
\includegraphics[trim=0cm 0cm 0.cm 0cm, width=0.5\textwidth]{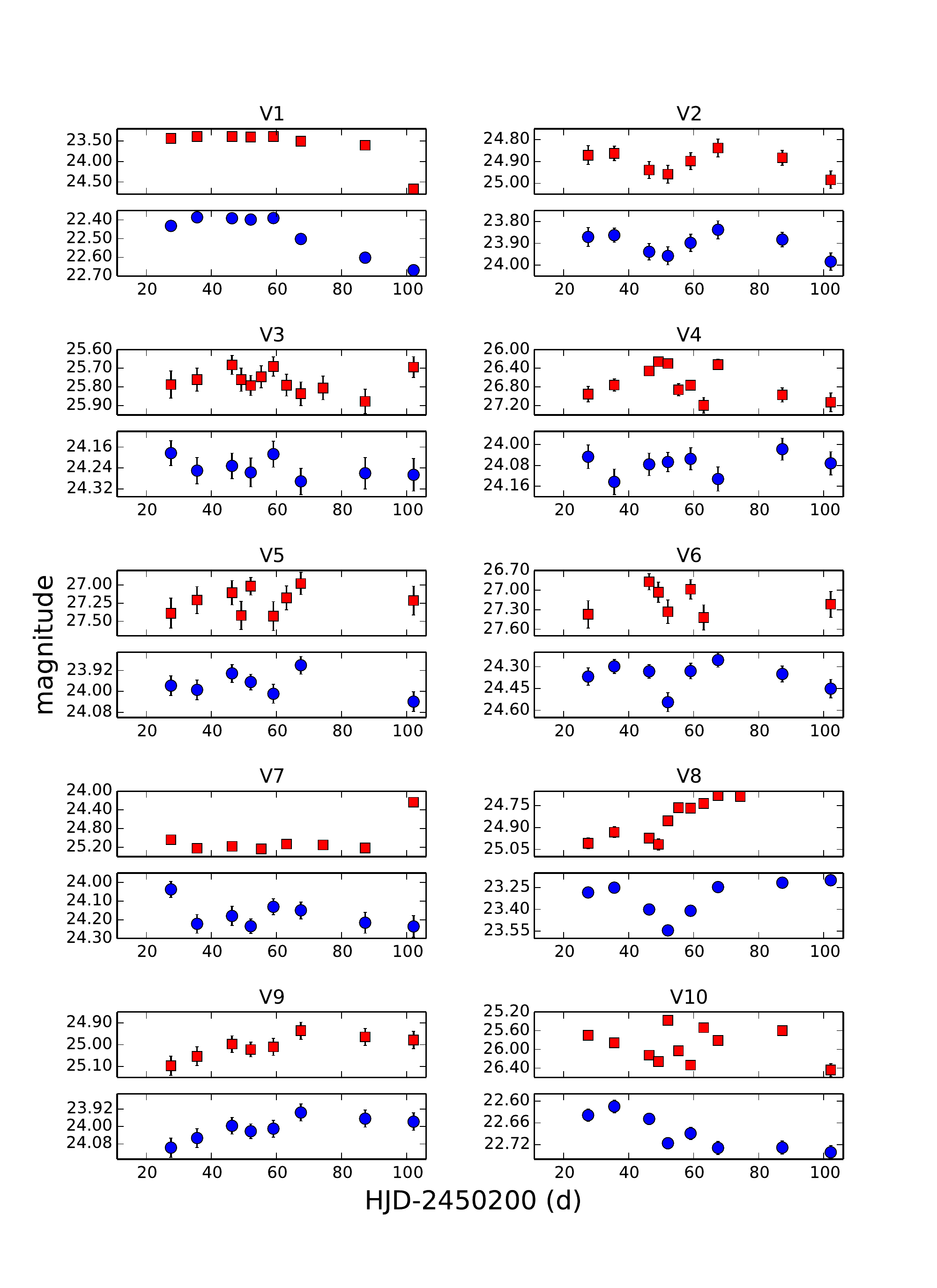} &
\includegraphics[trim=0cm 0cm 0.cm 0cm, width=0.5\textwidth]{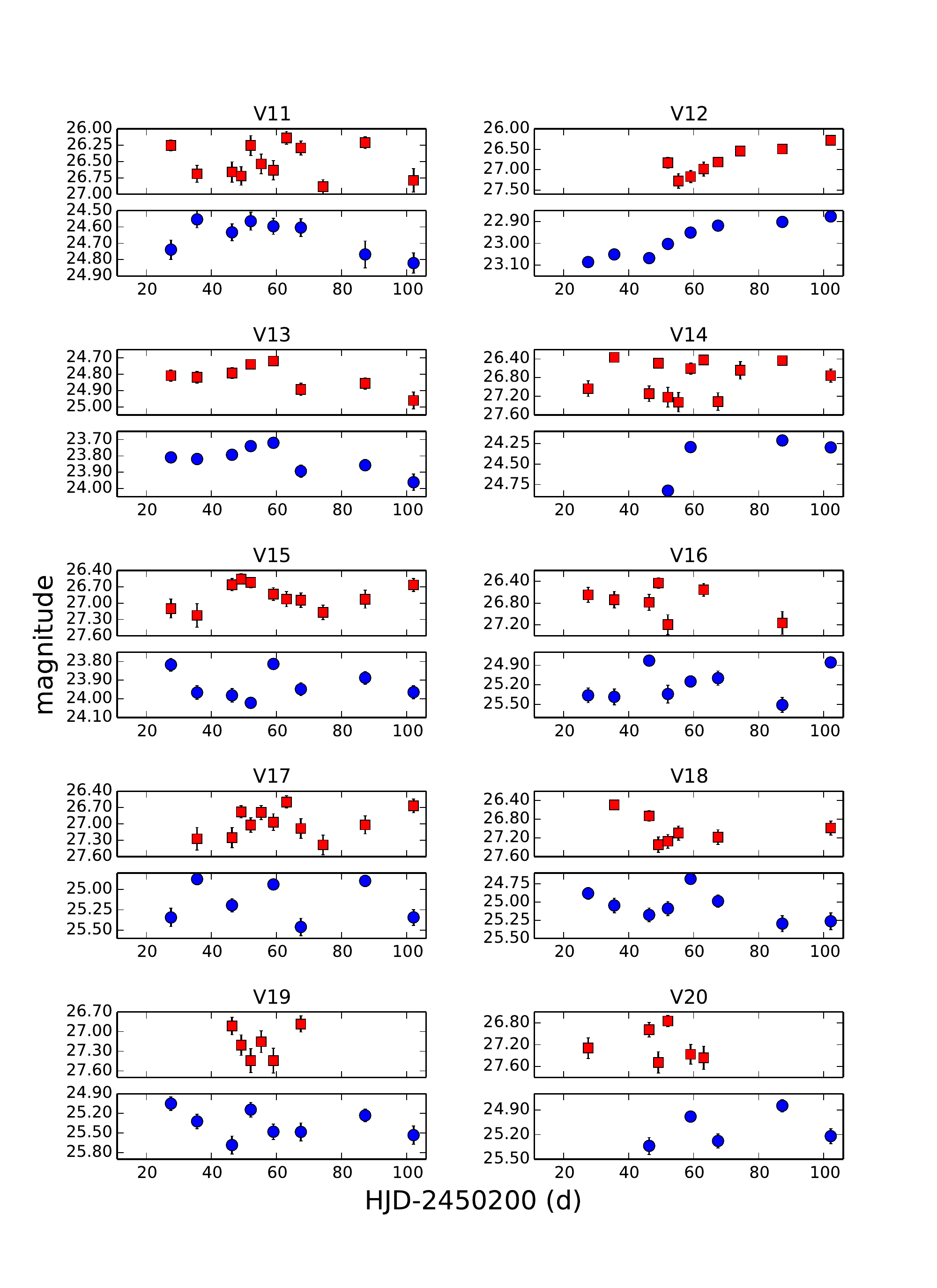}& \\
\includegraphics[trim=0cm 0cm 0.cm 0cm, width=0.5\textwidth]{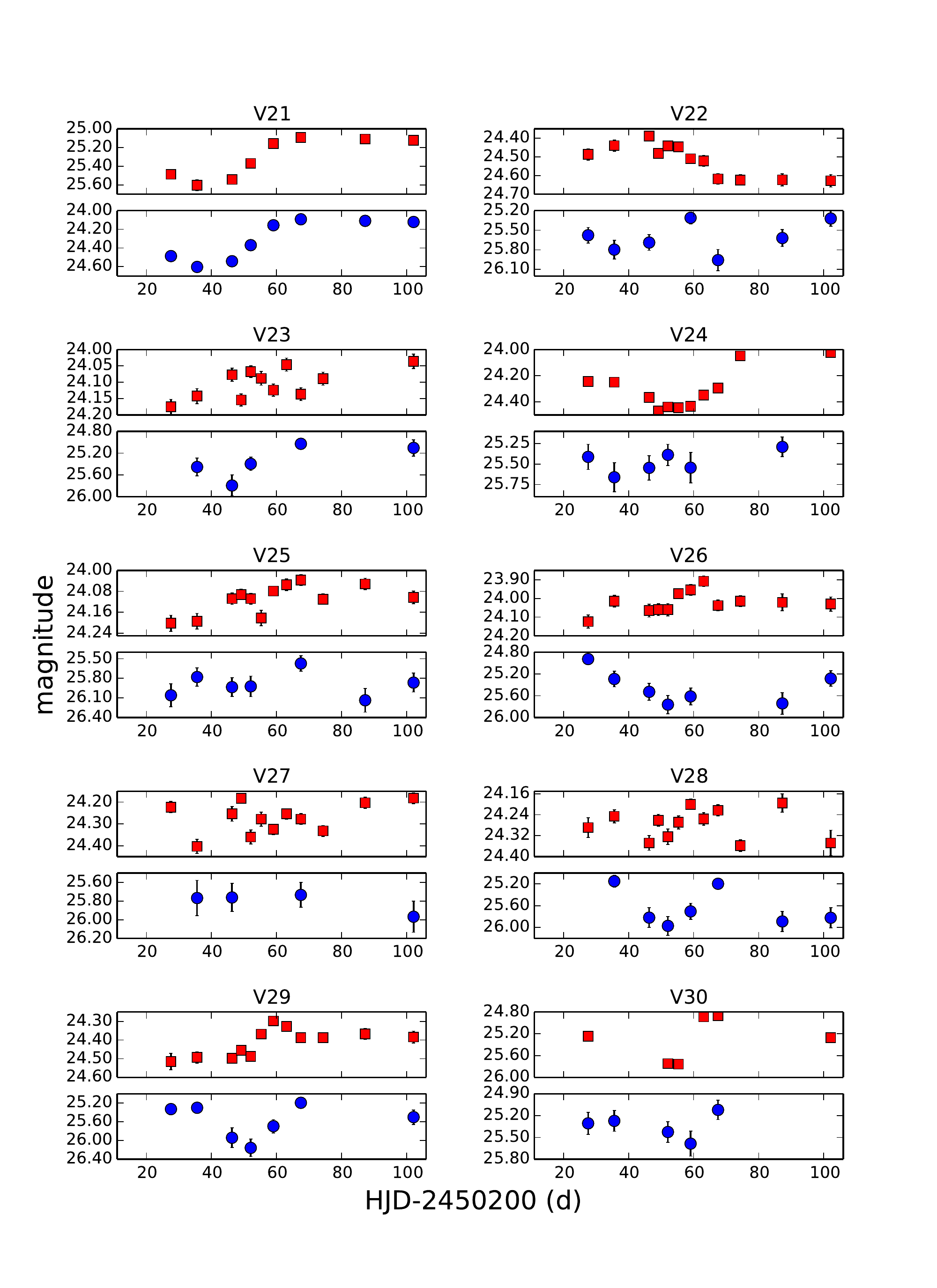}&
\includegraphics[trim=0cm 0cm 0.cm 0cm, width=0.5\textwidth]{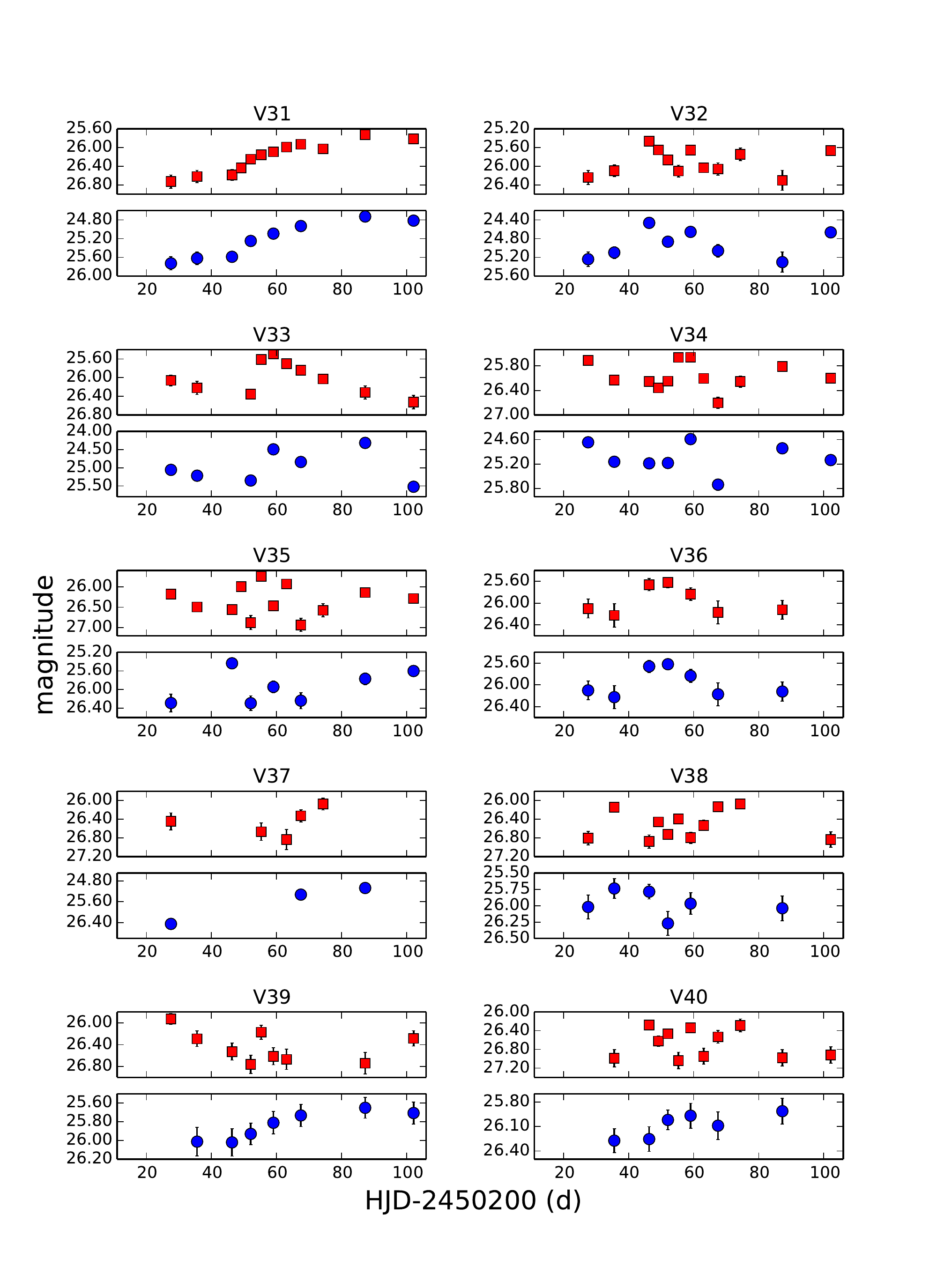}
\setcounter{figure}{8}
\end{tabular}
\caption{Light curves of the candidate variables in NGC 1326A. The V filter is shown in filled blue circles and the I filter in red squares.}
\label{lcs1326}
\end{figure*}

\begin{figure*}[h!tb]
\centering
\begin{tabular}{l c c c c c c c}
\hspace{0cm}
\includegraphics[trim=0cm 0cm 0.cm 0cm, width=0.5 \textwidth]{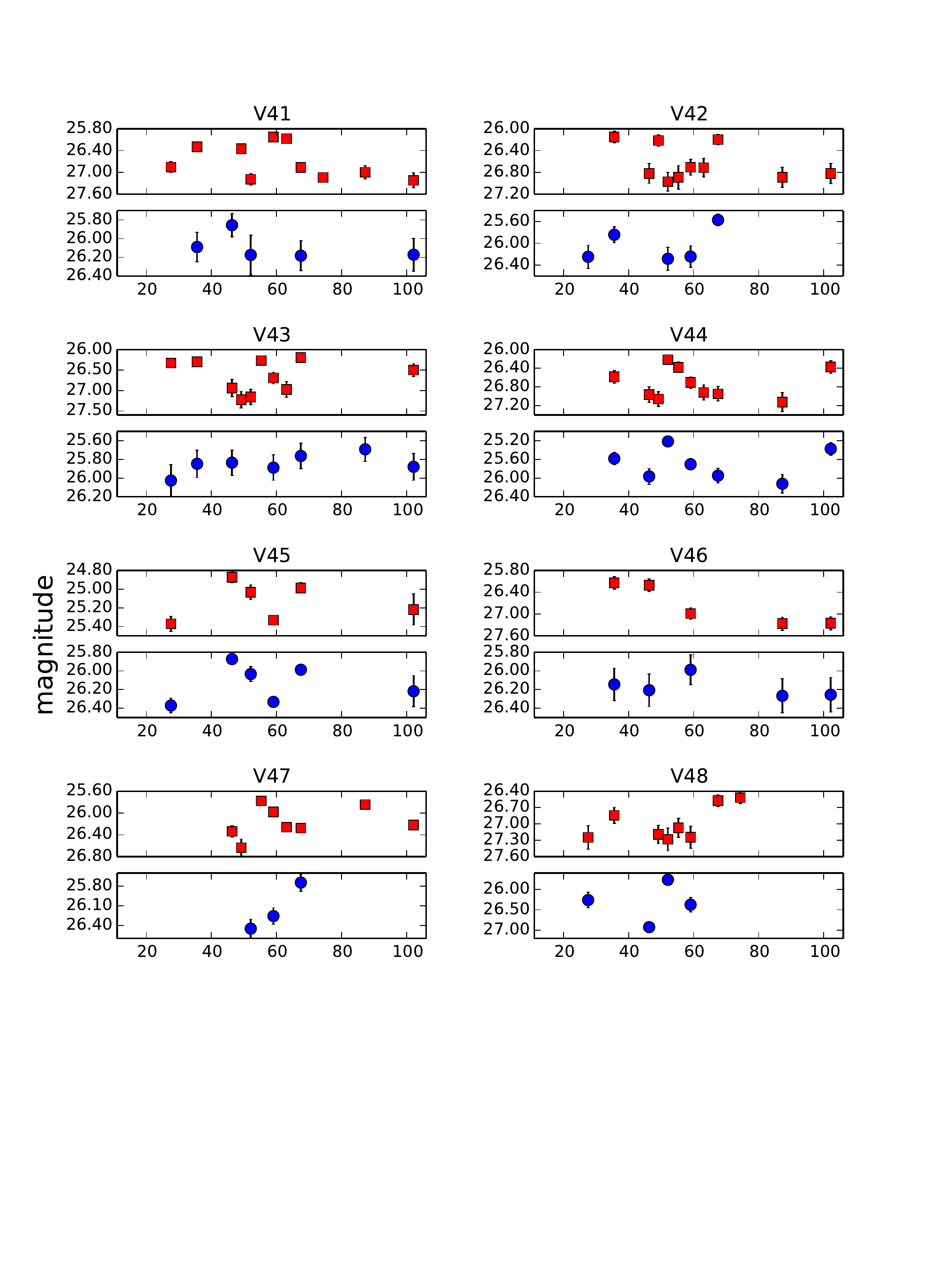}& 
\includegraphics[trim=0cm 0cm 0.cm 0cm, width=0.5\textwidth]{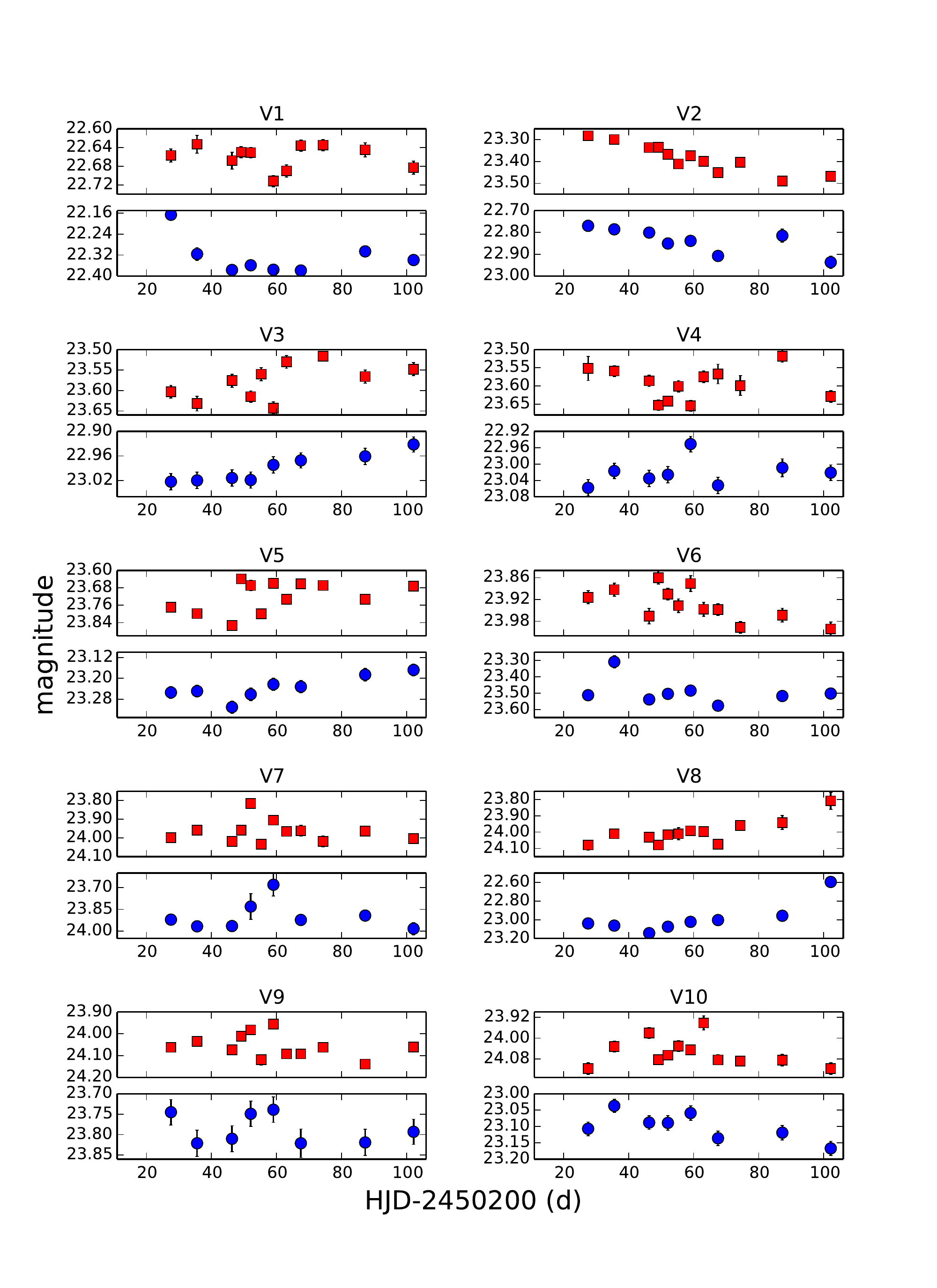}&\\
\includegraphics[trim=0cm 0cm 0.cm 0cm, width=0.5\textwidth]{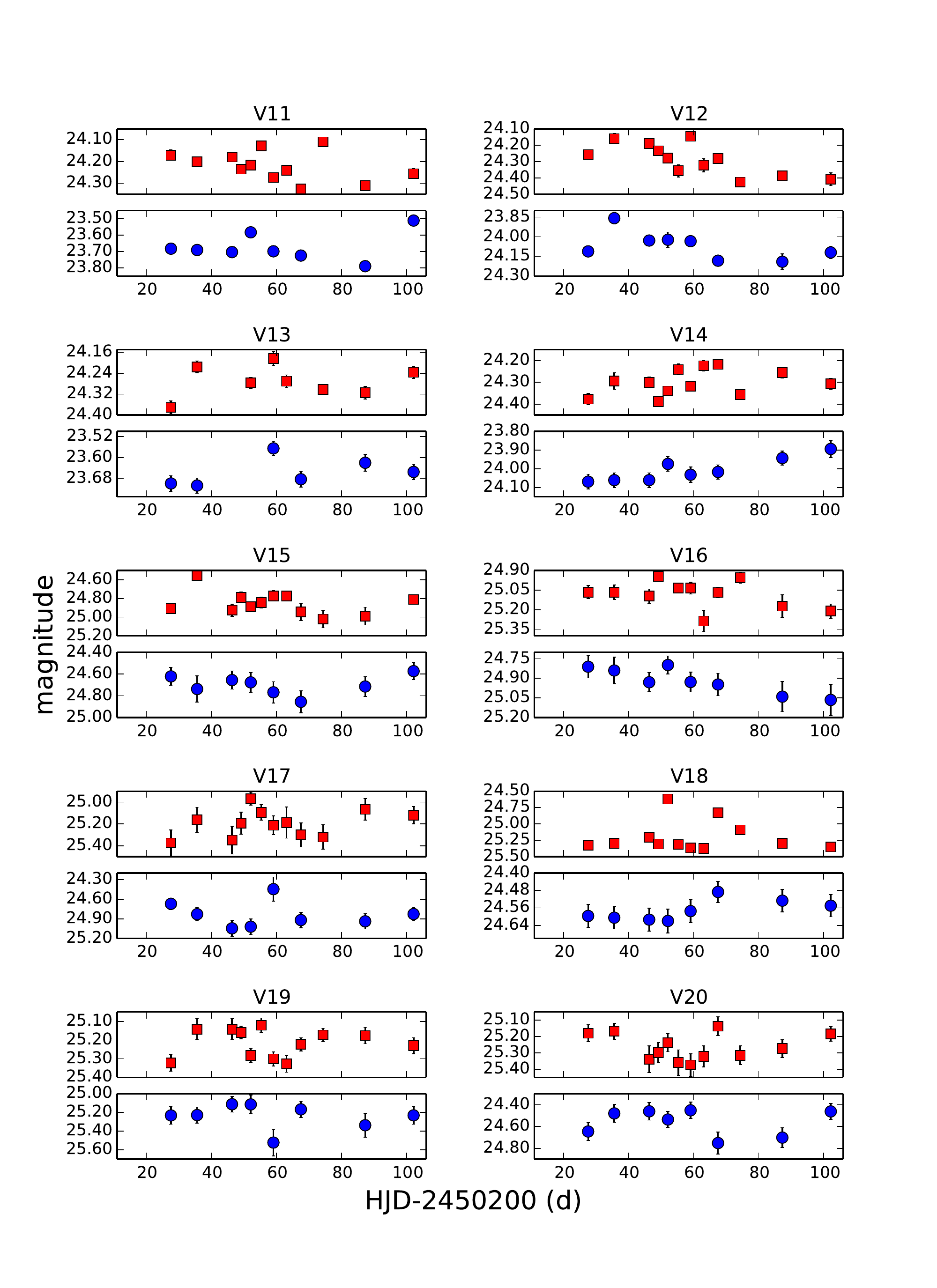}&
\includegraphics[trim=0cm 0cm 0.cm 0cm, width=0.5\textwidth]{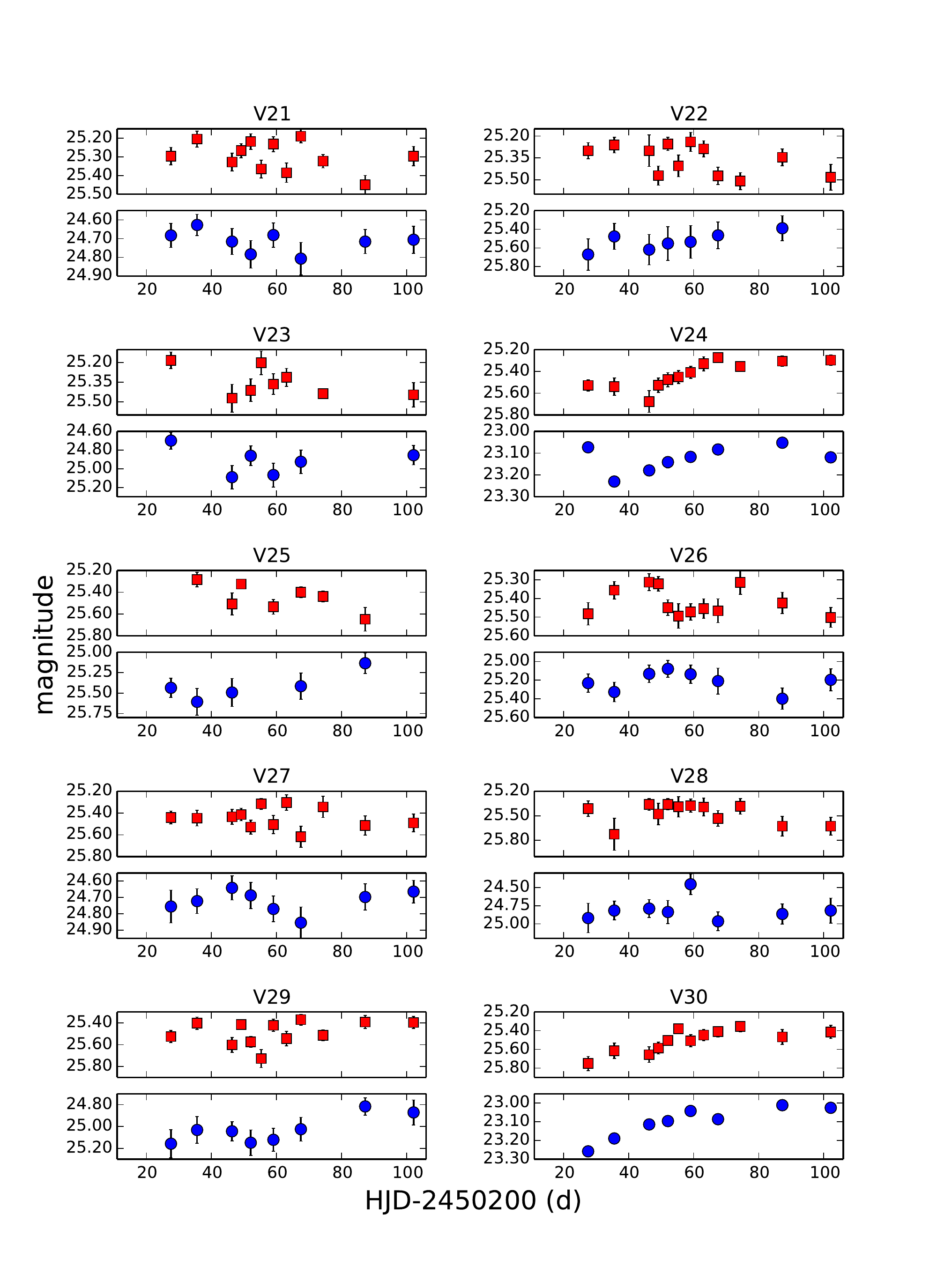}&
\setcounter{figure}{9}
\end{tabular}

\caption{Same as Fig.~\ref{lcs1326} but for NGC 1425.}
\label{lcs14251}
\end{figure*}

\newpage

\begin{figure*}[h!tb]
\centering
\begin{tabular}{l c c c c c c c}
\hspace{0cm}
\includegraphics[trim=0cm 0cm 0.cm 0cm, width=0.5 \textwidth]{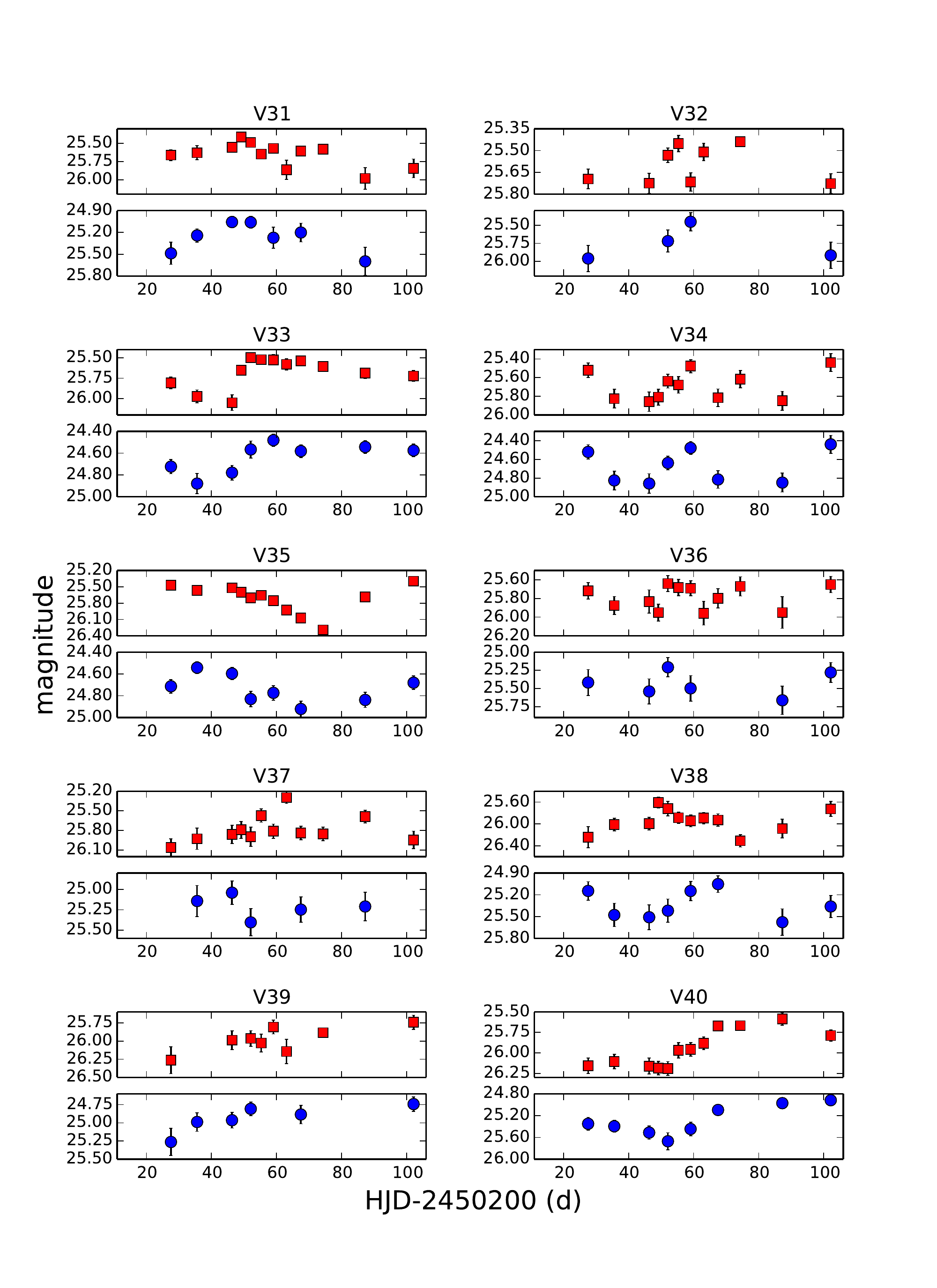}& 
\includegraphics[trim=0cm 0cm 0.cm 0cm, width=0.5\textwidth]{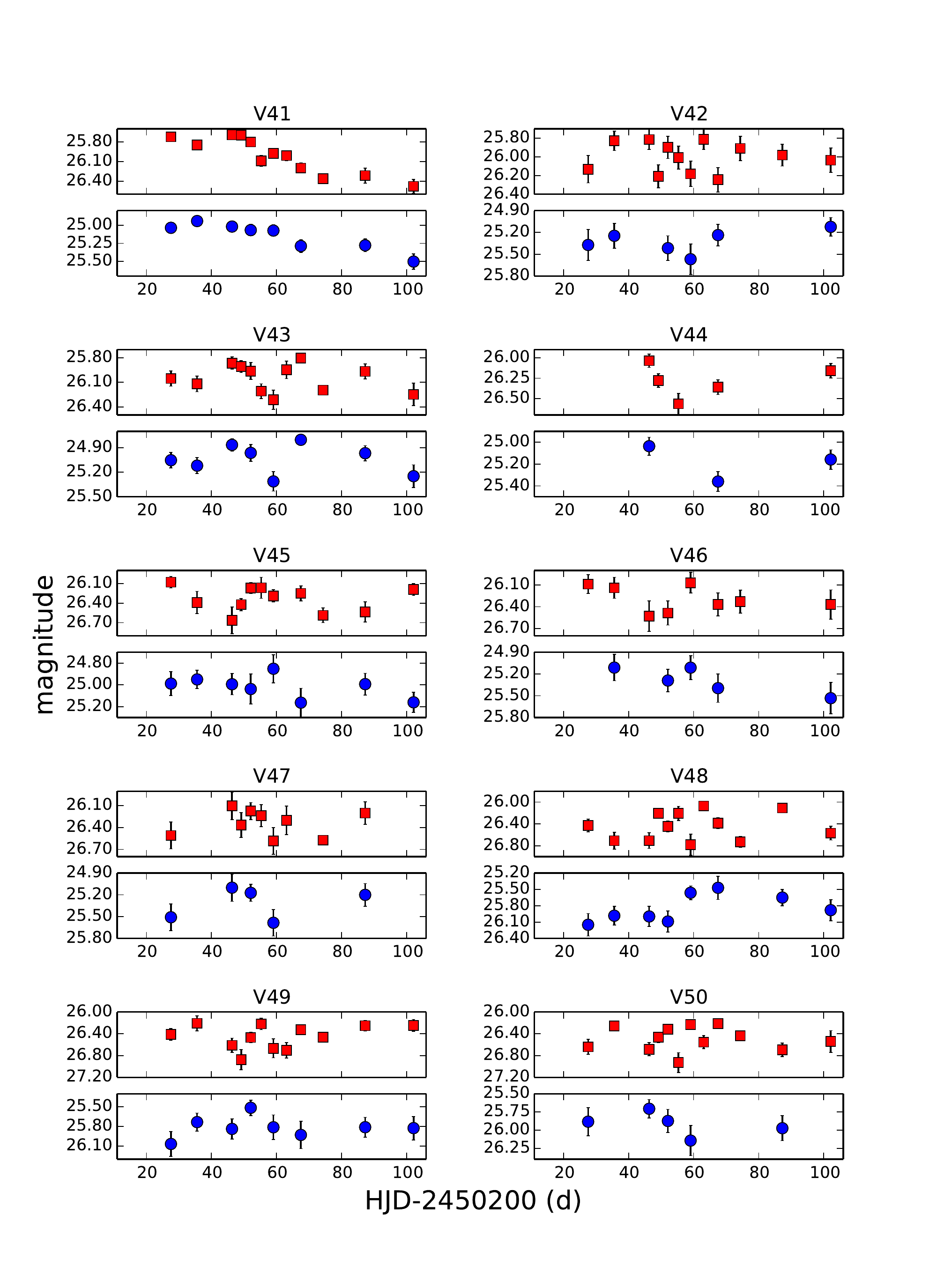}&\\
\includegraphics[trim=0cm 0cm 0.cm 0cm, width=0.5\textwidth]{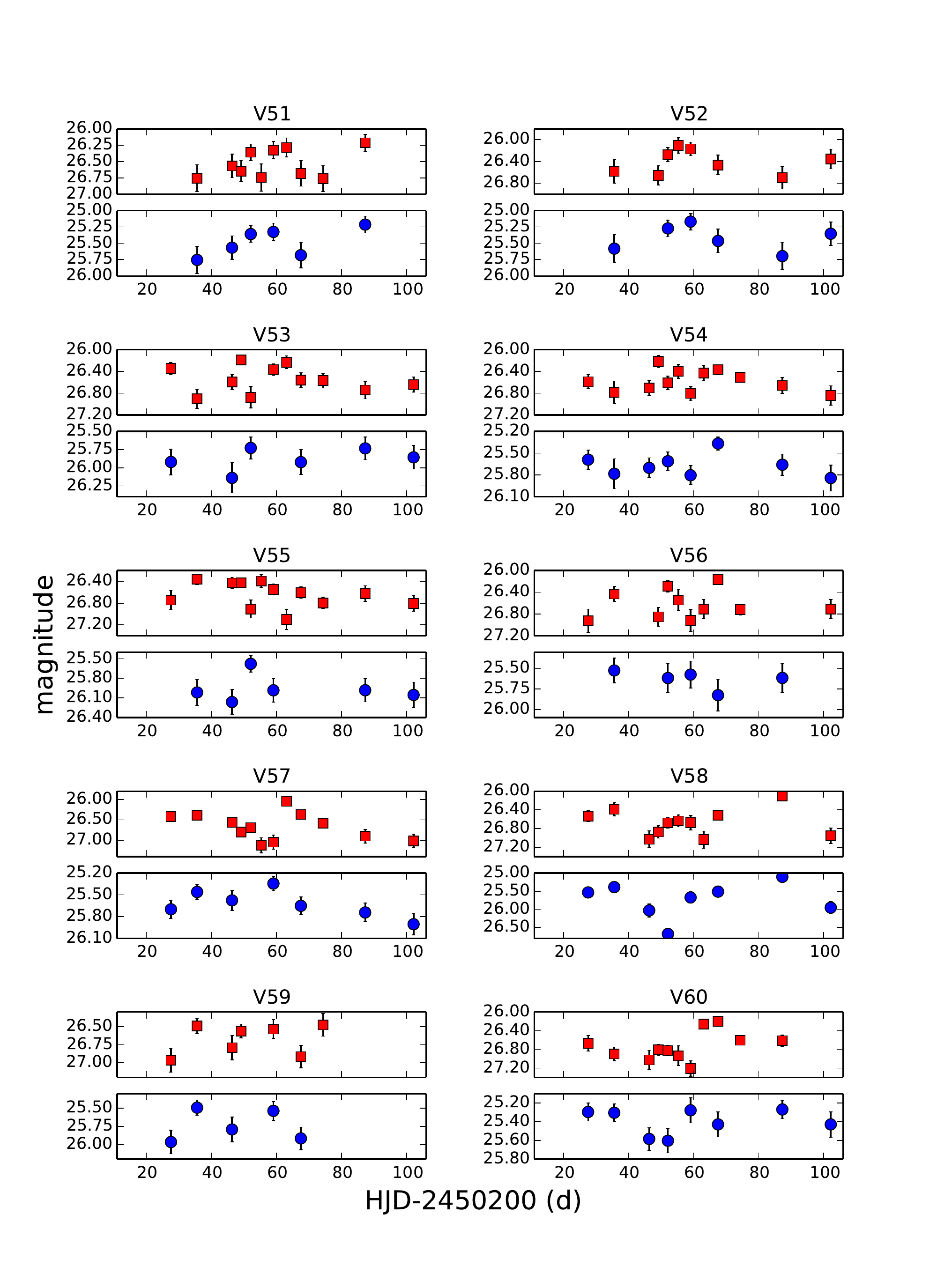}&
\includegraphics[trim=0cm 0cm 0.cm 0cm, width=0.5\textwidth]{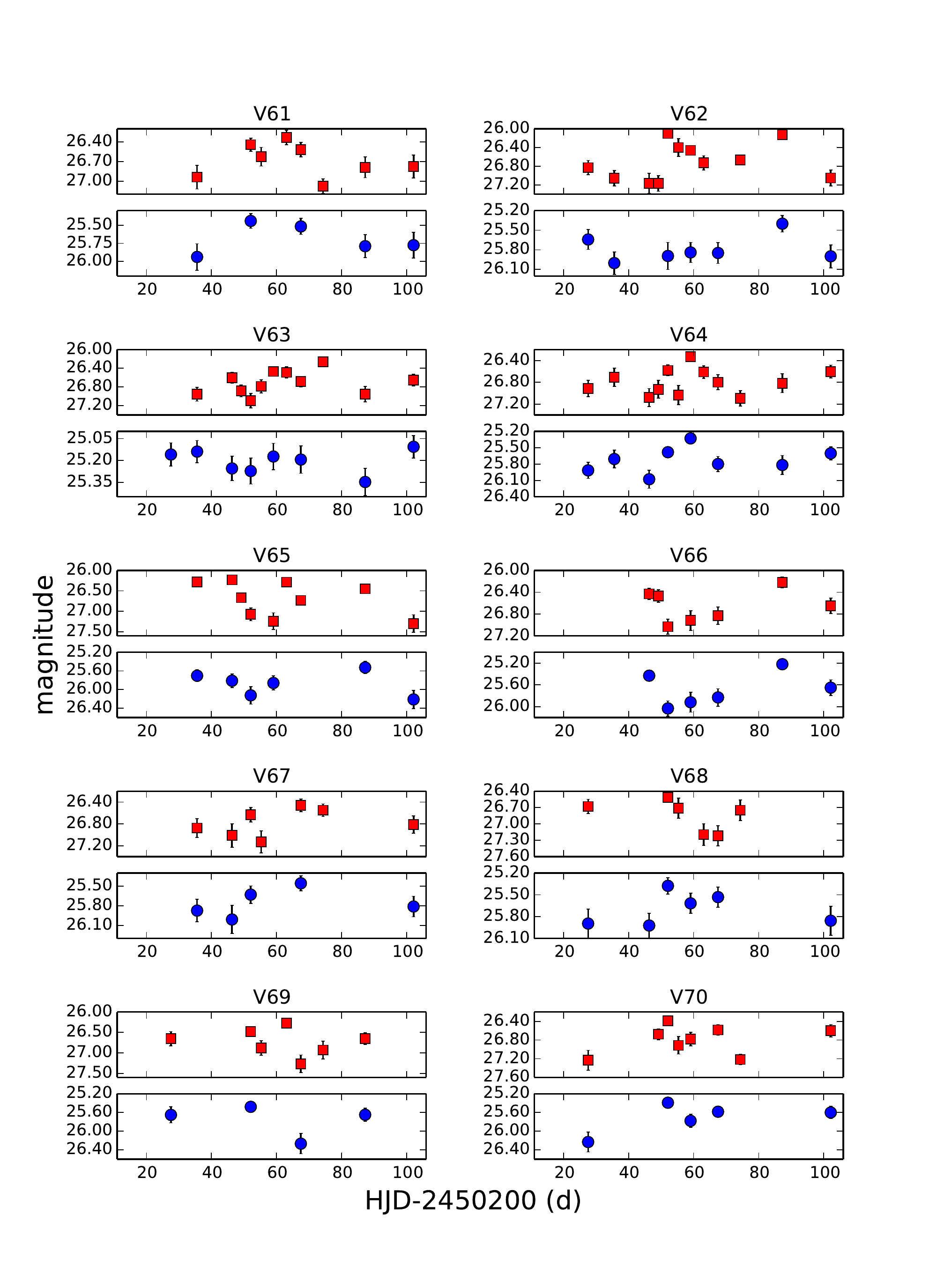}&
\setcounter{figure}{9}
\end{tabular}

\caption{continued}
\label{lcs14252}
\end{figure*}

\newpage

\begin{figure*}[h!tb]
\centering

\begin{tabular}{l c c c c c c c}
\hspace{0cm}
\includegraphics[trim=0cm 0cm 0.cm 0cm, width=0.5\textwidth]{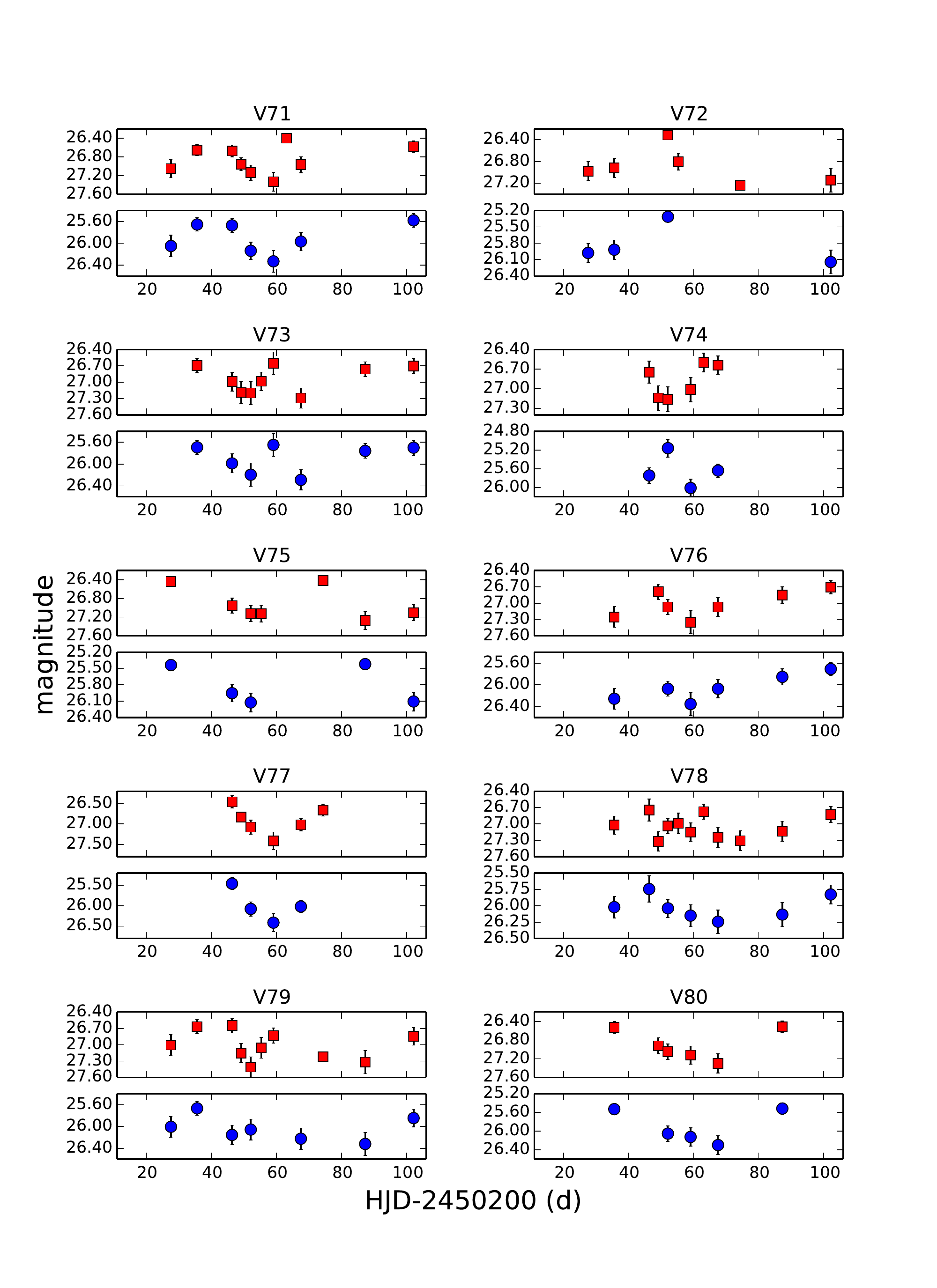}&
\includegraphics[trim=0cm 0cm 0.cm 0cm, width=0.5\textwidth]{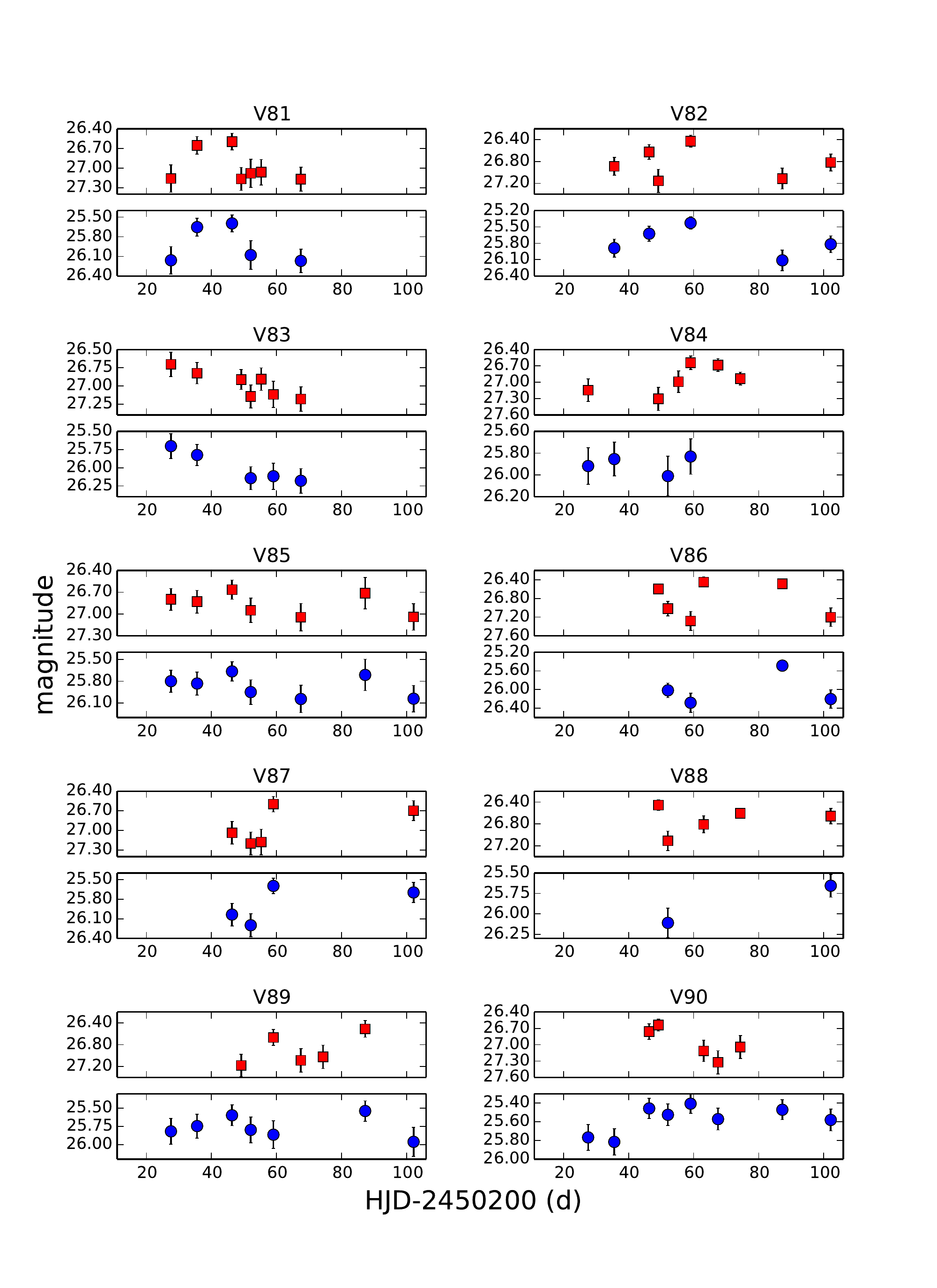}& \\
\includegraphics[trim=0cm 0cm 0.cm 0cm, width=0.5\textwidth]{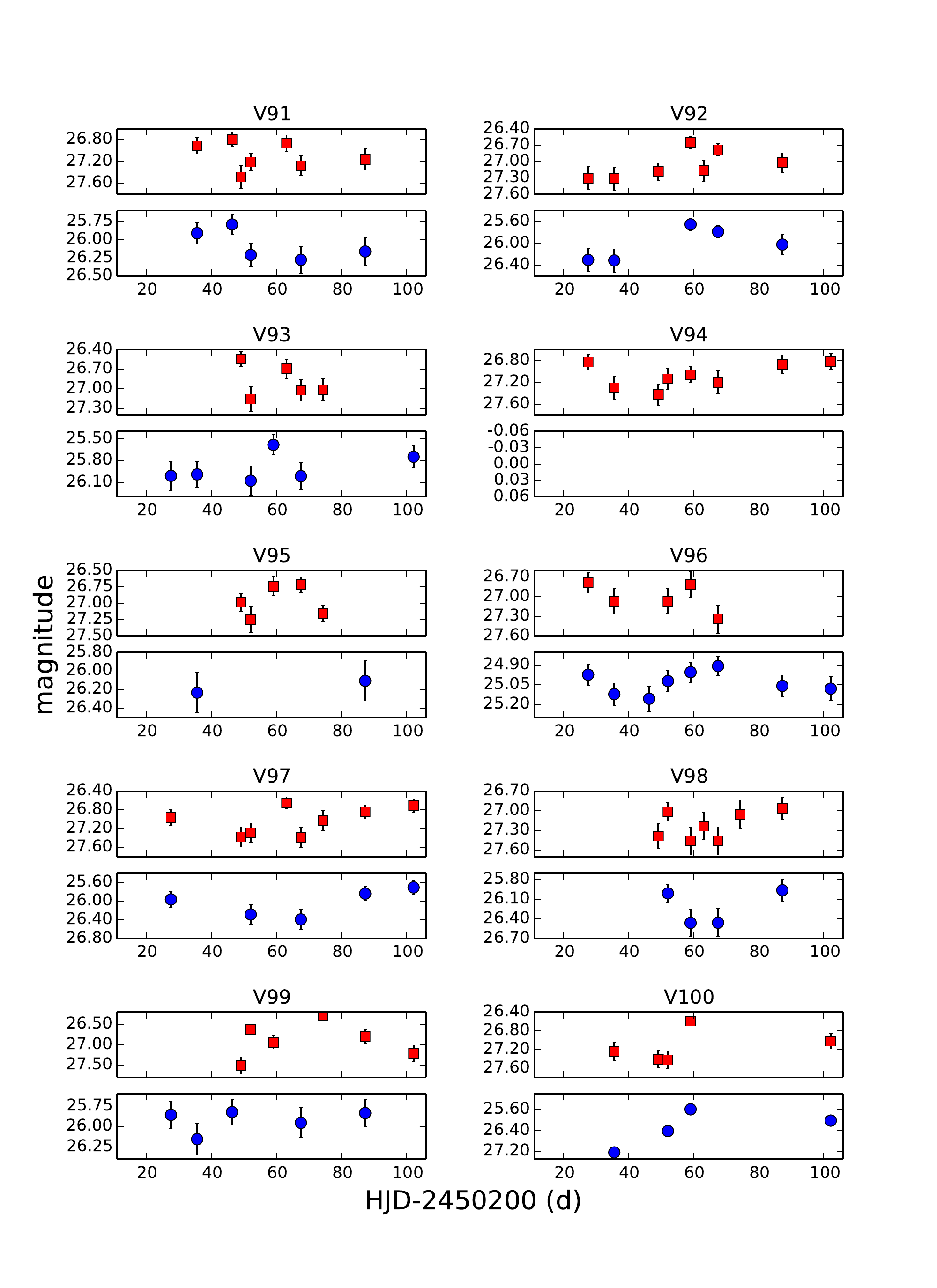}&
\includegraphics[trim=0cm 0cm 0.cm 0cm, width=0.5\textwidth]{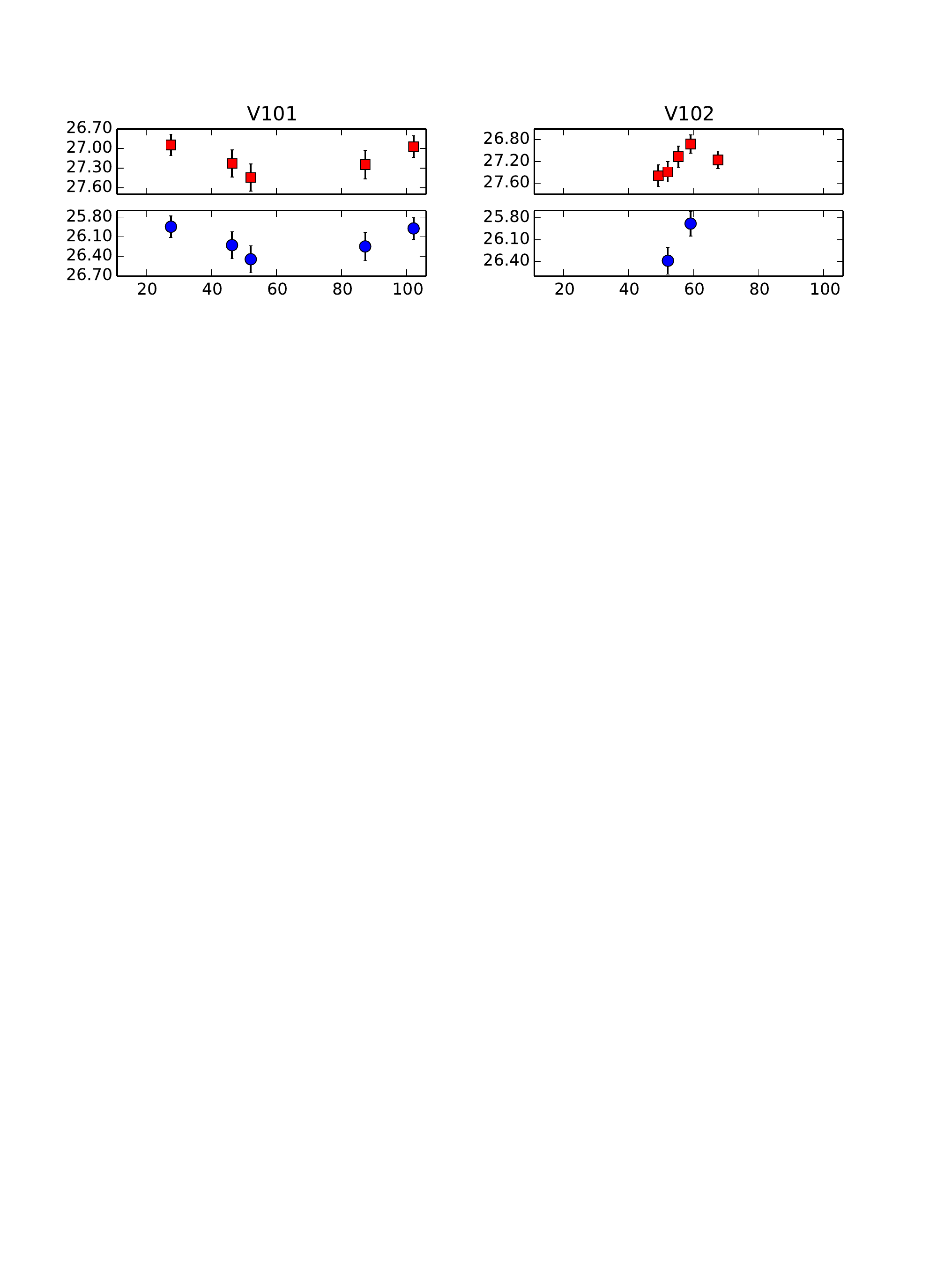}
\setcounter{figure}{9}
\end{tabular}
\caption{continued}
\label{lcs14253}
\end{figure*}

\clearpage
\begin{figure*}[h!tb]
\centering

\begin{tabular}{l c c c c c c c}
\hspace{0cm}
\includegraphics[trim=0cm 0cm 0.cm 0cm, width=0.5\textwidth]{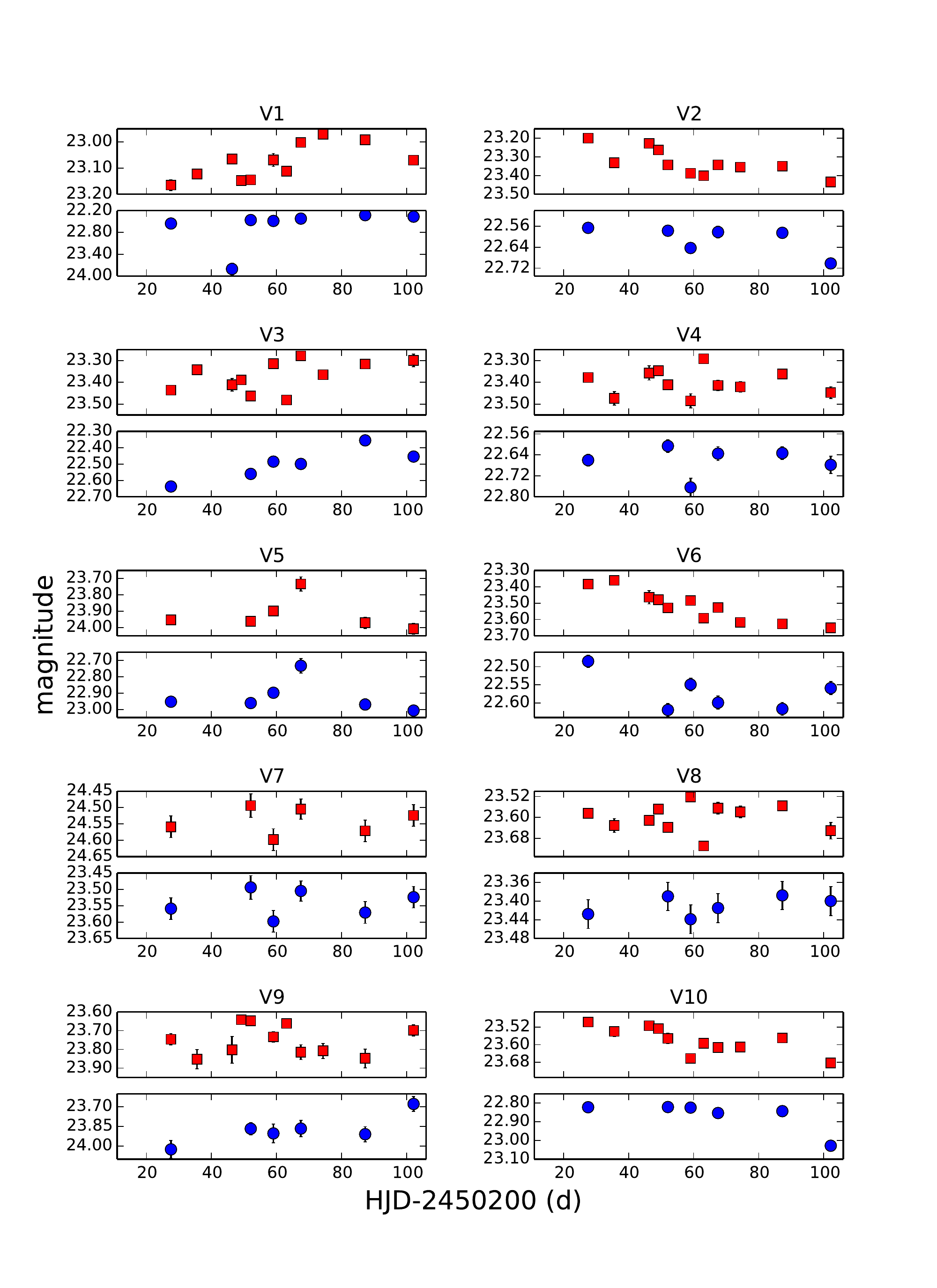} &
\includegraphics[trim=0cm 0cm 0.cm 0cm, width=0.5\textwidth]{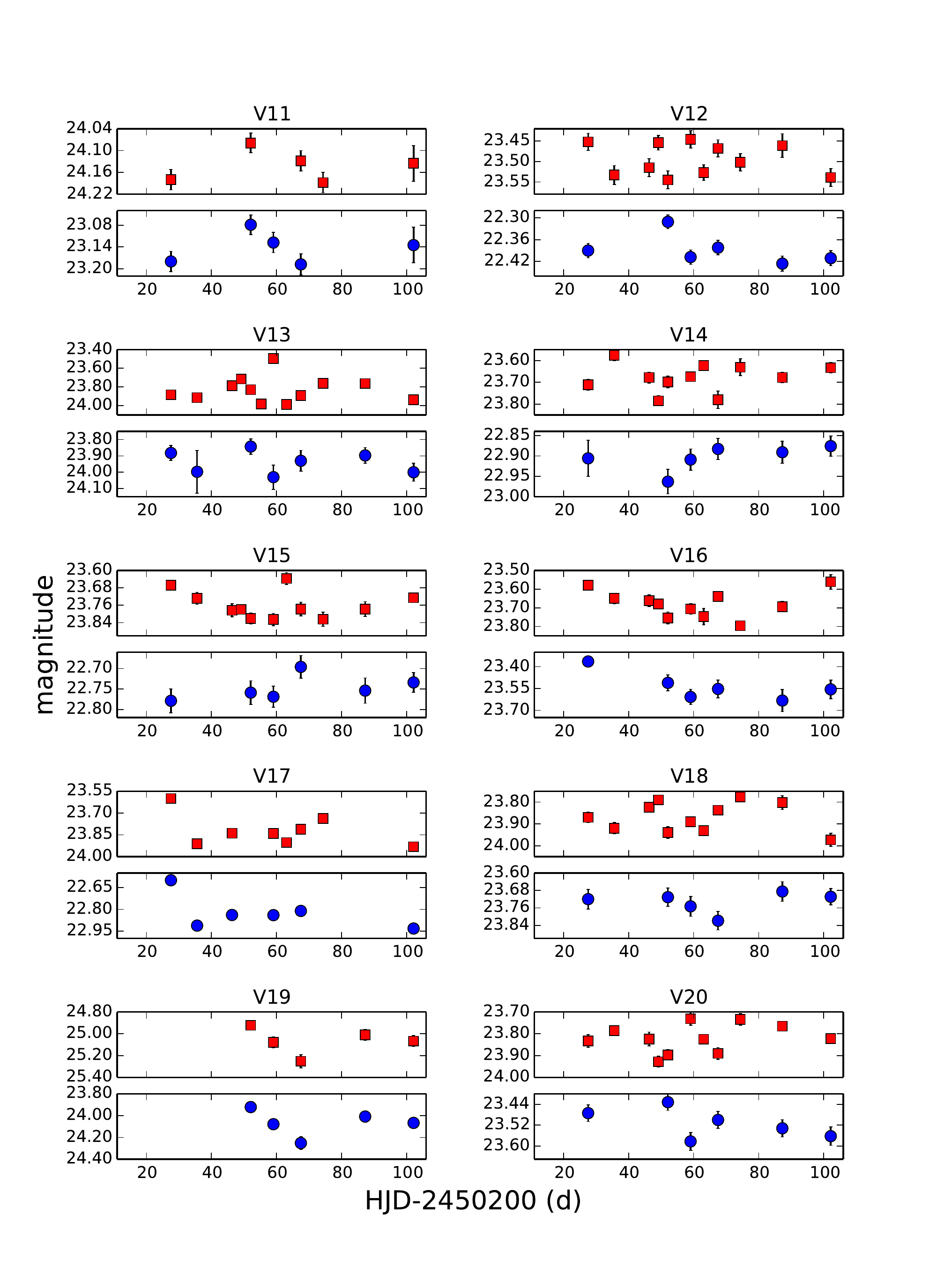}& \\
\includegraphics[trim=0cm 0cm 0.cm 0cm, width=0.5\textwidth]{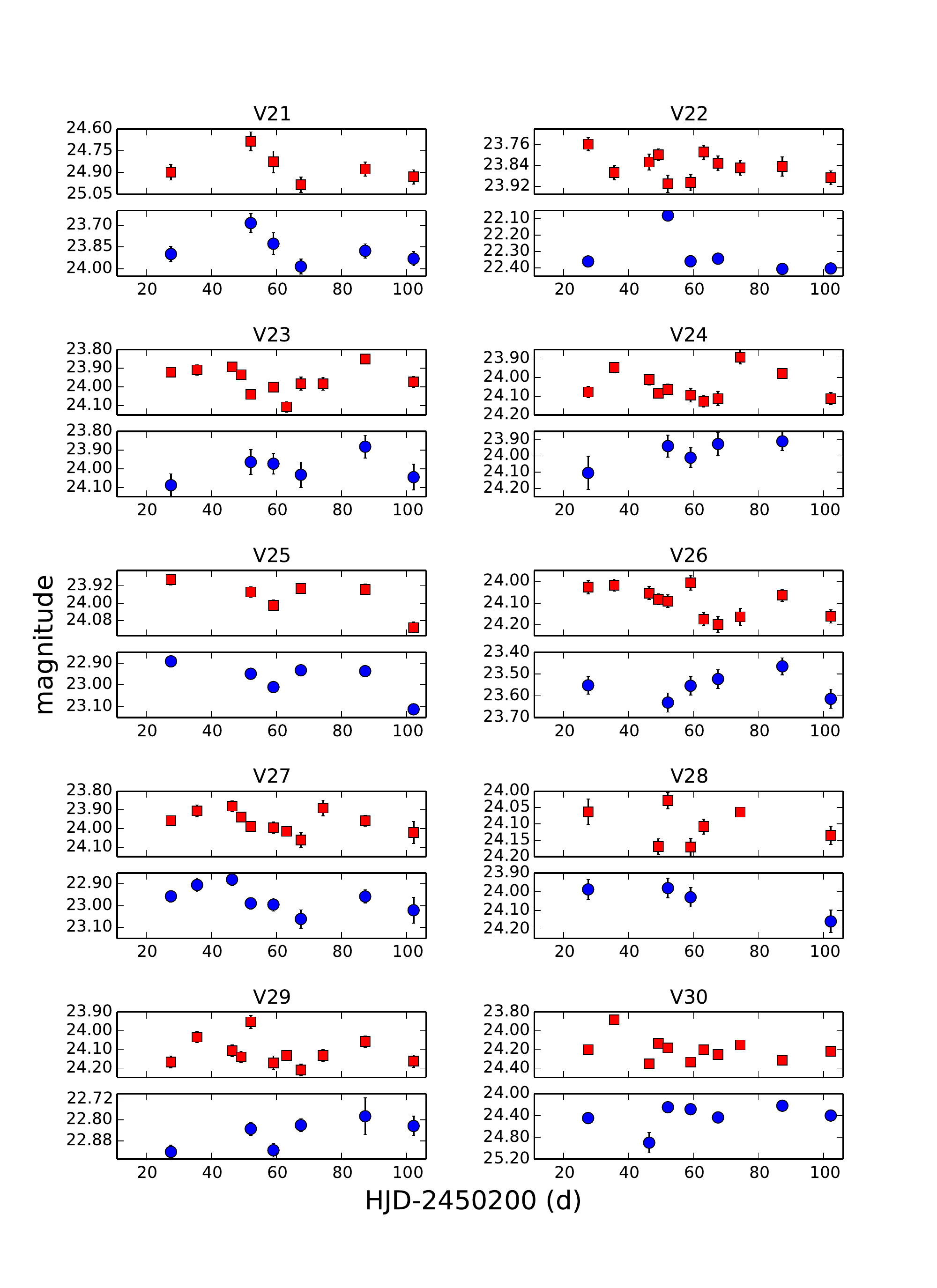}&
\includegraphics[trim=0cm 0cm 0.cm 0cm, width=0.5\textwidth]{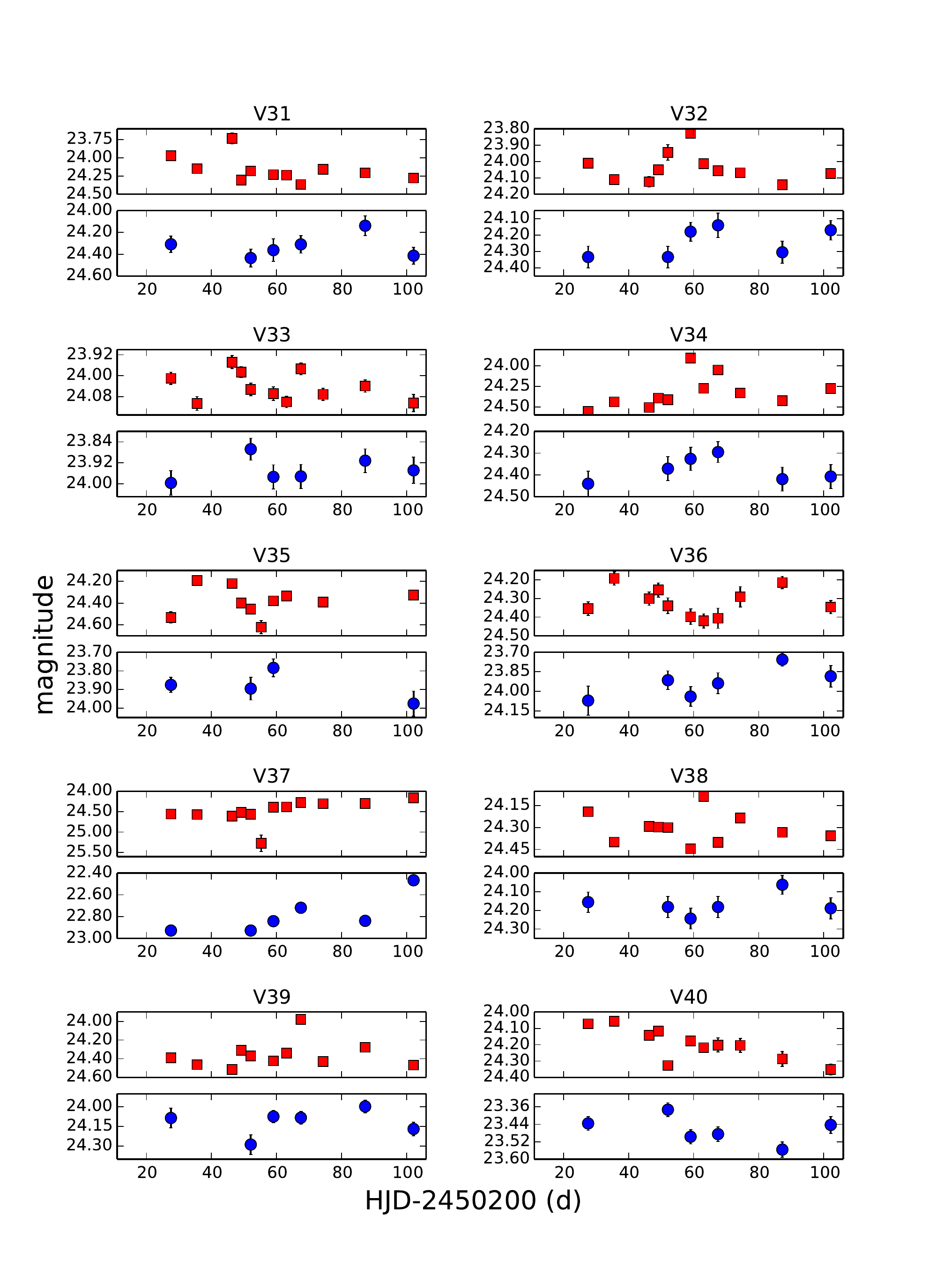}
\setcounter{figure}{10}
\end{tabular}

\caption{Same as Fig.~\ref{lcs1326} but for NGC 4548}
\label{lcs45481}
\end{figure*}

\clearpage
\begin{figure*}[h!tb]
\centering

\begin{tabular}{l c c c c c c c}
\hspace{0cm}
\includegraphics[trim=0cm 0cm 0.cm 0cm, width=0.5\textwidth]{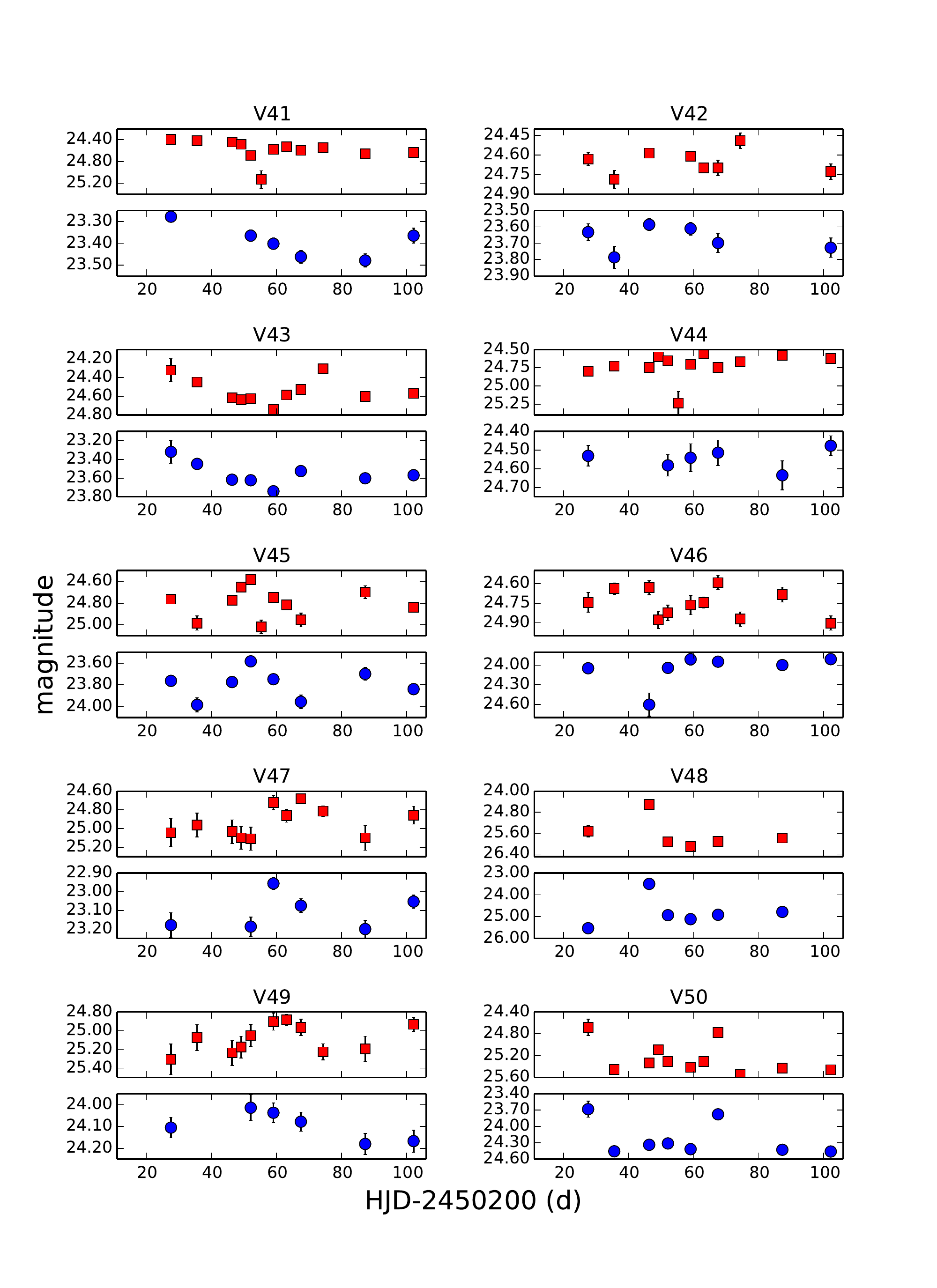} &
\includegraphics[trim=0cm 0cm 0.cm 0cm, width=0.5\textwidth]{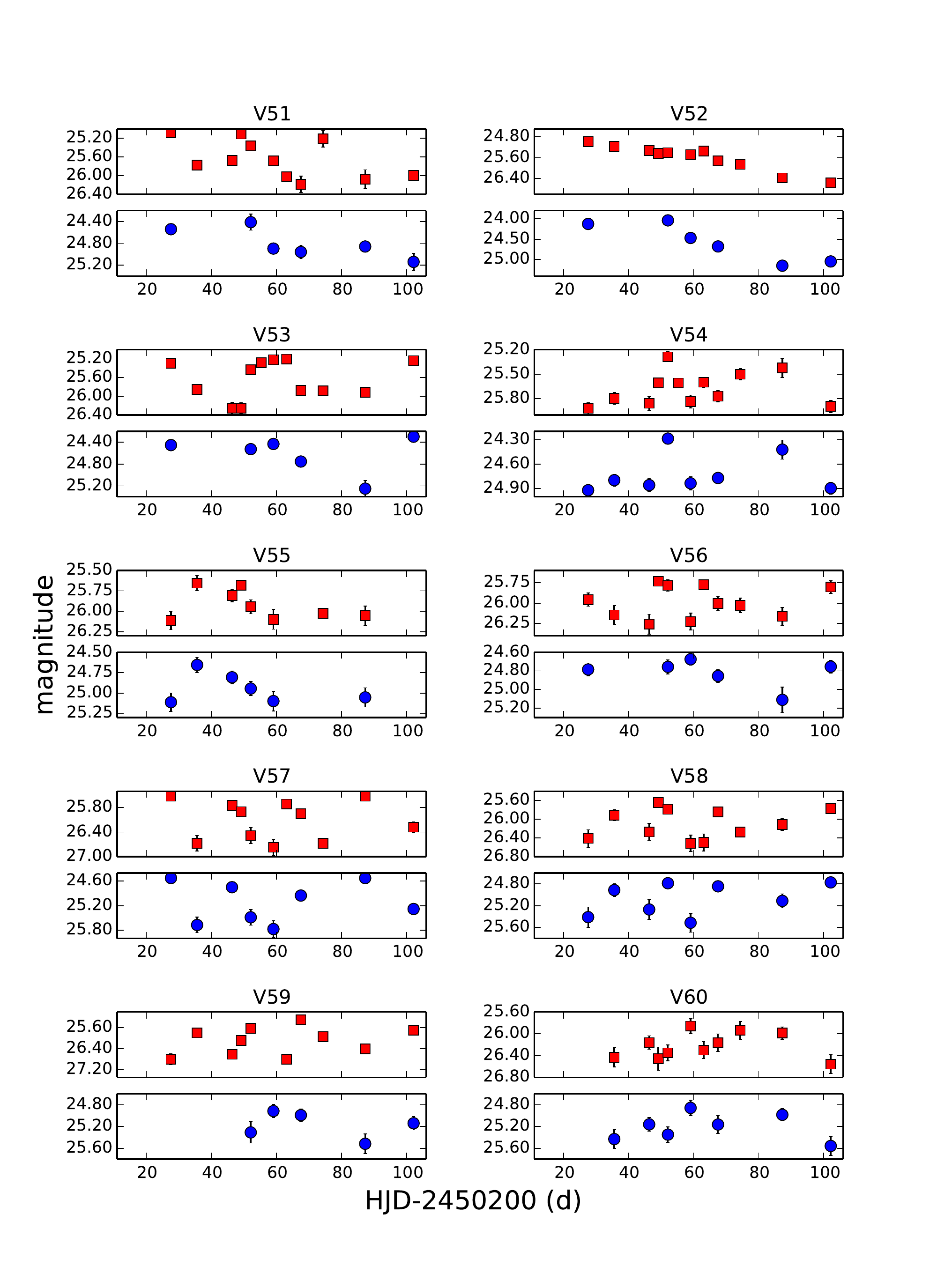}& \\
\includegraphics[trim=0cm 0cm 0.cm 0cm, width=0.5\textwidth]{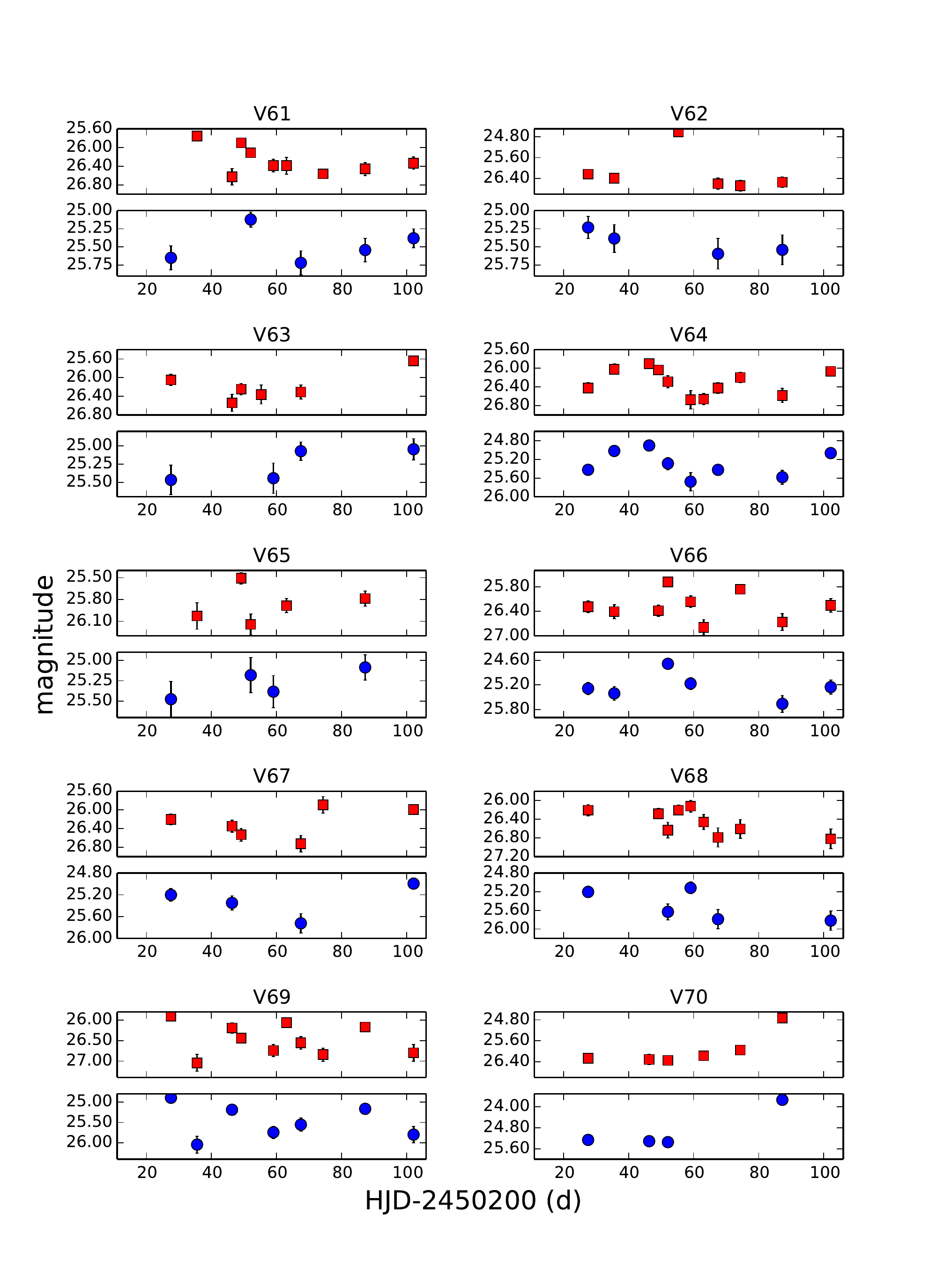}&
\includegraphics[trim=0cm 0cm 0.cm 0cm, width=0.5\textwidth]{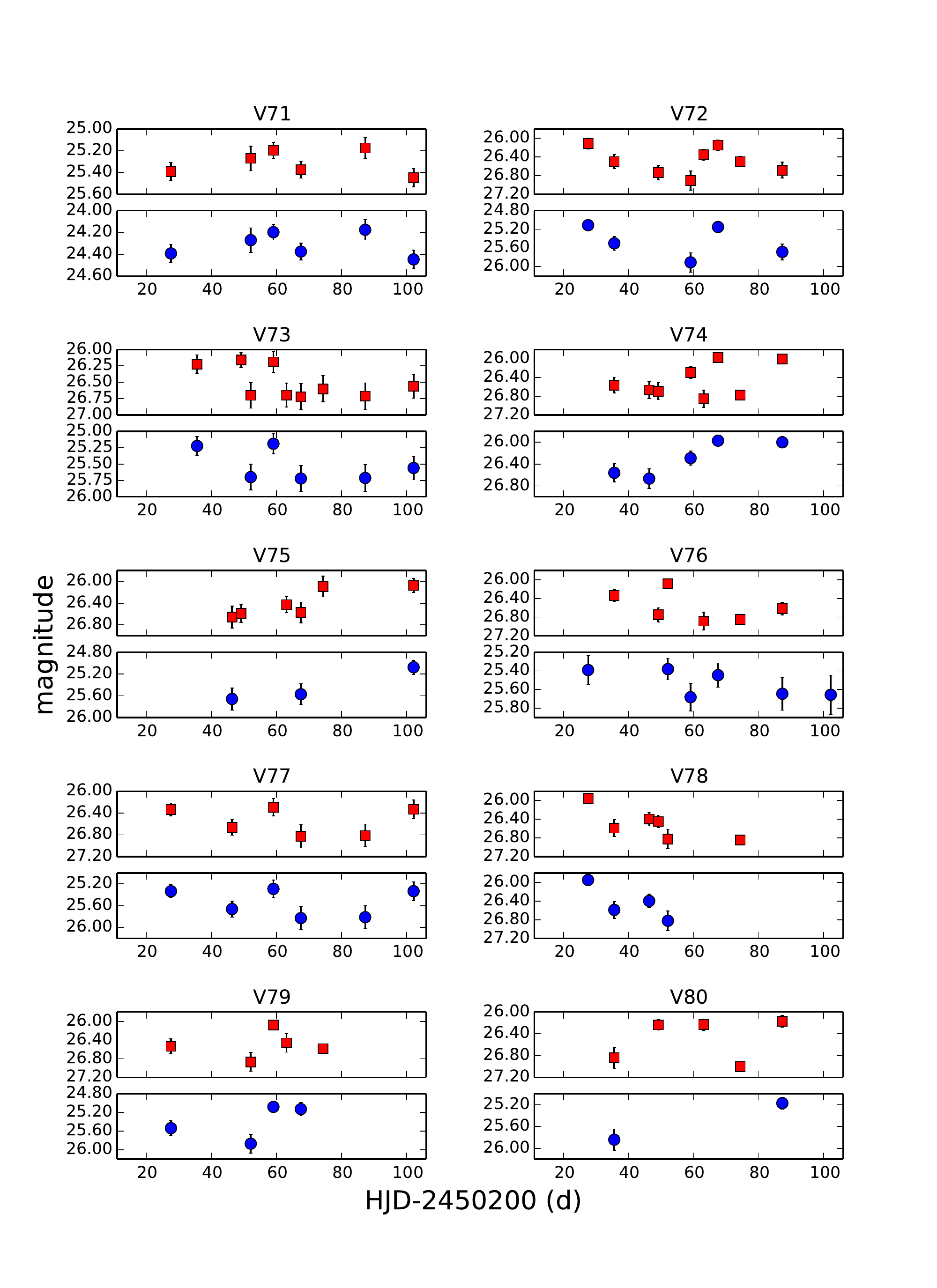}
\setcounter{figure}{10}
\end{tabular}

\caption{continued}
\label{lcs45482}
\end{figure*}

\clearpage
\begin{figure}[h!tb]
\vspace{0cm} 
\begin{tabular}{l c c }
\includegraphics[trim=0cm 0cm 0.cm 0cm, width=0.5\textwidth]{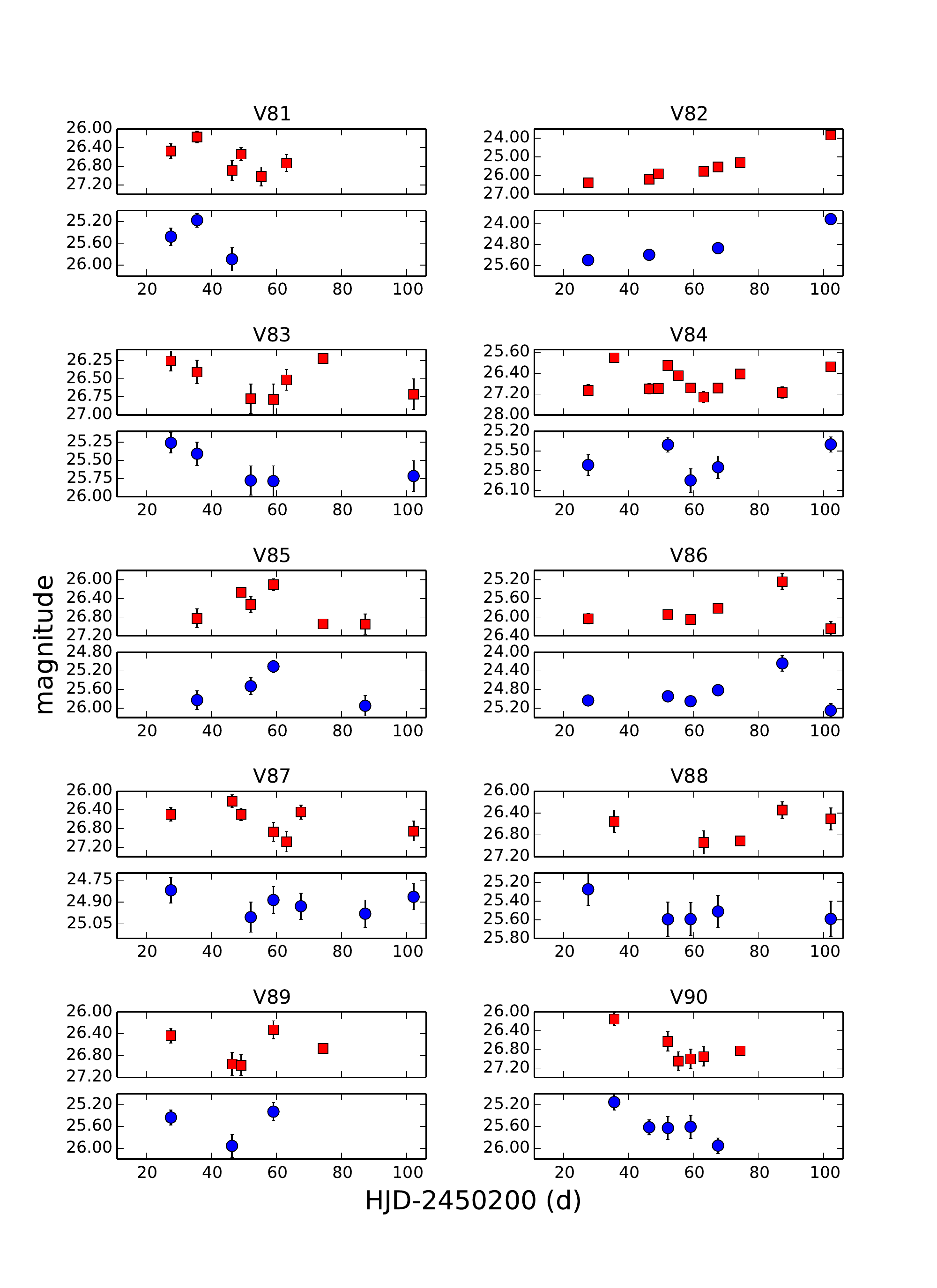} &
\includegraphics[trim=0cm 0cm 0.cm 0cm, width=0.5\textwidth]{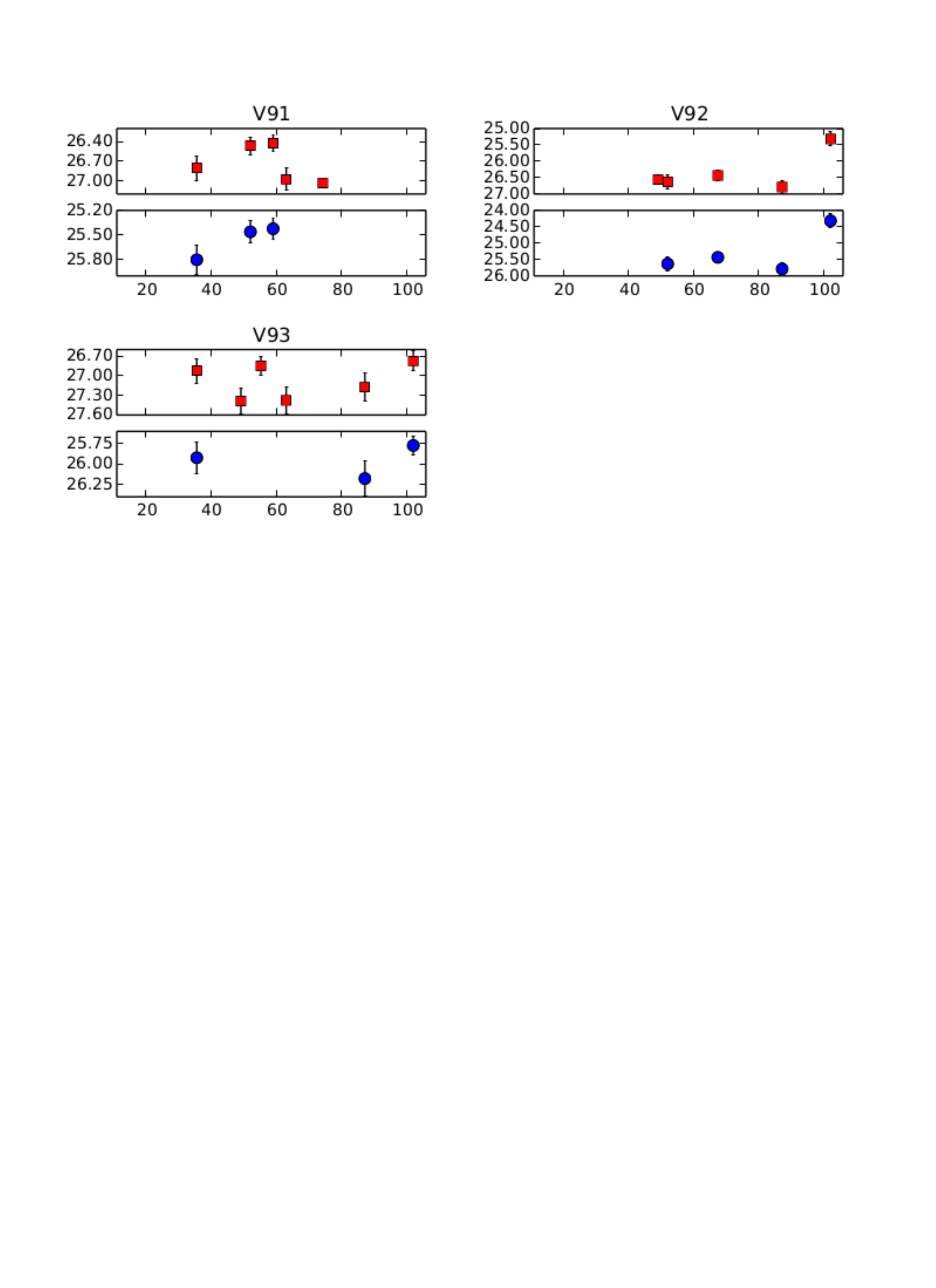}& \\
\setcounter{figure}{10}
\end{tabular}
\caption{continued}
\label{lcs45483}
\end{figure}

\end{document}